%% file: ccs-template.tex
\documentclass[letterpaper,twocolumn,10pt]{article}

\usepackage{usenix}

\makeatother
\usepackage{algpseudocode}
\usepackage[ruled,linesnumbered,boxed,commentsnumbered]{algorithm2e}
\usepackage{multirow} 

\usepackage{amssymb}
\usepackage{graphicx}
\usepackage{subfigure}
\usepackage{makecell}
\usepackage{threeparttable}
\usepackage{tikz}
\usepackage{amsmath}
\usepackage{appendix}
\usepackage{microtype}
\usepackage{float}
\begin{document}
%-------------------------------------------------------------------------------

\date{}

\title{High Recovery with Fewer Injections: Practical Binary Volumetric Injection Attacks against Dynamic Searchable Encryption}

\author{}
%{\rm Your N.\ Here}\\
%Your Institution
%\and
%{\rm Second Name}\\
%Second Institution
% copy the following lines to add more authors
% \and
% {\rm Name}\\
%Name Institution
%} % end author

\maketitle

\begin{abstract}
Searchable symmetric encryption enables private queries over an encrypted database, but it also yields information leakages. 
Adversaries can exploit these leakages to launch  
injection attacks {(Zhang et al., USENIX'16)} to recover the underlying keywords from queries. 
The performance of the existing injection attacks is strongly dependent on the amount of leaked information or injection.
In this work, we propose two new injection attacks, namely  BVA and BVMA, by leveraging a \textit{binary volumetric} approach.  
We enable adversaries to inject fewer files than the existing volumetric attacks by using the known keywords and reveal the queries by observing the volume of the query results.   
Our attacks can {thwart} well-studied defenses (e.g., threshold countermeasure, static padding) without exploiting the distribution of target queries and client databases.
{We evaluate the proposed attacks empirically in real-world datasets with practical queries.}  
The results show that our attacks can obtain a high recovery rate ($>80\%$) in the best case and a roughly $60\%$ recovery even under a large-scale dataset with a small number of injections ($<20$ files). 

\end{abstract}

\input{ccs-body-revised}

\bibliographystyle{plain}
\bibliography{ccs-sample}

\begin{appendices}
\section{{Summary of Notations and Concepts}}\label{AppenDesonPa}
    We denote $[n]$ as the set of integers $\{1,...,n\}$. For a set $S$, we use $\#S$ to refer to its cardinality. For a file $f$ (or file set $F$), we use $|f|_{w}$ ($|F|_{w}$) to represent its word count, also called file $size$. $\mathcal{A}\rightarrow{x}$ means that $x$ is the output of an algorithm $\mathcal{A}$. $\mathbb{N}$ denotes natural number. We use $\lambda$ to denote the security parameter. See the frequently used notations in Table \ref{Remark1}. 
    %We present the descriptions for leakage, injection, attack, and defense in Table \ref{Remark1}.
    
    \begin{table}[ht]
    	\center
    	
    	\caption{Notation and Concept} 
        	\label{Remark1}
        	\resizebox{\linewidth}{!}
        	{
        	\begin{tabular}{ll}
        		\hline
                {Notation} & Description\\
                \hline
        
        		$\mathbf{D}$ & An encrypted database.\\
        		
        		$\mathbf{W}$ & A keyword universe $\mathbf{W} = \{w_{1},w_{2},...,w_{m} \}$.\\
        		
        		$\mathbf{Q}$ & A sequence of queries $\mathbf{Q} = \{q_{1},q_{2},...,q_{l} \}$.\\
        		
        		$\mathbf{R}$ & The response to each query.\\
        		
        		$\mathbf{F}$ & A set of injected files.\\
        		
        		$|f|_w$ & Total word count for a file $f$.\\
        		
        		$file_{u}$ & Injected files containing a keyword $w_{u}$.\\
        
        		$\textit{offset}$ & Minimum file size for decoding.\\
        
        		$\gamma$ & Basic injection size for BVA.\\
        		
        		$k$ & Keyword partitions for multiple-round.\\
        
        		$m$ & Injection constant for single-round.\\
        		
        		$T$ & Size threshold of each file.\\
        		
        		$\mathbf{LP}$ & Leakage pattern.\\
        		
        		${\mathbf{RL}}$ & Response length ${\mathbf{RL}} = \{{rl}_{1},{rl}_{2},...,{rl}_{l} \}$.\\
        		
        		${\mathbf{RS}}$ & Response size ${\mathbf{RS}} = \{{rs}_{1},{rs}_{2},...,{rs}_{l} \}$.\\
        		
        		${\mathbf{Freq}}$ & Query frequency ${\mathbf{Freq}} = \{{freq}_{1},...,{freq}_{l} \}$.\\
        		
        		${\mathbf{Q}_{r}}$ & A query recovery set,  $\mathbf{Q_{r}}\subseteq \mathbf{Q}$.\\
        		\hline
                \hline
        		{Leakage Pattern} & Description\\
        		\hline
        		access pattern (\textit{ap}) & identifiers of the files matching a query.\\
        		
        		access injection pattern (\textit{aip}) & identifiers of injected files matching a query.\\
        		
        		search pattern (\textit{sp}) & whether a query is repeated.\\
        		
        		response length pattern (\textit{rlp}) & the number of (response) returned files.\\
        		
        		response size pattern (\textit{rsp}) & the total word count of (response) returned files.\\ 
        		
        		volume pattern (\textit{vp}) & include \textit{rlp} and \textit{rsp}.\\ 
        		\hline
        		\hline
        		{Injection Information} & Description \\
        		\hline
        		injection length (ILen) & the number of injected files.\\
        		
        		injection size (ISize) & the total word count of injected files.\\
        		
        		injection volume & include ILen and ISize.\\
        		\hline
        		\hline
        		{Attack} & Description \\
        		\hline
        	    passive attack & attacks only based on the observations.\\ %

        	    injection attack & attacks can actively inject files.\\ %
                
        	    volumetric injection attacks (VIAs) & injection attack relying on \textit{rlp} (or \textit{rsp}).\\
        	    
        	    \hline
        		\hline
        		{Defense} & Description \\
                \hline
        	    ORAM & oblivious retrieval to hide \textit{ap} and \textit{aip}.\\

        	    padding & index dummy files to obfuscate \textit{vp}.\\

        	    TC & limit file size to avoid large-size file injection.\\
        		\hline
        	\end{tabular}}
        \end{table}
   % \medskip

\section{{Searchable Symmetric Encryption}} \label{SectionSE} 
A standard dynamic SSE \cite{DBLP:conf/ccs/KamaraPR12} includes a polynomial-time algorithm, $Setup$, executed by the client and two protocols, $Query$ and $Update$, run between the client and the server.
\\
$\bullet$ $Setup(\lambda)$: This probabilistic algorithm takes the security parameter $\lambda$ as the input and outputs $(K,\delta;\mathbf{D})$, where $K$ and $\delta$ are the secret key and state of the client, respectively; and $\mathbf{D}$ is an empty encrypted database stored in the server.
\\	
$\bullet$ $Update(K,\delta,op,(w,id);\mathbf{D})$: The update protocol carries $K$, $\delta$, $op\in \{add,delete\}$, a keyword-file pair $(w,id)$, and $\mathbf{D}$ as input. 
It outputs a new client state $\delta'$ and an encrypted database $\mathbf{D}'$ after the ``add/delete" operation.
\\	
$\bullet$ $Query(K,\delta,q;\mathbf{D})$: This protocol takes the client secret key $K$, state $\delta$, and query $q$ as input, in which the server takes the database $\mathbf{D}$ as input. 
It returns $\mathbf{D}(q)$ as the query’s response.

In the query protocol, the client only retrieves the response identifiers. 
Later, it can perform an extra interaction with the server to obtain the encrypted files with the identifiers. 

\section{Definition for Attack Round}\label{AppenRound}
The multiple-round \cite{DBLP:conf/eurosp/PoddarWLP20} and search attacks \cite{DBLP:conf/ndss/BlackstoneKM20} recover one target query via many rounds of injections and observations. 
Recall that the adversary in the context of the multiple-round attack ``forces" the client to replay the same query repeatedly, in which ``completing a replay of the client query" is regarded as ``one attack \textit{round}". 
The search attack requires the adversary to carry out a sufficient amount of observations on the client's queries after a successful single file injection. 
In this context, an attack \textit{round} is referred to as ``the adversary completes a single file injection". 
In other attacks for multiple queries, the adversary aims to reveal the queries through an immediate observation right after multiple files injection. 
Here, completing a batch of file injections is also seen as an attack round.
We define the attack \textit{round} as follows.
    \newtheorem{myDef}{Definition}
    \begin{myDef}
        \label{label}
        Let ${OB}$ be a non-empty set (except the $OB_{0}$) of continuous observation operations and ${INJ}$ denote a non-empty set containing continuous ``inject" operations. 
        We denote $\Gamma=(OB_{0},INJ_{1},OB_{1},INJ_{2},OB_{2},...INJ_{t},OB_{t})$ as all the pairwise operations for an injection attack. Then, the total number of the attack round is $t$.
    \end{myDef}
    For example, in the search attack with a $\mathbf{W}$, the attack process is $\Gamma=(OB_{0}$, $INJ_{1}$, $OB_{1}$, $...$, $INJ_{\lceil\log\#\mathbf{W}\rceil}, OB_{\lceil\log\#\mathbf{W}\rceil})$. 
    Thus, the number of attack round is $\lceil\log\#\mathbf{W}\rceil$. 
    As for other attacks against multiple queries (e.g., single-round, decoding, and our attacks) their attack process is as $\Gamma=(OB_{0},INJ_{1},OB_{1})$ and thus, they only require \textit{one attack round}.

\section{An Example: BVA $vs.$ Decoding}\label{AppenComp}    
    
\begin{figure}[!t]
    	\subfigure[BVA injection]
    	{
    		\begin{minipage}{.42\linewidth}
    			
    			\centering
    			\includegraphics[width=\textwidth]{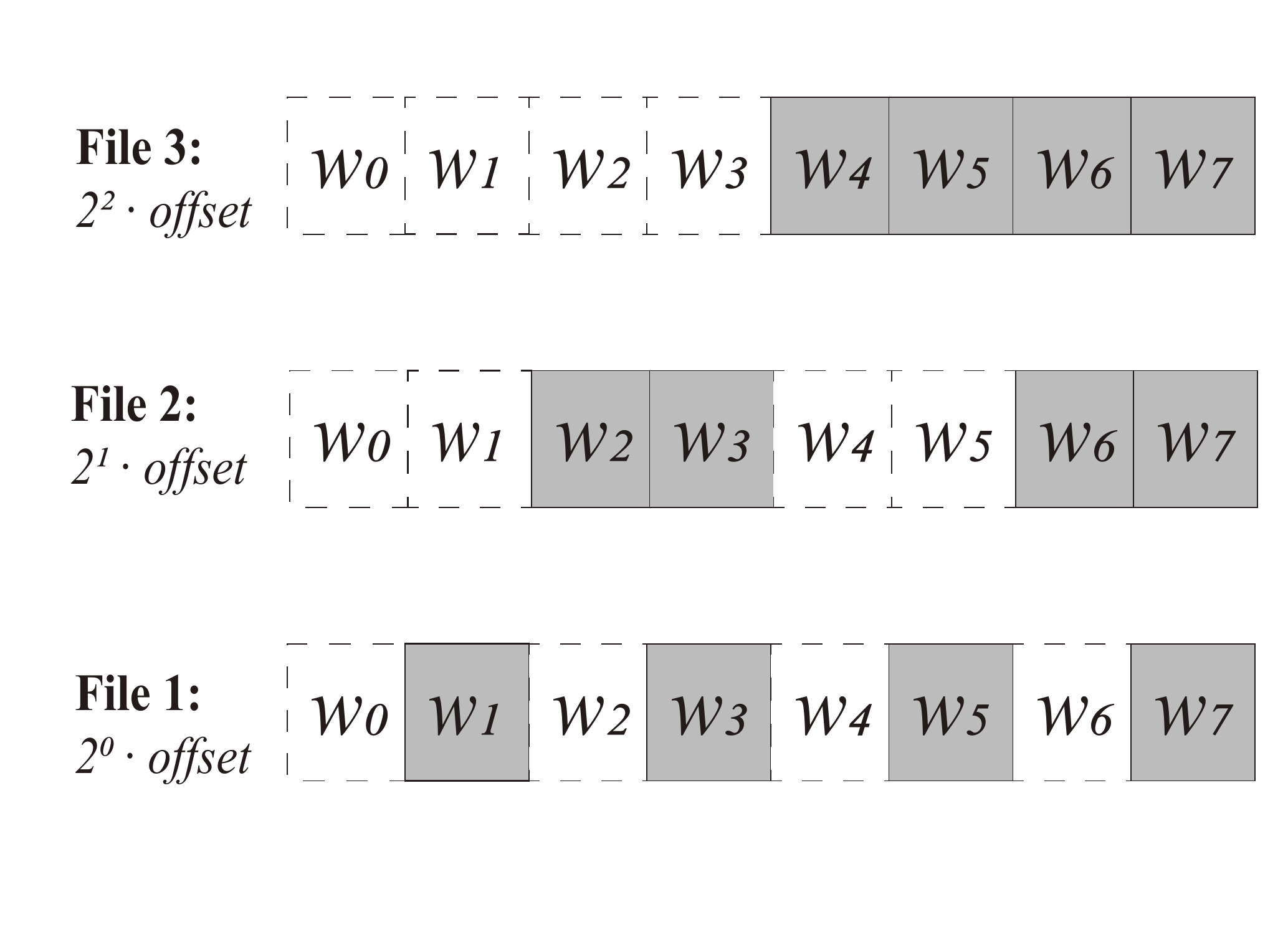}
    		\end{minipage}
    		
    	}
    	\subfigure[Decoding injection \cite{DBLP:conf/ndss/BlackstoneKM20}]{
    		\begin{minipage}{.42\linewidth}
    			
    			\centering
    			\includegraphics[width=\textwidth]{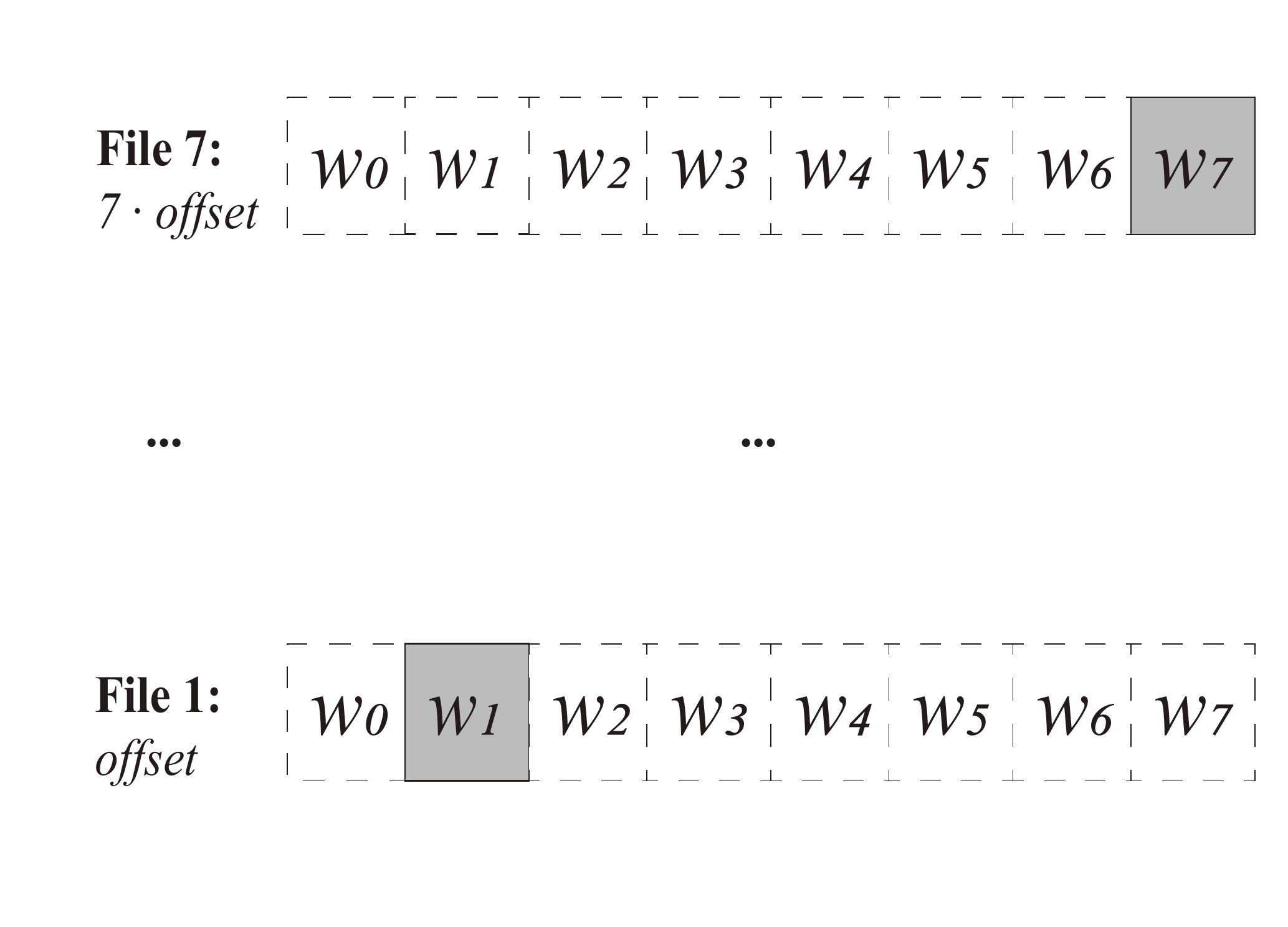}
    		\end{minipage}
    		
    	}
        \caption{An example: the adversary knows  $\#\boldsymbol{{W}}=8$. }
    	\label{Our-Decoding}
    \end{figure}

    In the example (see Figure \ref{Our-Decoding}), we assume there exists a special case where $\gamma=\textit{offset}$ so that the BVA and decoding attack can achieve the same recovery rate.  
    We see that the BVA only needs three files (each with a size of $2^{k-1}\cdot \textit{offset}$) as compared to the decoding attack injecting seven files (each with a size of $(t-1)\cdot \textit{offset}$), where $k\in [3]$ and $t\in [8]$. 
    Specifically, the BVA yields $7\cdot{\textit{offset}}$ injection size, while the decoding attack is $4\times$ of the cost of the BVA, i.e., $28\cdot{\textit{offset}}$. 
    
\section{{Effect of Search Pattern on BVMA}}\label{AppenBVMASP}
%The BVMA can definitely boost the recovery rate by exploiting the sp. 
We tested the effect of the sp on the recovery rate of the BVMA (see Figure \ref{BVMASPFIGURE}). 
%"BVMA\_SP" indicates that the attack uses the sp, while "BVMA\_NoSP" is without using it. 
By exploiting the sp (i.e. ``BVMA\_SP"), the attack slightly improves the recovery, around $8\%$ (under $100\%$ keywords leakage);   
without the sp (i.e. ``BVMA\_NoSP"), it still provides $81\%$ recovery. 

\section{Queries with Different Distributions }\label{AppenUQ}
\begin{figure}[!t]
	\centering
	\subfigure[Impact of sp on BVMA.]
	{
		\begin{minipage}{.415\linewidth}
			\centering			
            \includegraphics[width=\linewidth]{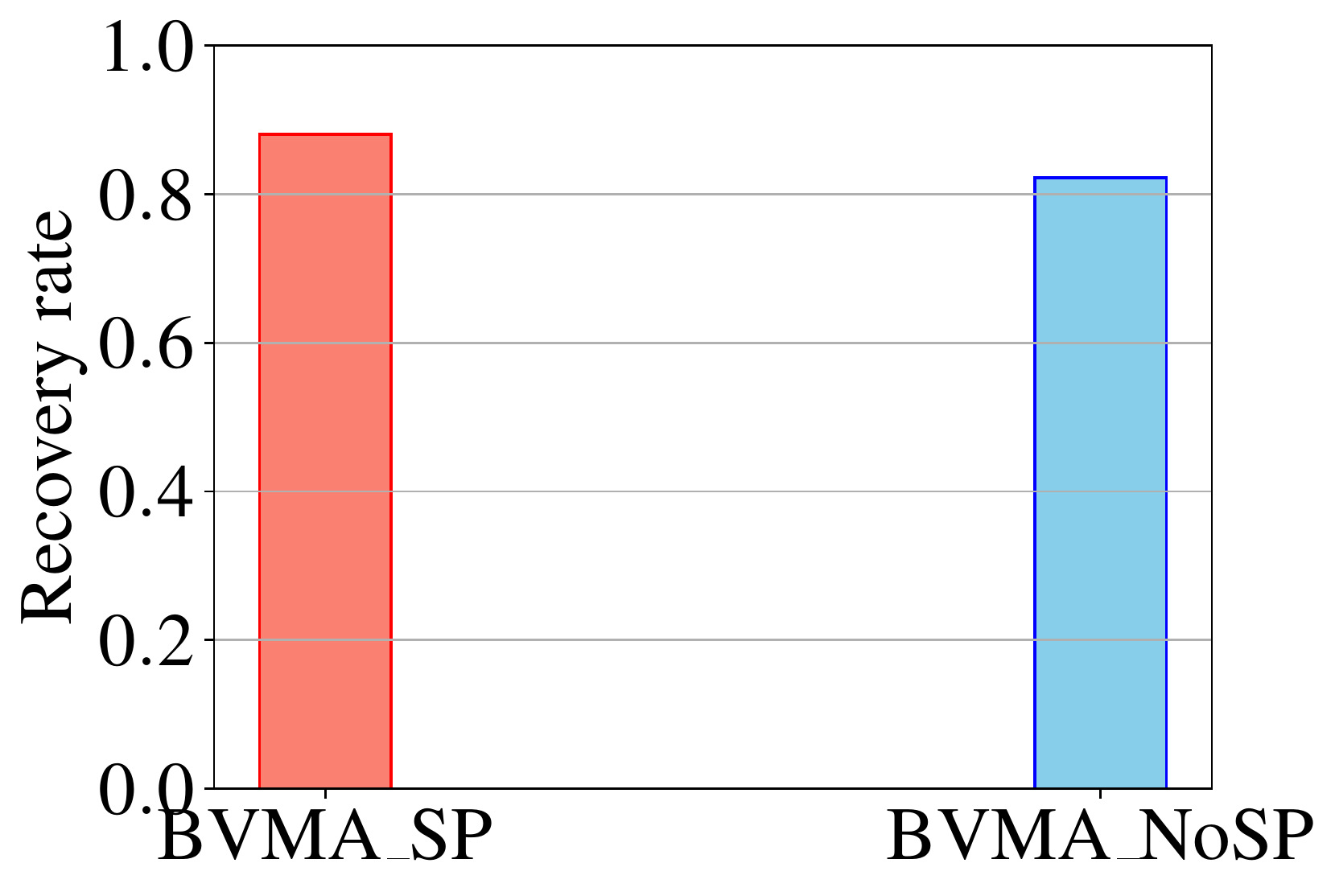}
            \label{BVMASPFIGURE}
		\end{minipage}
	}
	\subfigure[BVA and BVMA under different query distributions.]
	{
		\begin{minipage}{.39\linewidth}
			\centering
			\includegraphics[width=\linewidth]{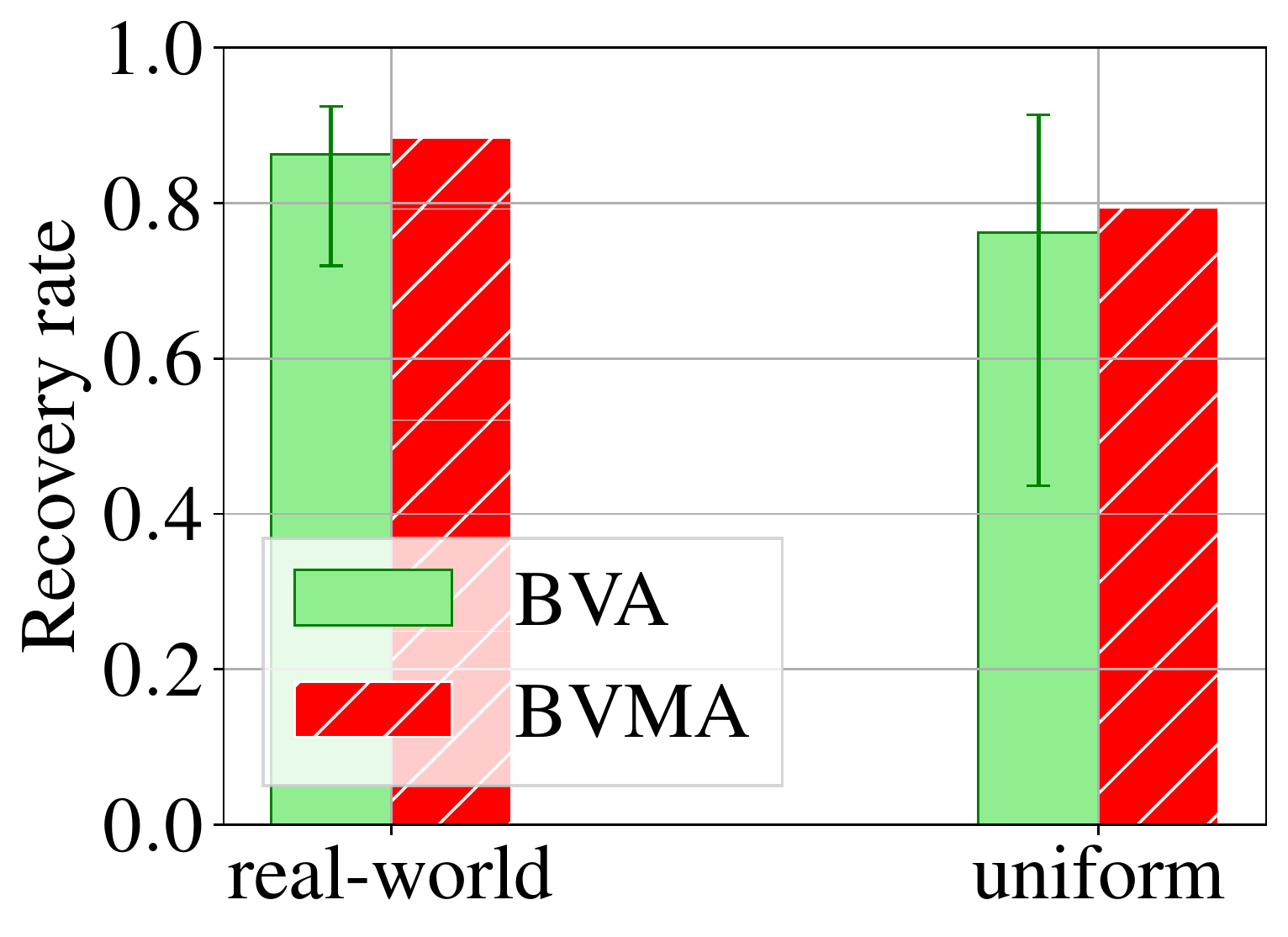}
			\label{TrendUniformEnron}
		\end{minipage}
	}
	\caption{Supplementary experiments of BVA and BVMA.}
	\label{EBVABVMA}
    \end{figure}

    {We used Google Trends and PageViews Toolforge to simulate the queries in the real-world distribution. 
    Alternatively, one may use other query distributions, such as uniform query. 
    In this section, we tested the recovery rate under two query distributions: real-world and uniform} (see Figure \ref{TrendUniformEnron}).  
    We assumed that the adversary knows $\mathbf{W}$ and has spent at least 8 weeks on observations in Enron.  
    We set error bars for the BVA to evaluate the recovery when $\gamma\in[\#\mathbf{W}/2,\textit{offset}/4]$.

    The recovery rate of the BVA varies moderately (about a $20\%$ gap) under the real-world query, but fluctuates (e.g., $<50\%$ recovery in the worst case and $>90\%$ in the best case) under the uniform distribution. 
    This indicates that the BVA with a small $\gamma$ (e.g., $\gamma=\#\mathbf{W}/2$) cannot perform stably in uniform query. 
    The average recovery rates on the two distributions only differ approximately $10\%$. 
    The gap could nearly disappear if we set $\gamma$ to be sufficiently large (e.g., $\gamma=\textit{offset}/4$) to yield the max. recovery under both distributions. 
    %The max. recovery of two query distributions are almost the same. 
    Recall that besides the frequency information, the BVMA relies on two types of volume patterns for query recovery. 
    The adversary can distinguish queries and exclude {the incorrect keywords} by exploiting the patterns.  
    The uniform query (hiding frequency information) only harms the recovery of the BVMA by $10\%$ as compared to the real-word query. 
    
    The distributions do not seriously affect our attacks' average recovery rate ($<10\%$).   
    The attacks perform slightly better on the real-world distribution than the uniform. 
    This is because under the real-world query, the adversary can always recognize its target query of which frequency is higher than others. 
    In contrast, the uniform distribution makes the adversary difficult to distinguish the target query, incurring the drop of the recovery rate.

    %In conclusion, different query distributions may slightly influence the average recovery rate of our attacks.

\section{Comparison in Lucene and Wikipedia}\label{AppenCLW}

\begin{figure*}[!t]
	\centering
	
	\subfigure[Recovery accuracy]
	{\label{LuceneTrendRerVol:Rer}
		\begin{minipage}{.26\linewidth}
			\centering
			\includegraphics[width=\linewidth]{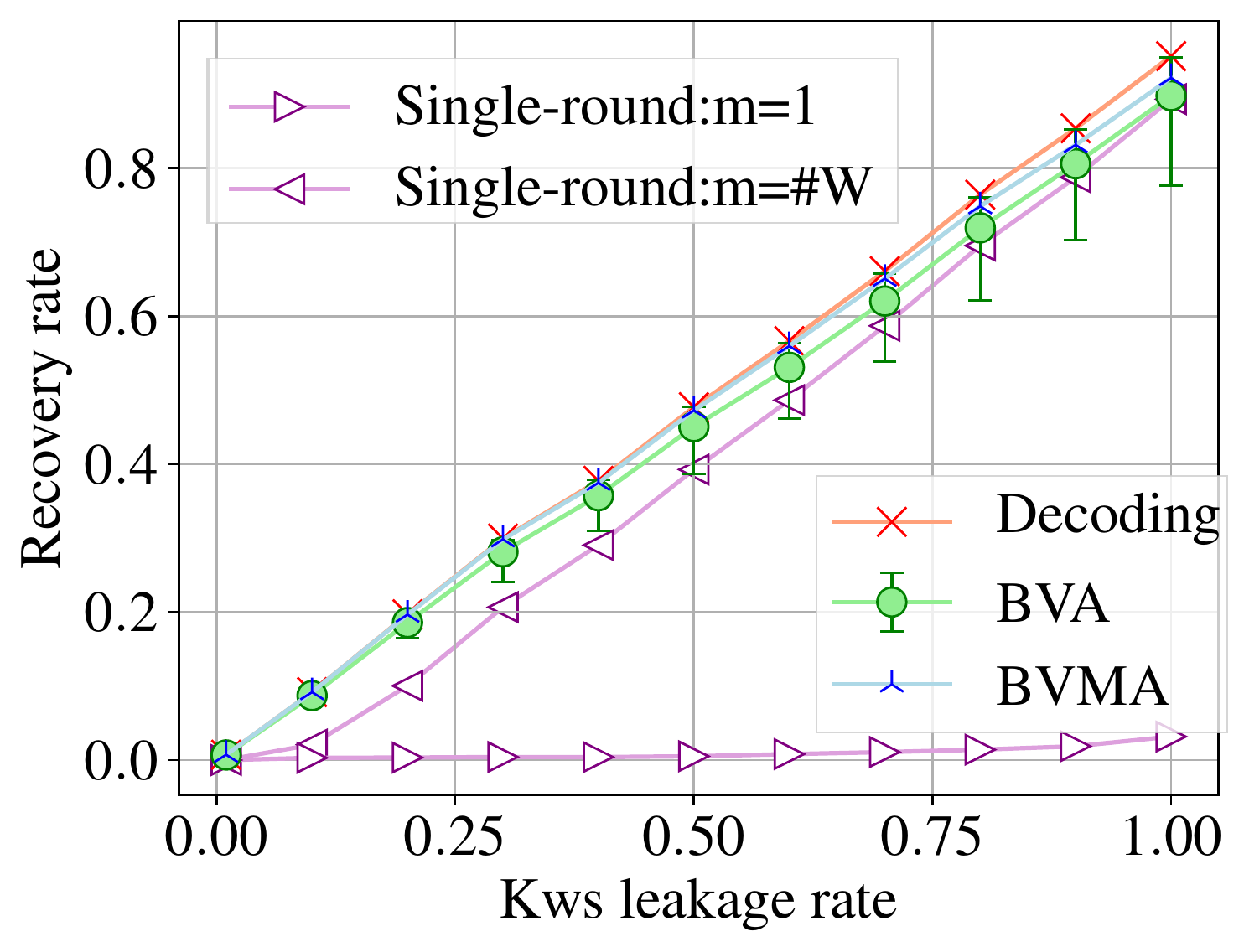}
		\end{minipage}
	}
	\subfigure[Injection length]
	{\label{LuceneTrendRerVol:ILen}
		\begin{minipage}{.26\linewidth}
			\centering
			\includegraphics[width=\linewidth]{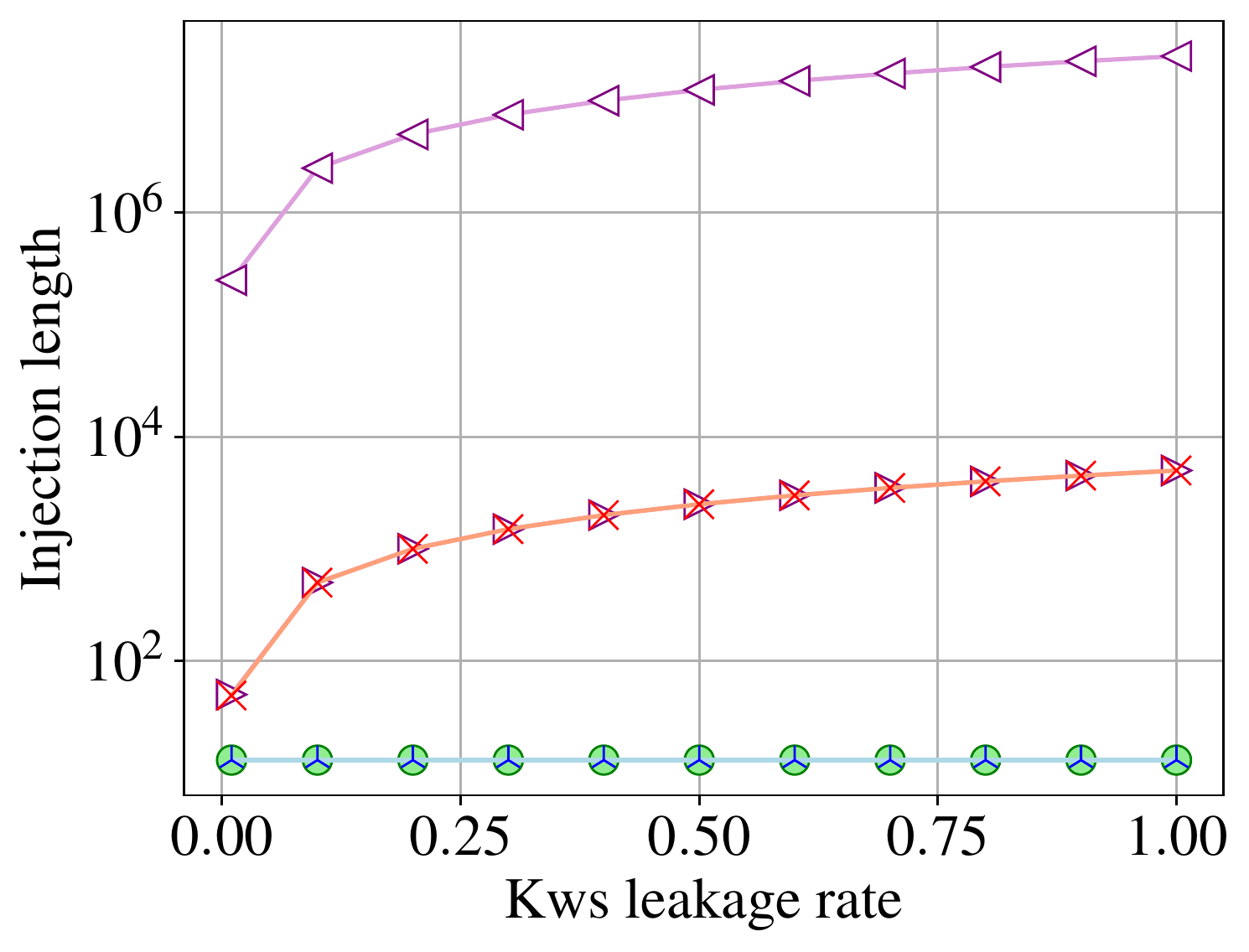}
		\end{minipage}
	}
	\subfigure[Injection size]
	{\label{LuceneTrendRerVol:ISize}
		\begin{minipage}{.26\linewidth}
			\centering
			\includegraphics[width=\linewidth]{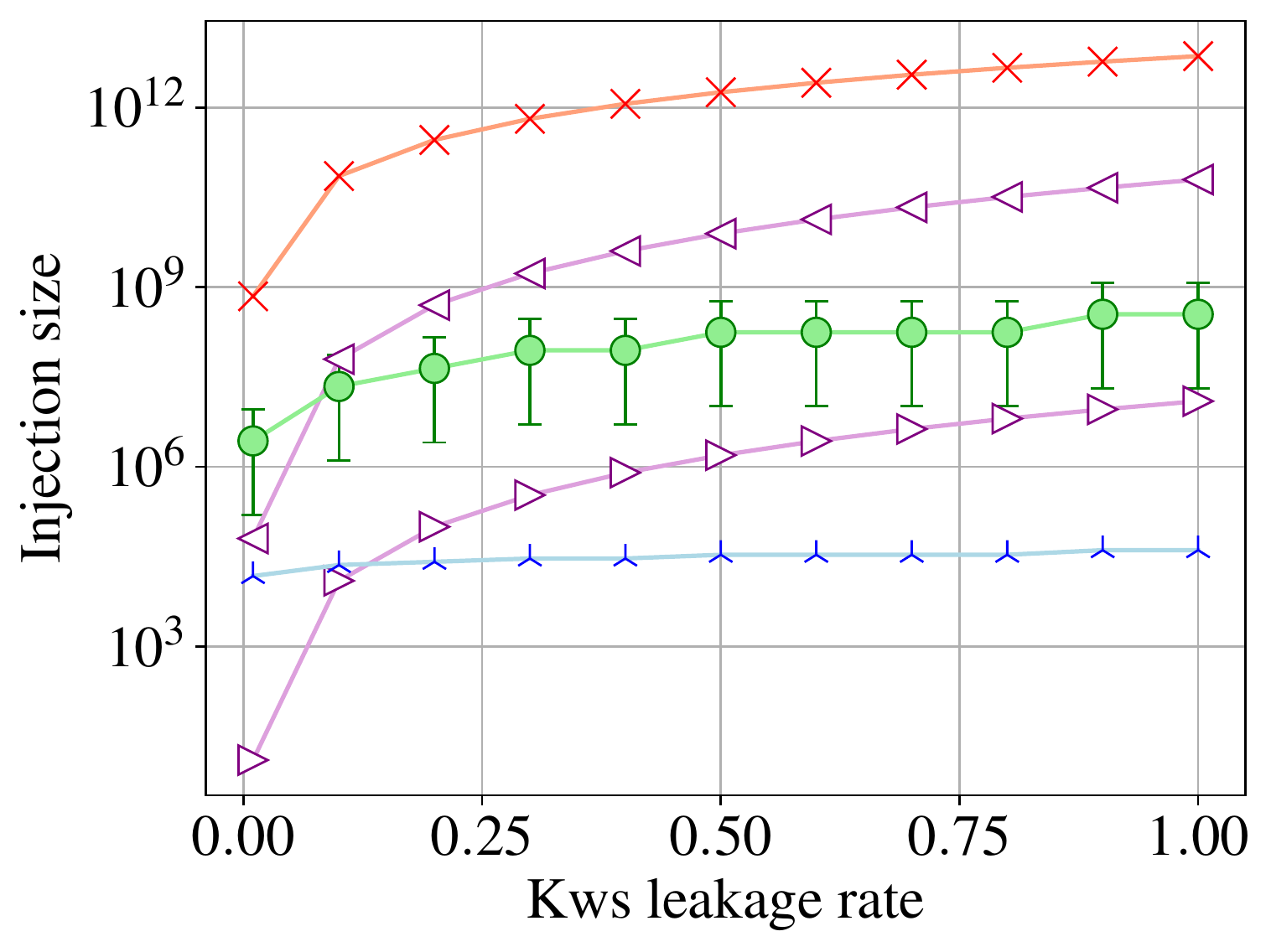}
		\end{minipage}
	}
	\caption{Comparison on the recovery rate, injection length, and injection size with different Kws leakages in \textbf{Lucene}.}
	\label{LuceneTrendRerVol}
\end{figure*}

    We show the comparison among the single-round, decoding and our attacks in Lucene and Wikipedia in Table \ref{RunningTimeLucene}-\ref{RunningTimeWiki} and Figure  \ref{LuceneTrendRerVol}-\ref{WikiTrendRerVol}.
    The single-round attack requires the least time cost ($<1s$) but yields the worst recovery (when $m=1$) or injection length (when $m==\#W$) in Lucene.  
    The decoding attack and our attacks are at a comparable performance level in terms of recovery and running time. 
    % , less than $10\%$ advantage on average in Lucene
    In Wikipedia, our attacks achieve around $60-80\%$ recovery in which the BVA and BVMA take around 3 and 15 mins respectively.
    For injection length, other attacks inject $10^{2}\times$ more files than ours. % 
    The BVMA only requires $10^{6}$ injection size in Wikipedia. 
    To maintain a similar level of recovery (e.g., $Kws = 0.25$), the decoding and single-round attacks (with $m=\#\mathbf{W}$) cost at least $10^{6}\times$ more injection size than the BVMA. 
    We see that the BVMA in Wikipedia (up to $60\%$ recovery) does not perform as well as in Enron and Lucene.  
    %{The BVMA provides around  in Wikipedia, whilst it performs well in .   %indicates that it may not be applicable to large-scale databases. 
    If it performs on the same injection size as the BVA (but still much $<$ that of the decoding and single-round attacks), its recovery will be close to the BVA's. %}

% % AppenBVMASP, AppenCLW, AppenDYSEAL        
\section{{Attacks against Extended SEAL}}\label{AppenDYSEAL}
Extended SEAL fills the total \textit{rlp} of the corresponding keyword into $x^{t}$ after every batch update.
Table \ref{DynamicSEALT} presents our attacks against the extended SEAL.   
The results indicate that the extension can properly resist VIAs ($<1\%$ Rer) by sacrificing a huge amount of overhead. 
For example, when $x=2$, to counter our attacks with 12 injected files, the extended SEAL should use around $33,000$ extra files ($2,730\times$ of the injected files) to obfuscate the injection volume. 
The cost of I-Query also increases by $1.8\times$. %
%{\color{blue}We leave the optimization on the overhead of the extended SEAL as an open problem.} 

\begin{algorithm}[t] 
	\footnotesize
    \caption{Optimization against ShieldDB} 
    \label{AlgoABVIA}
    \SetKwFunction{FMain}{OP\_against\_ShieldDB}
    \SetKwProg{In}{procedure}{}{end}
    \In{\FMain{${\widetilde{Q}}, {W}, {Q}, \alpha, t$}}{
        \tcp{Baseline}
        Pick up the keywords $CW[i,j]$ which represents the $j$th keyword in the $i$th cluster {\tcp*{adversary can obtain this information based on other knowledge or additional injection}}%, and identify each cluster with response length ${RLC}$\; 
        % $C[0],..., C[\#{W}/\alpha]$
 
        Observe the leakage information ${\widetilde{RS}}$ as in \hyperref[BVA-Baseline]{\textit{BVA.Baseline}}$({\widetilde{Q}})$\;
        
        \tcp{Injection}
        ${Inj\_F}\gets\emptyset$\;
        Select $t$ co-prime numbers $\Gamma[1],...,\Gamma[t]$ satisfying that $\{\Gamma[k]\geq{\#{W}/(2\alpha)}\ |\ k\in[t]\}$\;
		\For{$j=1 \to t$}{
		    inject files ${F}$ by calling algorithm \hyperref[BVA-Injection]{\textit{BVA.Injection}}($CW[:,j]$) but replace the injection parameter $\gamma$ with $\Gamma[j]$\;
		    ${Inj\_F}={Inj\_F}\cup{F}$\;
		}
        
        \tcp{Padding}
        Pad ${Inj\_F}$ to ${IPad\_F}$ according to the padding strategy\;
        Upload ${IPad\_F}$ to server\;
        Record the total response length ${TRL}$ of each keyword after injection and padding {\tcp*{adversary can obtain this information with the help of ${CW},{Inj\_F}$}}
        
        \tcp{Recovery}
        initialize an empty set ${Q}_{r}$\;
        gather the new observed response size ${RS}$, response length ${RL}$ for victim's target queries ${Q}$\; 
		
		\For{$i=1 \to \#{{{RS}}}$}{
		    ${CA}\gets\emptyset$\;
		    add keywords to ${CA[:,:]}$ from ${CW}[j,:]$ satisfying that ${TRL}[j]\equiv{{RL}[i]},j\in[\#{CW}]$\; 
		    
			find $CW[v,u]$ satisfying that ${{rs}}_{i}-v\cdot \Gamma[u] =\widetilde{rs}_{j}$ for $CW[v,u]\in{CA},\exists{\widetilde{rs}_{j}\in{\widetilde{RS}}}$\;
			add $CW[v,u]$ to ${Q}_{r}$\;
		}
		\KwRet ${Q}_{r}$\;    
    }
    
\end{algorithm}
\section{{Optimization against ShieldDB}}\label{AppenOA}
\begin{figure*}[!t]
	\centering
	\subfigure[Recovery accuracy]
	{\label{WikiTrendRerVol:Rer}
		\begin{minipage}{.26\linewidth}
			\centering
			\includegraphics[width=\linewidth]{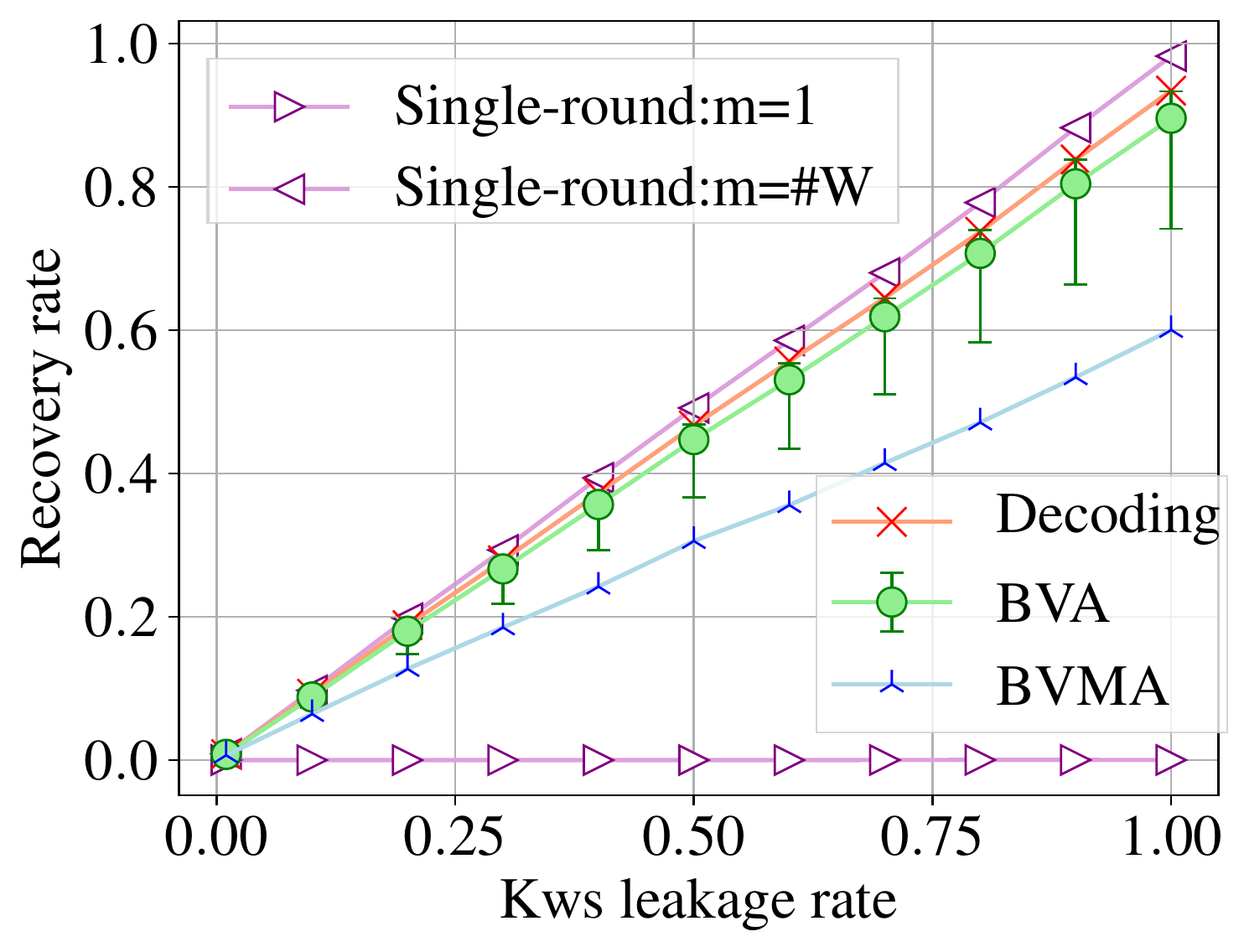}
		\end{minipage}
	}
	\subfigure[Injection length]
	{\label{WikiTrendRerVol:ILen}
		\begin{minipage}{.26\linewidth}
			\centering
			\includegraphics[width=\linewidth]{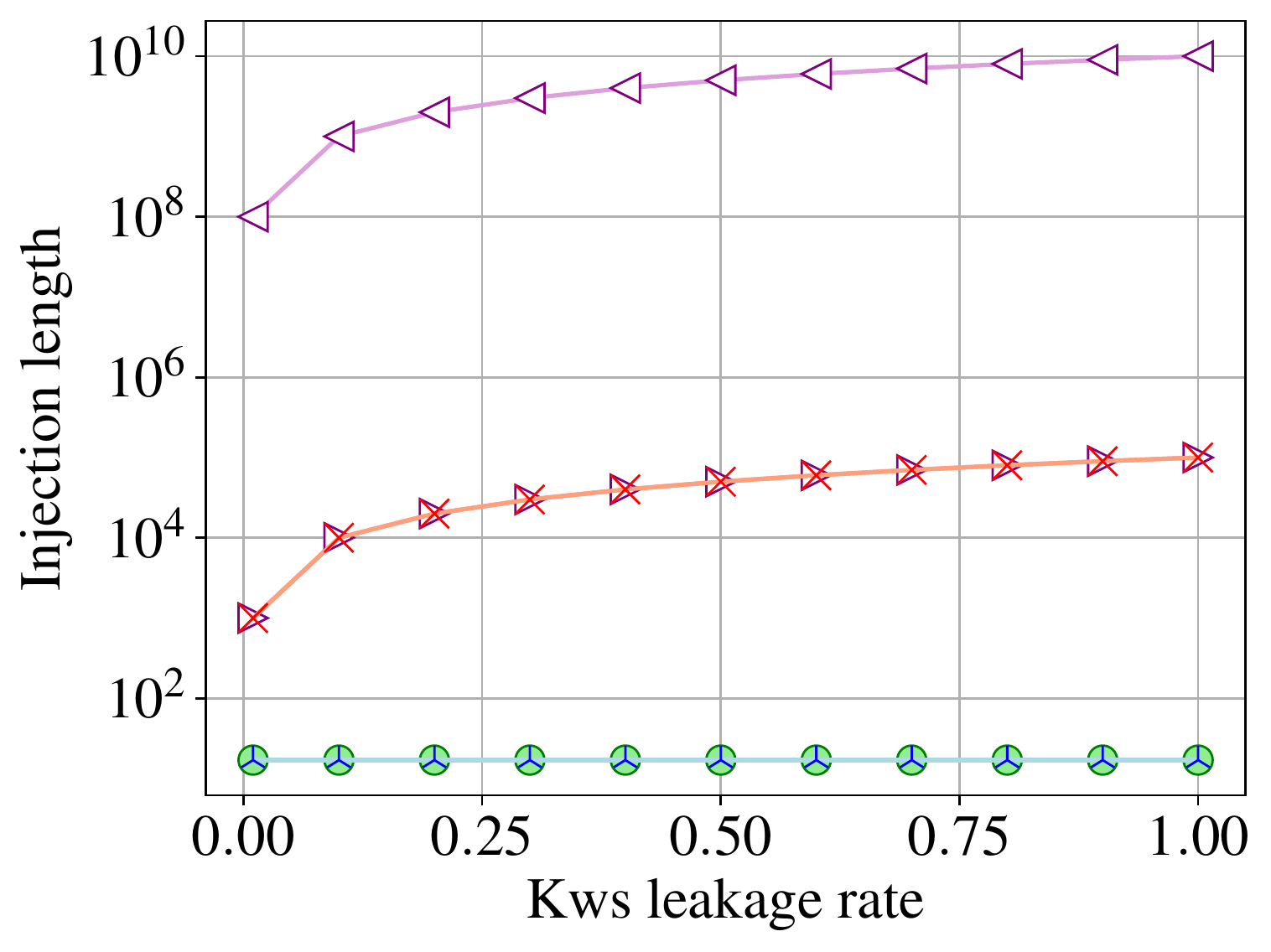}
		\end{minipage}
	}
	\subfigure[Injection size]
	{\label{WikiTrendRerVol:ISize}
		\begin{minipage}{.26\linewidth}
			\centering
			\includegraphics[width=\linewidth]{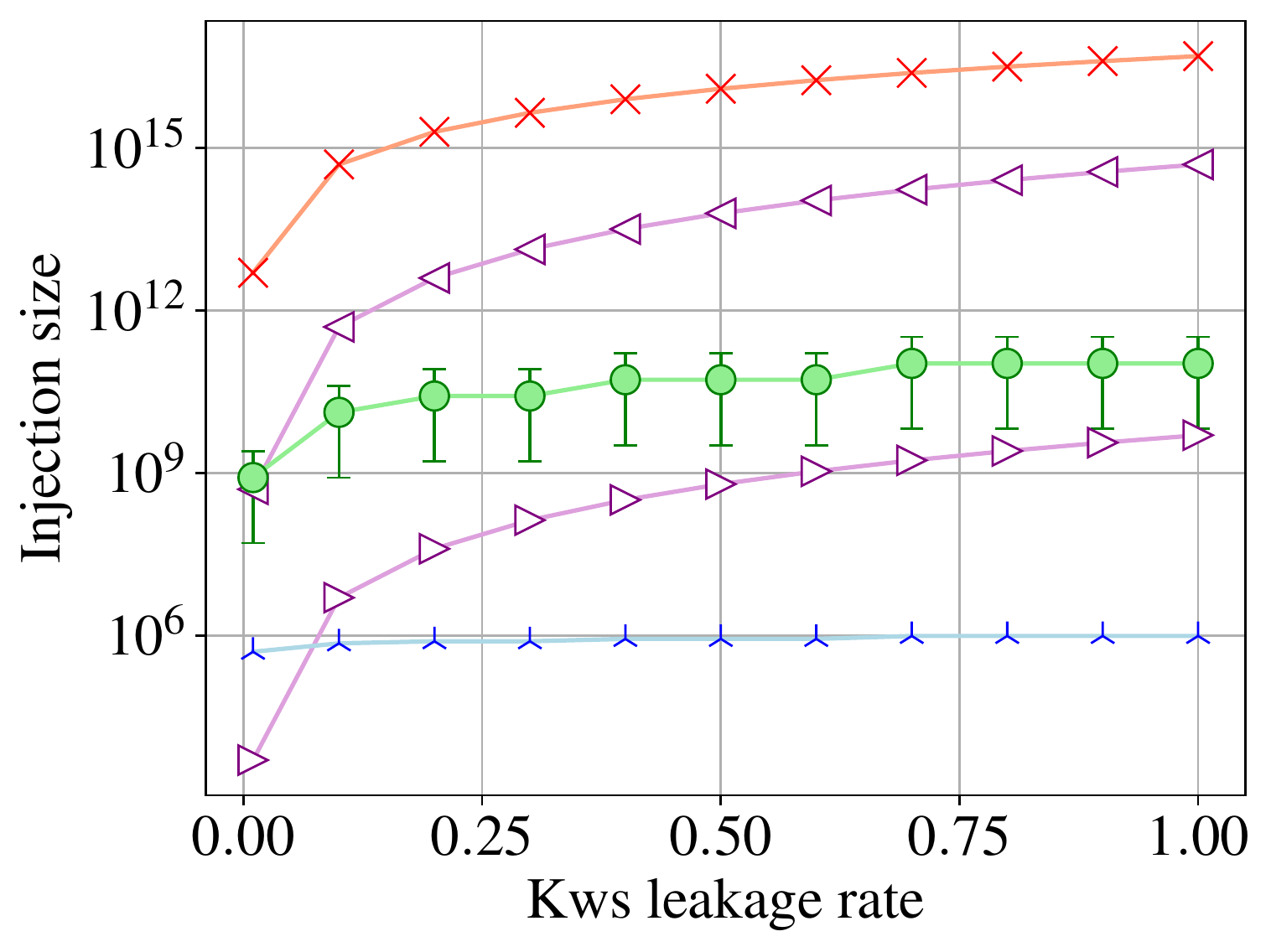}
		\end{minipage}
	}
	\caption{Comparison on the recovery rate, injection length, and injection size with different Kws leakages in \textbf{Wikipedia}.}
	\label{WikiTrendRerVol}
\end{figure*}

    We developed an optimization (see Algorithm \ref{AlgoABVIA}) against ShieldDB. 
    The optimization is an extension of the BVA and BVMA. 
    It starts with the BVA injection method (coding injection) and further optimizes the method by keyword grouping. 
    Following the BVMA, the optimization uses the \textit{rlp} to determine the keyword clustering.
    There are three main stages: baseline, injection, and recovery. 
    In the baseline phase, the algorithm determines the cluster $CW$ at which each keyword is located (as described previously). 
    Here, each row  $CW[i,:]$ represents a set containing all the keywords in cluster $i$, and each column $CW[:,j]$ represents the keyword set under the same position ($j)$ of different clusters. 
    The algorithm records the \textit{rlp} referred to as $\widetilde{RL}$ and \textit{rsp} as $\widetilde{RS}$. 
    In the injection phase, it chooses $t\ (t\leq\alpha)$ groups of keywords, in which each group contains the keywords in a column of $CW$. 
    It selects different $\gamma$ (from the set $\Gamma$) for the groups to launch the binary encoding file injection (which is similar to the BVA).  
    Such a grouping approach ensures that the target keywords in the same cluster share the same number of injected files. 
    In the recovery phase, given a target query $q_{i}$, the algorithm first identifies the candidate clusters by using $q_{i}$'s \textit{rlp} and then collects the correct keywords $CW[v,u]$, from the clusters, satisfying the following: 
    $
        {{rs}}_{i}-v\cdot \Gamma[u] =\widetilde{rs}_{j},if \  \exists{\widetilde{rs}_{j}\in{\widetilde{RS}}}, \nonumber
    $
    where $rs_{i}$ is the \textit{rsp} of $q_{i}$, and $\Gamma$ is the injection parameter set. 
    Since the optimization is based on both BVA and BVMA, we could predict from Figure \ref{WikiTrendRerVol} that its performance is close to that of the BVA, e.g., $>80\%$ recovery in Wikipedia.

    \begin{table*}[!t]
	\caption{Running time of recovery algorithm for $10\times1,000$ queries in Lucene.} %{\color{blue}Note the cost of the BVA is within the time range by varying $\gamma=[\#\mathbf{W}/2, \textit{offset}/4]$.}} 
	\label{RunningTimeLucene}
	\centering
	
	\setlength{\tabcolsep}{3.6mm}{\begin{tabular}{lccccc}
		\hline
		 & Decoding & Single-round $(m=1)$ & Single-round $(m=\#{W})$ & BVA & BVMA\\
		\hline
		Running time $(s)$ & 3.54 & 0.01 & 0.02 & (2.53, 3.15) &33.10\\
		\hline
		% {\color{red}[what do you mean???average??mean???]}
	\end{tabular}}
    \end{table*}

\begin{table*}[!t]
	\caption{Running time of recovery algorithm for $10\times5,000$ queries in Wikipedia. }
	\label{RunningTimeWiki}
	\centering
	
	\setlength{\tabcolsep}{3.6mm}{\begin{tabular}{lccccc}
		\hline
		 & Decoding & Single-round $(m=1)$ & Single-round $(m=\#{W})$ & BVA & BVMA\\
		\hline
		Running time & $3min$ $42s$ & $0.05s$ & $0.06s$ & ($2min$ $52s$, $3min$ $37s$) & $15min$ $34s$\\
		\hline
		
	\end{tabular}}
    \end{table*}
    
\begin{table*}[!t]

    \begin{threeparttable} 
	    \caption{Performance under the extended SEAL. The overhead ($\times$ /  No.) represents (the increased overhead as compared to no padding; and the number of files). Setup\&Fill and Inj\&Fill are the storage overheads on the database setup and file injection, respectively. S-Query and I-Query are the query bandwidth overheads before and after Inj\&Fill, respectively.}  
		\label{DynamicSEALT}
		\centering
		%\resizebox{\textwidth}{12mm}
		%\resizebox{\linewidth}{!}
		\setlength{\tabcolsep}{5mm}{
        \begin{tabular}{lcccccc}
			\hline
			% \multirow{2}{*}{Name}，2为所占的行数，此语句可以使得内容垂直居中
			% \multicolumn{2}{c|}{Flag}，2为所占的列数，格式由第二个{}控制
			% 
			%\cline{2-3} %指本行的2,3列画横线
			\multirow{2}{*}{Extended SEAL} & \multicolumn{4}{c}{Overhead ($\times$ / No.)} & \multicolumn{2}{c}{Recovery rate $\%$}\\
			\cline{2-7}
			%\cline{2-8}
			
			& Setup\&Fill & S-Query & Inj\&Fill & I-Query & BVA & BVMA \\ 
			\hline
			no padding  & $0$ / $30k$ & $0$ / $730$ & $0$ / $12$ (injection) & $0$ / $735$ & $70$ & $87$\\
			\hline
			$x=2$ & $0.5$ / $45k$ & $0.4$ / $1,058$ & $2,730$ / $33k$ & $1.8$ / $2,106$ & $<1$ & $<1$\\
		    \hline
			$x=3$ & $0.5$ / $44k$ & $0.8$ / $1,331$ & $3280$ / $40k$ & $4.5$ / $3,981$ & $<1$ & $<1$\\			
			\hline
			$x=4$ & $1.6$ / $79k$ & $1.2$ / $1,595$ & $16,384$ / $197k$ & $7.7$ / $6,316$ & $<1$ & $<1$\\
			%IKK %\cite{DBLP:conf/ndss/IslamKK12} & ap & \checkmark & $\times$ & -- & -- & --\\
			%\hline
			\hline
		\end{tabular}
}
    \end{threeparttable} 
\end{table*}

\section{Other Countermeasures}
    Besides the clustering-based padding strategies \cite{DBLP:conf/ccs/CashGPR15, BostCLuster, ShieldDB, TwoLayer, DBLP:conf/uss/DemertzisPPS20}, one may consider using other defense systems.  
    For example, some PIR (e.g., DORY \cite{DBLP:conf/osdi/DautermanFLPS20}) and ORAM (e.g., \cite{DBLP:conf/crypto/KamaraMO18, DBLP:conf/eurocrypt/GeorgeKM21}) related techniques focus on protecting the \textit{ap} and \textit{sp} (besides the \textit{vp}). 
    These solutions could not be as simple and natural as a direct padding (on \textit{vp}).  
    {Chen et al. \cite{DBLP:conf/infocom/ChenLRZ18} obfuscated the \textit{ap} and \textit{vp} by adding random false-positive and false-negative files, but this approach cannot protect the \textit{sp}.}
    Patel et al. \cite{DBLP:conf/ccs/PatelPYY19} and Wang et al. \cite{DBLP:conf/ccs/WangSLQ022} introduced the volume-hiding encrypted multi-maps with low server storage.
    Shang et al. \cite{DBLP:conf/ndss/ShangOPK21} proposed to hide the \textit{sp} by obfuscating the search token. 
    As these technologies do not protect the leakage in the dynamic context, they cannot work properly against VIAs. 
    Kamara and Moataz \cite{DBLP:journals/iacr/KamaraM18} investigated the dynamic volume-hiding system. 
    But their approach does not support the client to make \textit{atomic} update, e.g., adding and deleting a \textit{single} keyword-file pair; and it also requires a high query bandwidth. 
    \cite{DBLP:journals/iacr/AmjadPPYY21,DBLP:journals/iacr/ZhaoWL21} extended KM \cite{DBLP:journals/iacr/KamaraM18} and PPYY \cite{DBLP:conf/ccs/PatelPYY19} to propose fully dynamic volume-hiding encryption systems, respectively. 
    They can resist most of the query recovery attacks with a price that the query complexity is proportional to the maximum \textit{rlp}  $(O(rlp_{max}))$. %, thus greatly reduces their efficiency.
    %However, as shown by Oya et al. \cite{DBLP:conf/uss/OyaK21}, with the padding mode of PPYY, the adversary can still achieve more than $20\%$ recovery rate through adaptive volume-matching attacks. In a dynamic scenario, the weakness of \cite{DBLP:journals/iacr/AmjadPPYY21} may be further amplified because the adversary can obtain additional information through injection.
    %
    Xu et al. \cite{TwoLayer} combined the file padding (to hide file size) and the traditional index padding (to hide response length). 
    Through clustering and refilling files, their design ensures that files in a cluster remain the same size, which obfuscates the size of files.
    % and may resist VIAs {\color{blue} [e.g., what???our attacks???]}
    However, the adversary can simply counter this file padding strategy by generating a sufficient amount of injected files, in which each file is set to the same size. 
    We say that an effective countermeasure to VIAs should be $hybrid$ and $probabilistic$, i.e., being able to hide both file size and response length by random (or differentially private) noisy padding. 
    
    It is also interesting to see that there are some systems, e.g., \cite{Cetus, Concealer}, applying trusted hardware, e.g., SGX, to hide the volume leakage. 
    We say that is orthogonal to this work.

\section{Attack against Frequent Updates}\label{Section AFU}
We optimized BVA to limit the impact of updates while maintaining the $O(log\#W)$ injection length. 
But this could increase the injection size. 
Please see the details below, and we'll put the heuristic strategy in discussions on the paper. 
\begin{itemize}
\item 1) Baseline: the adversary observes the queries' result and finds the largest rsp (denoted as $rsp_{max}$). 
It predicts (or sets) the total size of all files that the user may update during the attack (denoted as $Tsize_{upd}$).

\item 2) Injection: the adversary sets $\gamma>rsp_{max}+Tsize_{upd}\wedge\gamma\geq\#W/2$, and then injects files according to BVA's binary injection strategy;

\item 3) Recovery: for a target query q, the adversary observes its injected (and updated) response size (denoted as $rsp_{q}$), then it can recover query $q$ as the keyword $w_{\lfloor rsp_{q}/\gamma\rfloor}$..
\end{itemize}

\begin{figure*}[h]
	\centering
	\subfigure[All updates are add operations]
	{
		\begin{minipage}{.3\linewidth}%
			\centering
			\includegraphics[width=\linewidth]{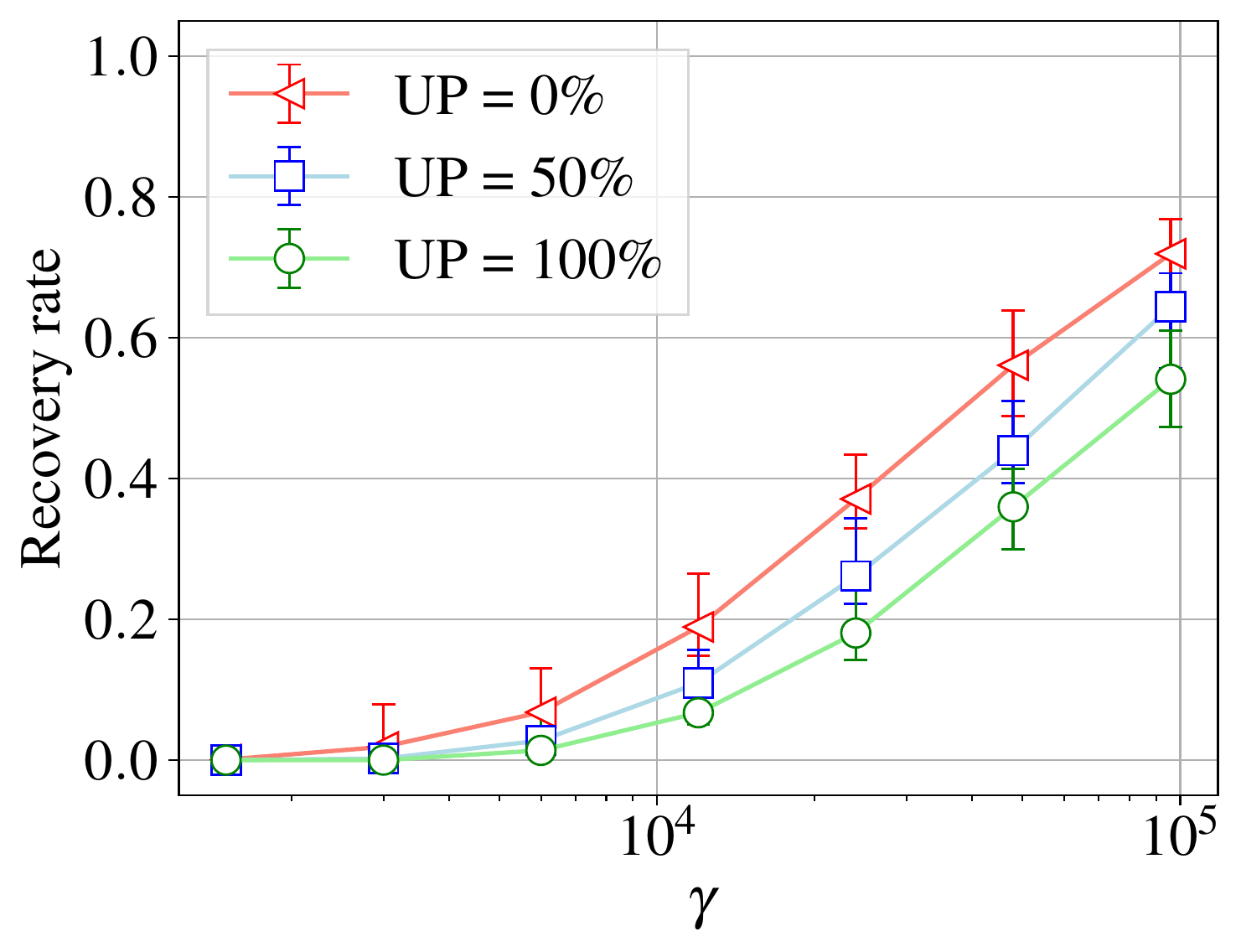}
            \label{UAWGAdd}
		\end{minipage}
	}
	\subfigure[Each update randomly selects add or delete]
	{
		\begin{minipage}{.3\linewidth}
			\centering
			\includegraphics[width=\linewidth]{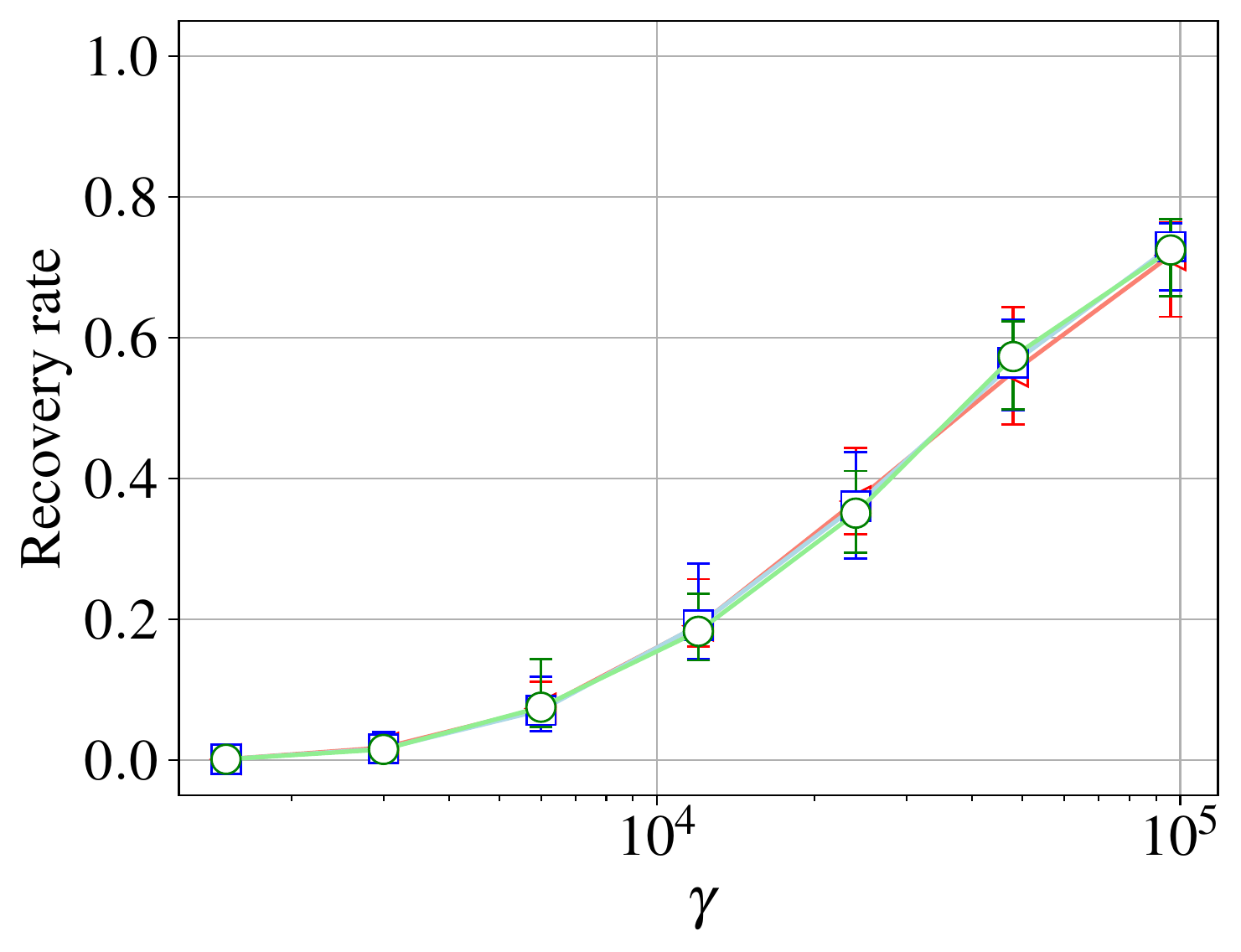}
            \label{UAWGUniform}
		\end{minipage}
	}
	\subfigure[All updates are delete operations]
	{
		\begin{minipage}{.3\linewidth}
			\centering
			\includegraphics[width=\linewidth]{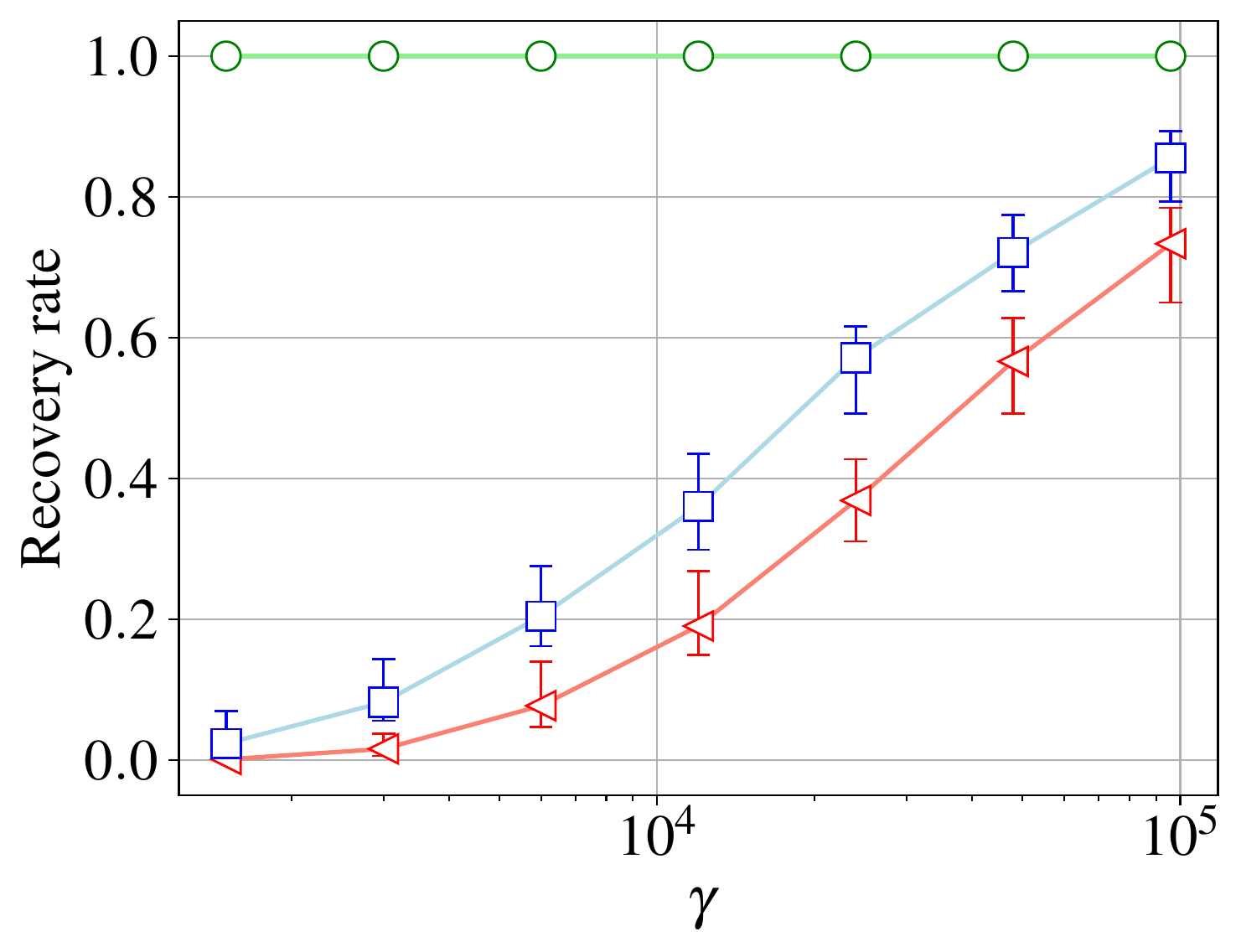}
            \label{UAWGDelete}
		\end{minipage}
	}
	\caption{The modified attack's recovery rate with $\gamma$.}
	\label{AccWithGamma}
\end{figure*}

\begin{figure*}[h]
	\centering

	\subfigure[All updates are add operations]
	{
		\begin{minipage}{.3\linewidth}%
			\centering
			\includegraphics[width=\linewidth]{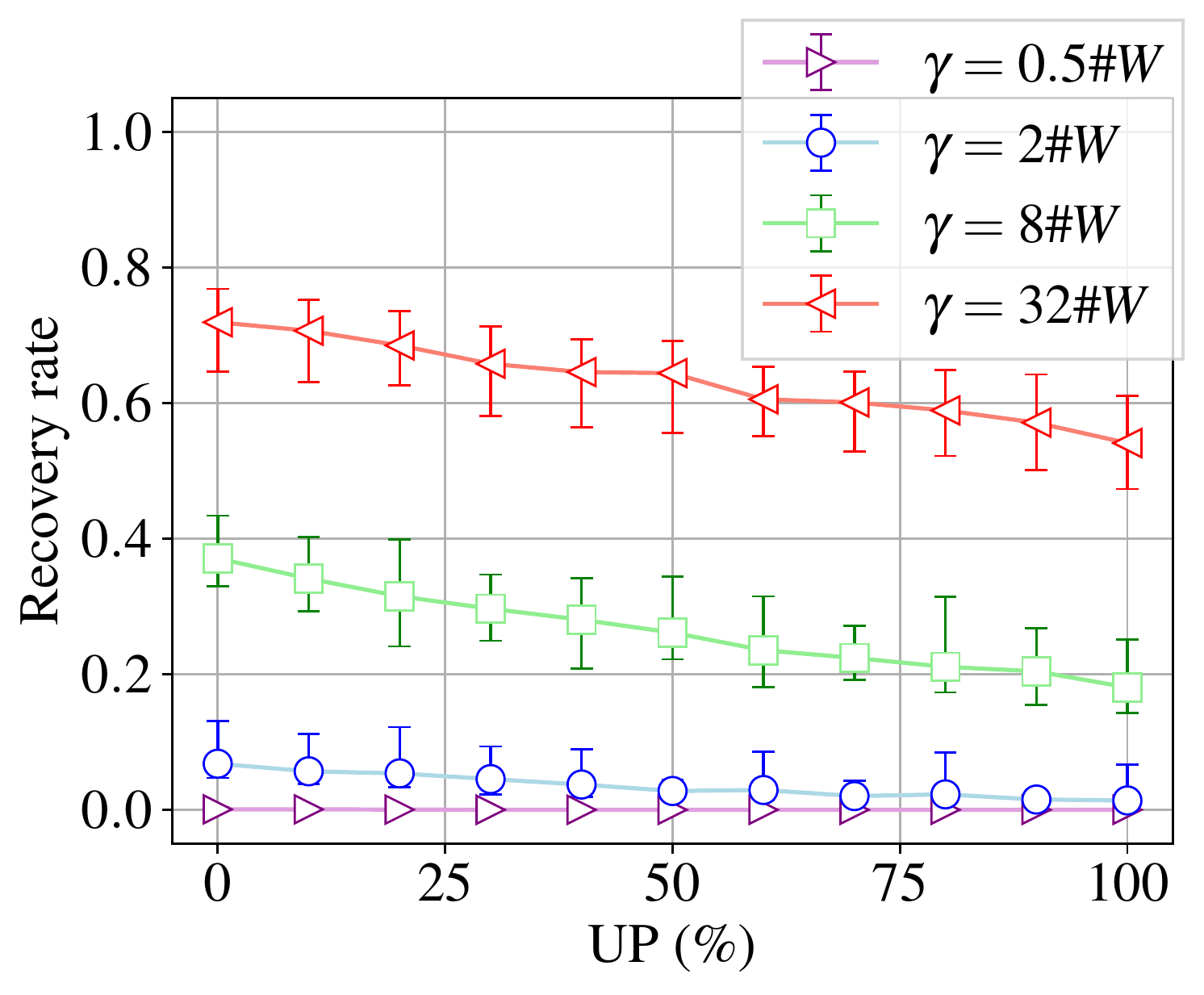}
            \label{UAWUAdd}
		\end{minipage}
	}
	\subfigure[Each update randomly selects add or delete]
	{
		\begin{minipage}{.3\linewidth}
			\centering
			\includegraphics[width=\linewidth]{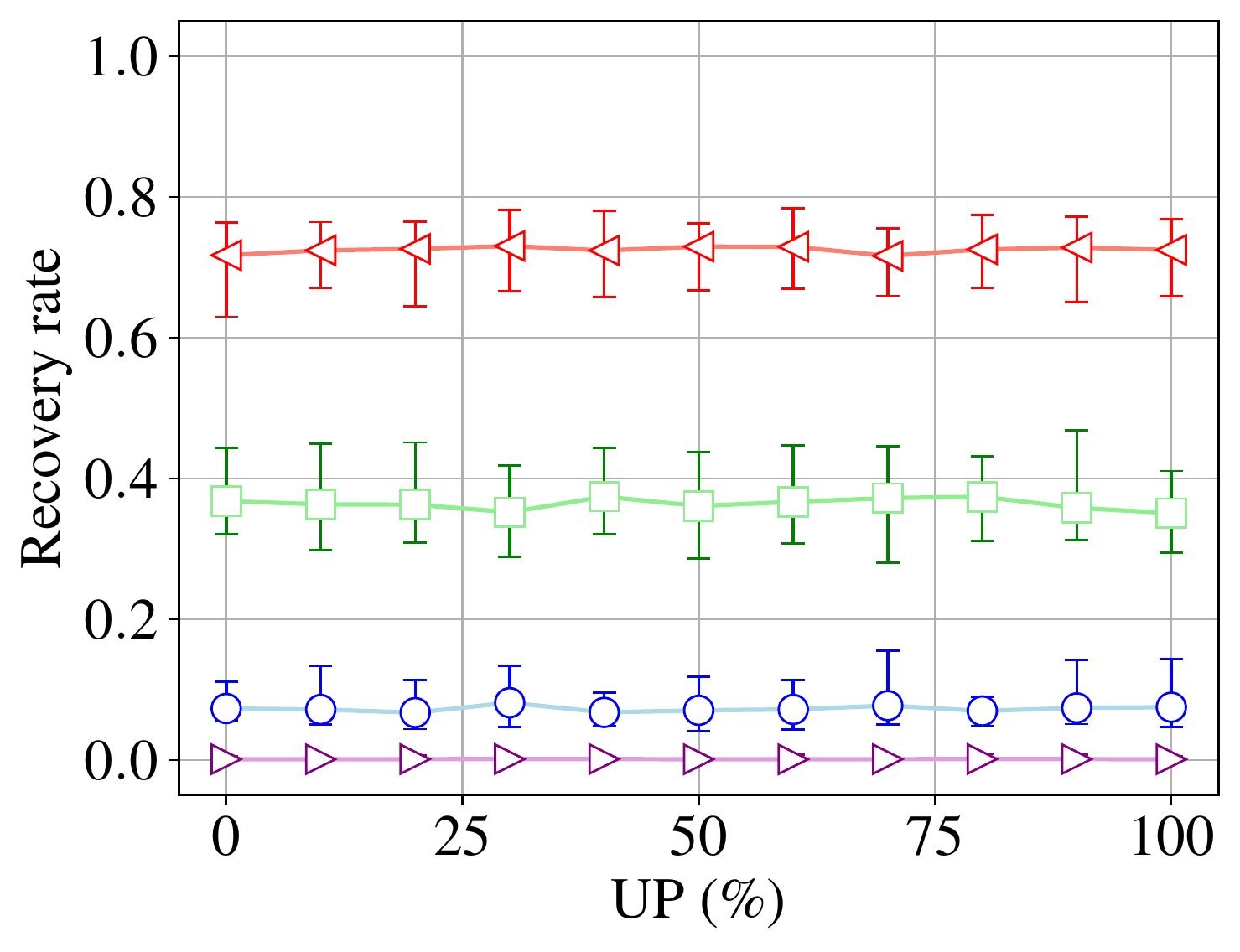}
            \label{UAWUUniform}
		\end{minipage}
	}
	\subfigure[All updates are delete operations]
	{
		\begin{minipage}{.3\linewidth}
			\centering
			\includegraphics[width=\linewidth]{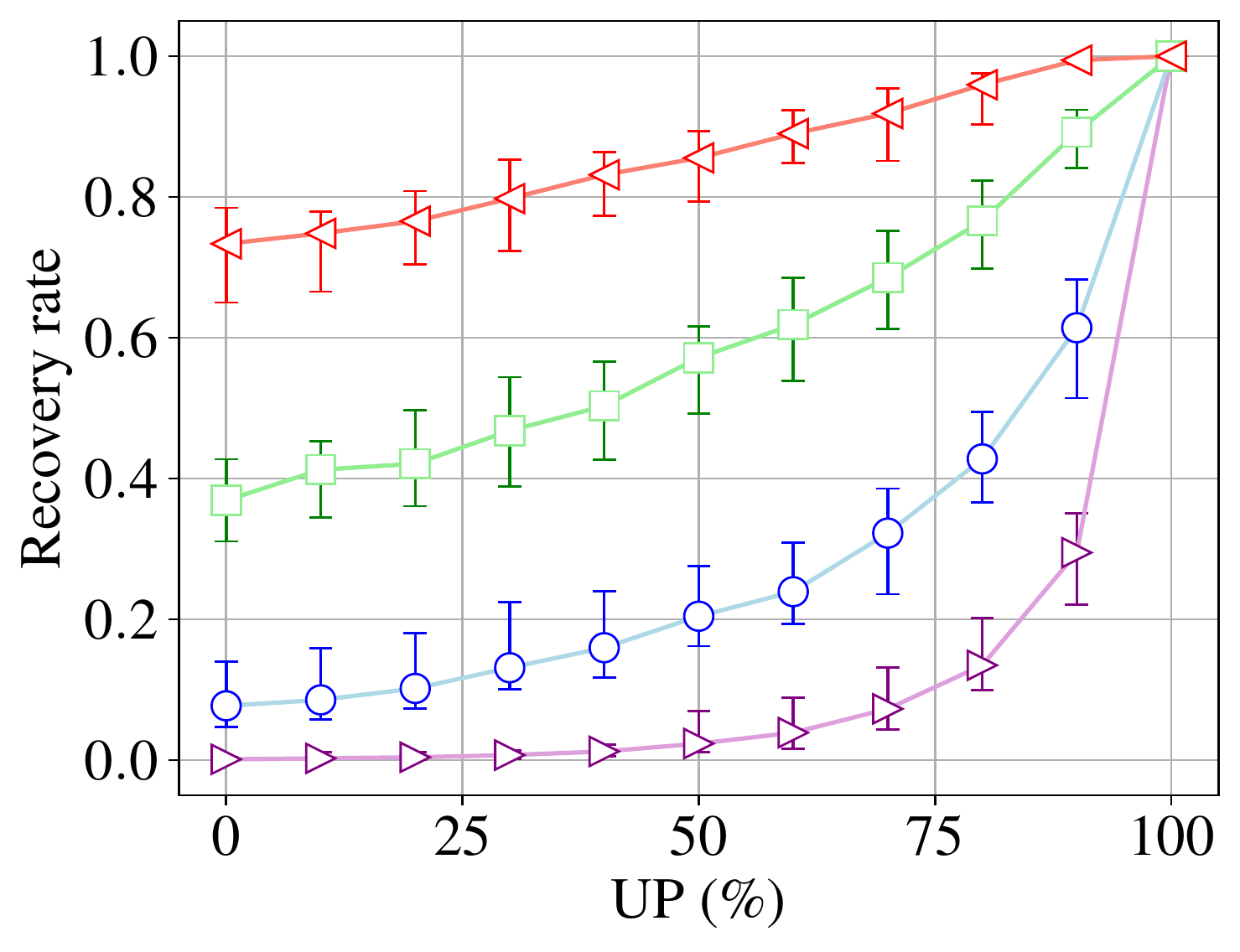}
            \label{UAWUDelete}
		\end{minipage}
	}
	\caption{The modified attack's recovery rate with update percentage (UP).}
	\label{AccWithUP}
\end{figure*}

\begin{figure*}[h]
	\centering

	\subfigure[All updates are add operations]
	{
		\begin{minipage}{.3\linewidth}%
			\centering
			\includegraphics[width=\linewidth]{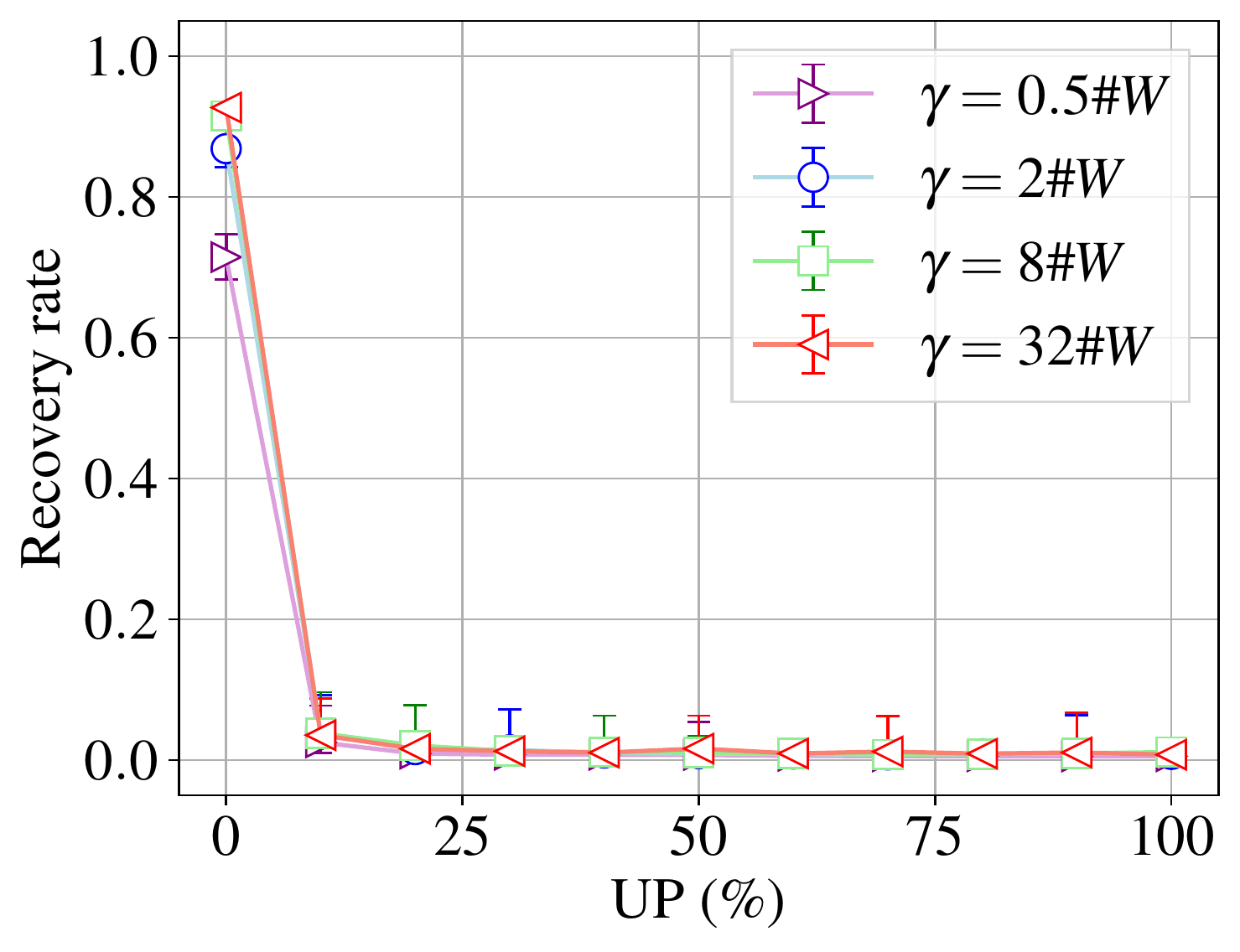}
            \label{BVAAdd}
		\end{minipage}
	}
	\subfigure[Each update randomly selects add or delete]
	{
		\begin{minipage}{.3\linewidth}
			\centering
			\includegraphics[width=\linewidth]{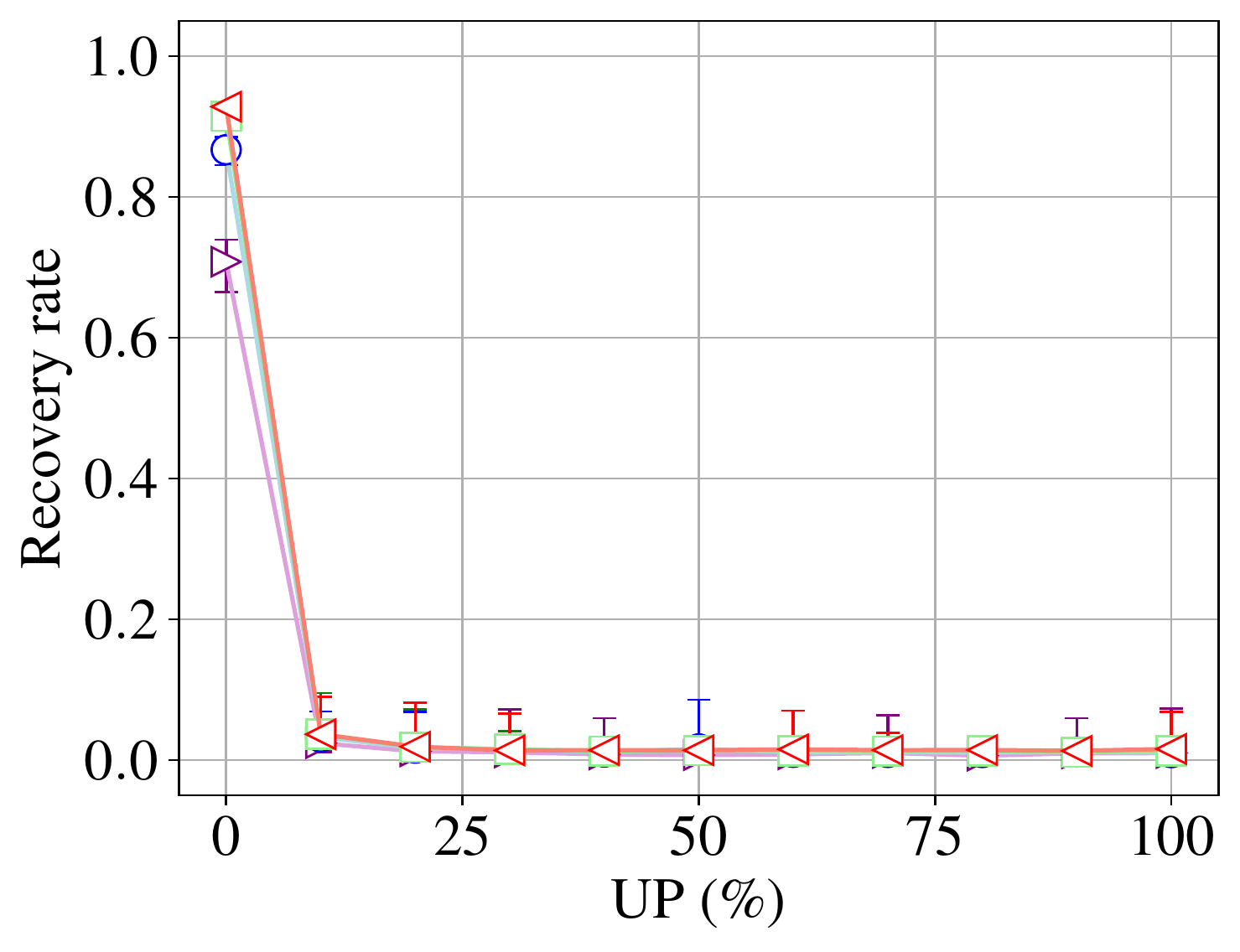}
            \label{BVAUniform}
		\end{minipage}
	}
	\subfigure[All updates are delete operations]
	{
		\begin{minipage}{.3\linewidth}
			\centering
			\includegraphics[width=\linewidth]{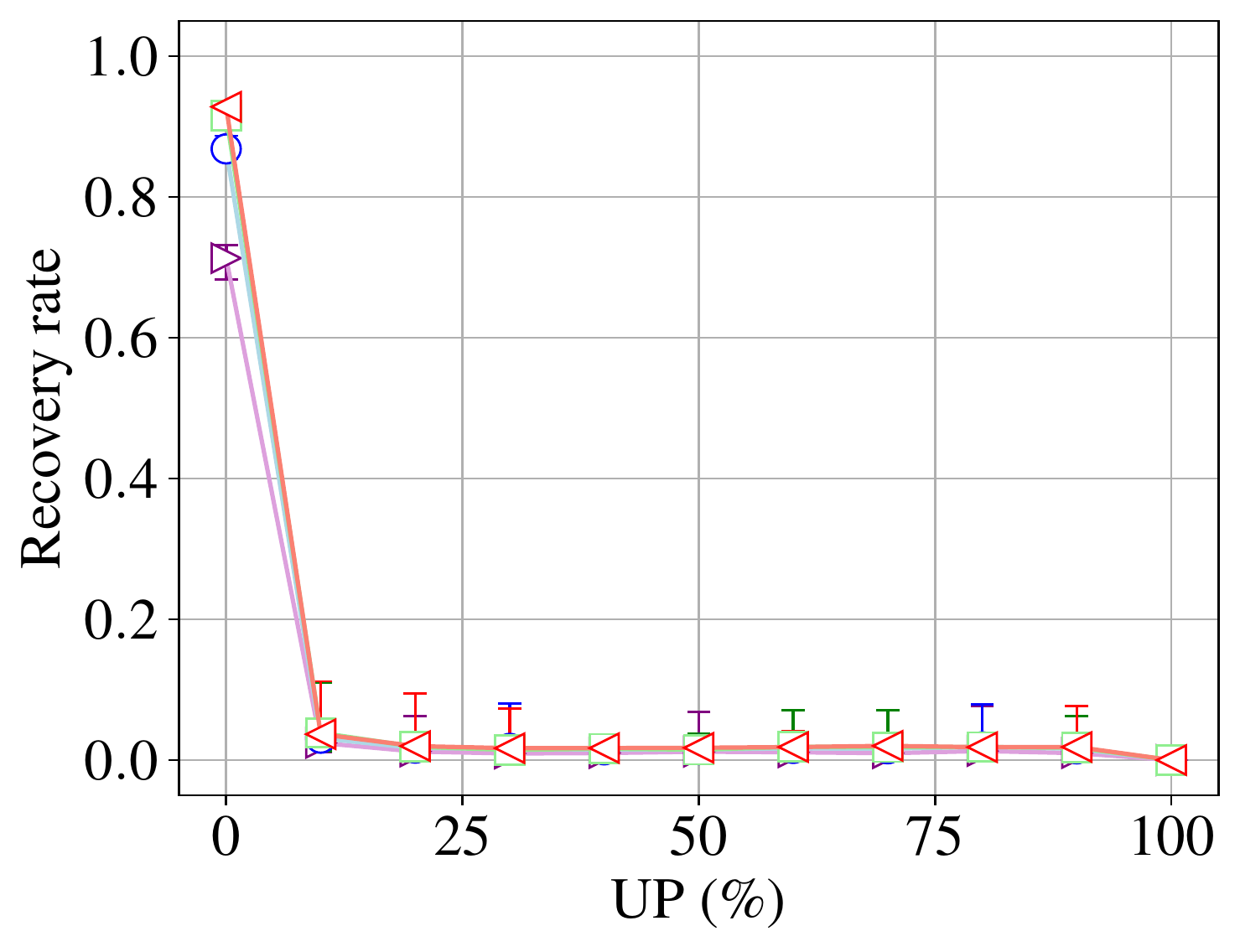}
            \label{BVADelete}
		\end{minipage}
	}
	\caption{BVA's recovery rate with $UP$.}
	\label{BVAAccWithUP}
\end{figure*}

\begin{figure*}[h]
	\centering

	\subfigure[Recovery rate]
	{
		\begin{minipage}{.3\linewidth}%
			\centering
			\includegraphics[width=\linewidth]{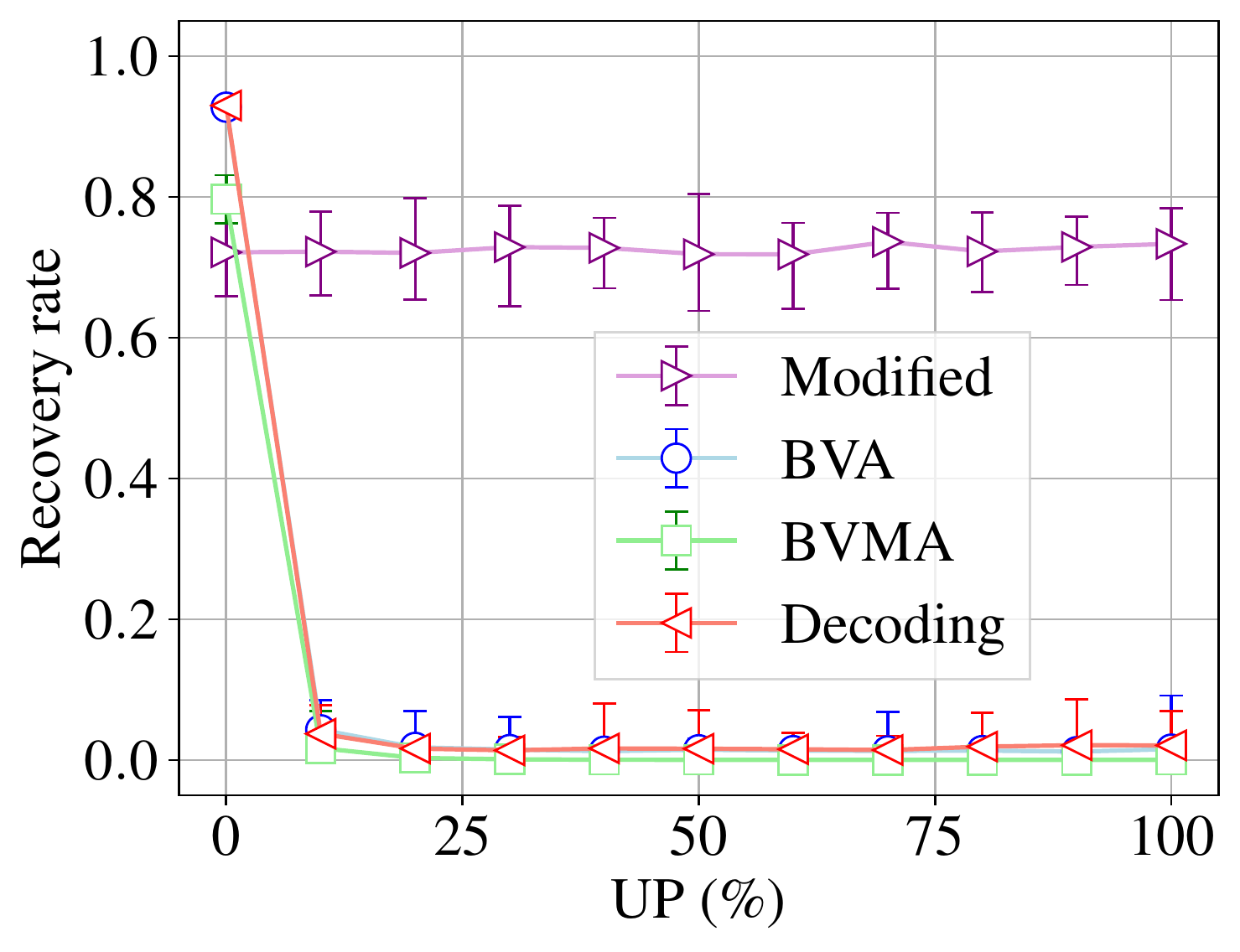}
            \label{ComAcc}
		\end{minipage}
	}
	\subfigure[Injection length]
	{
		\begin{minipage}{.3\linewidth}
			\centering
			\includegraphics[width=\linewidth]{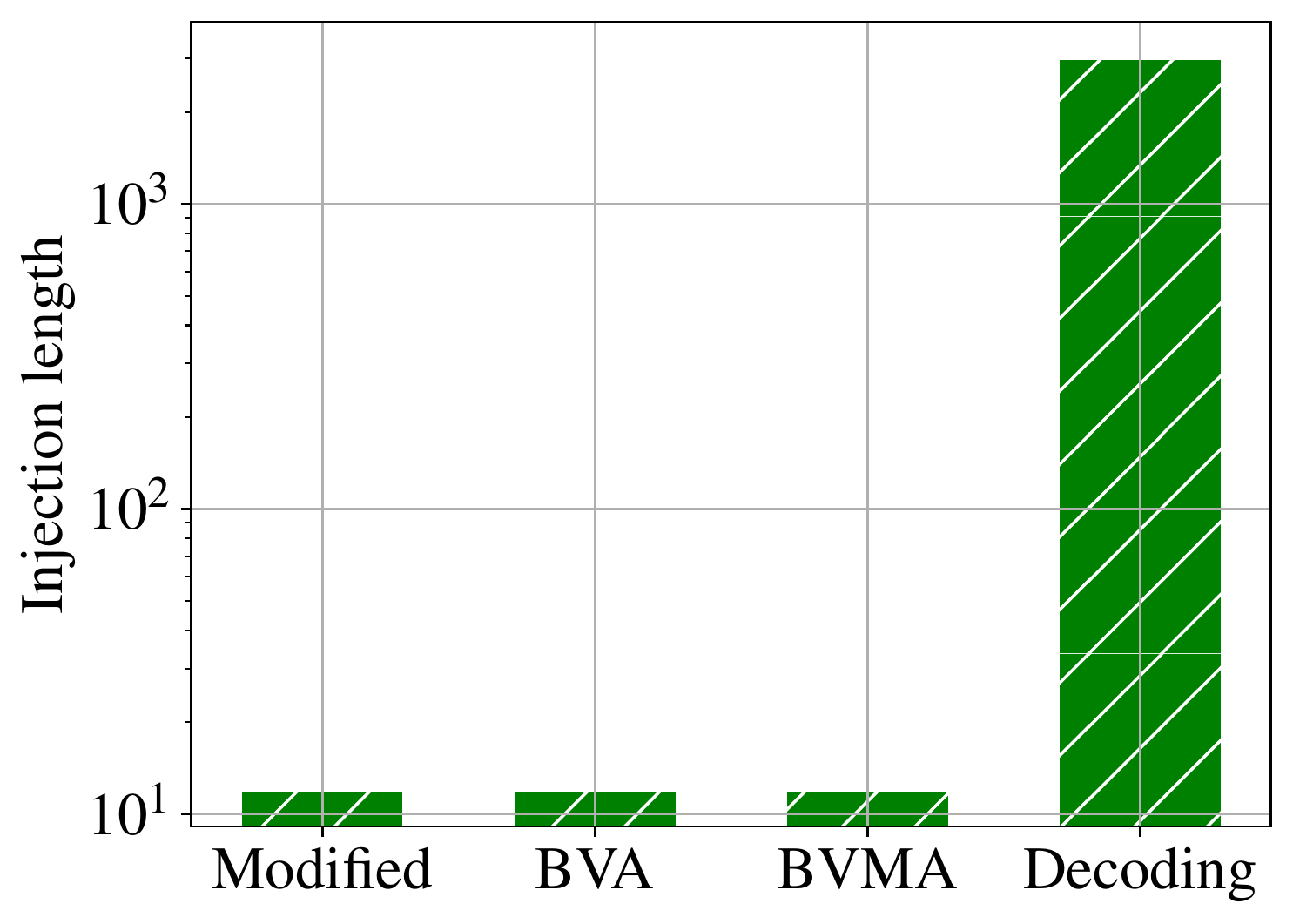}
            \label{ComIlen}
		\end{minipage}
	}
	\subfigure[Injection size]
	{
		\begin{minipage}{.3\linewidth}
			\centering
			\includegraphics[width=\linewidth]{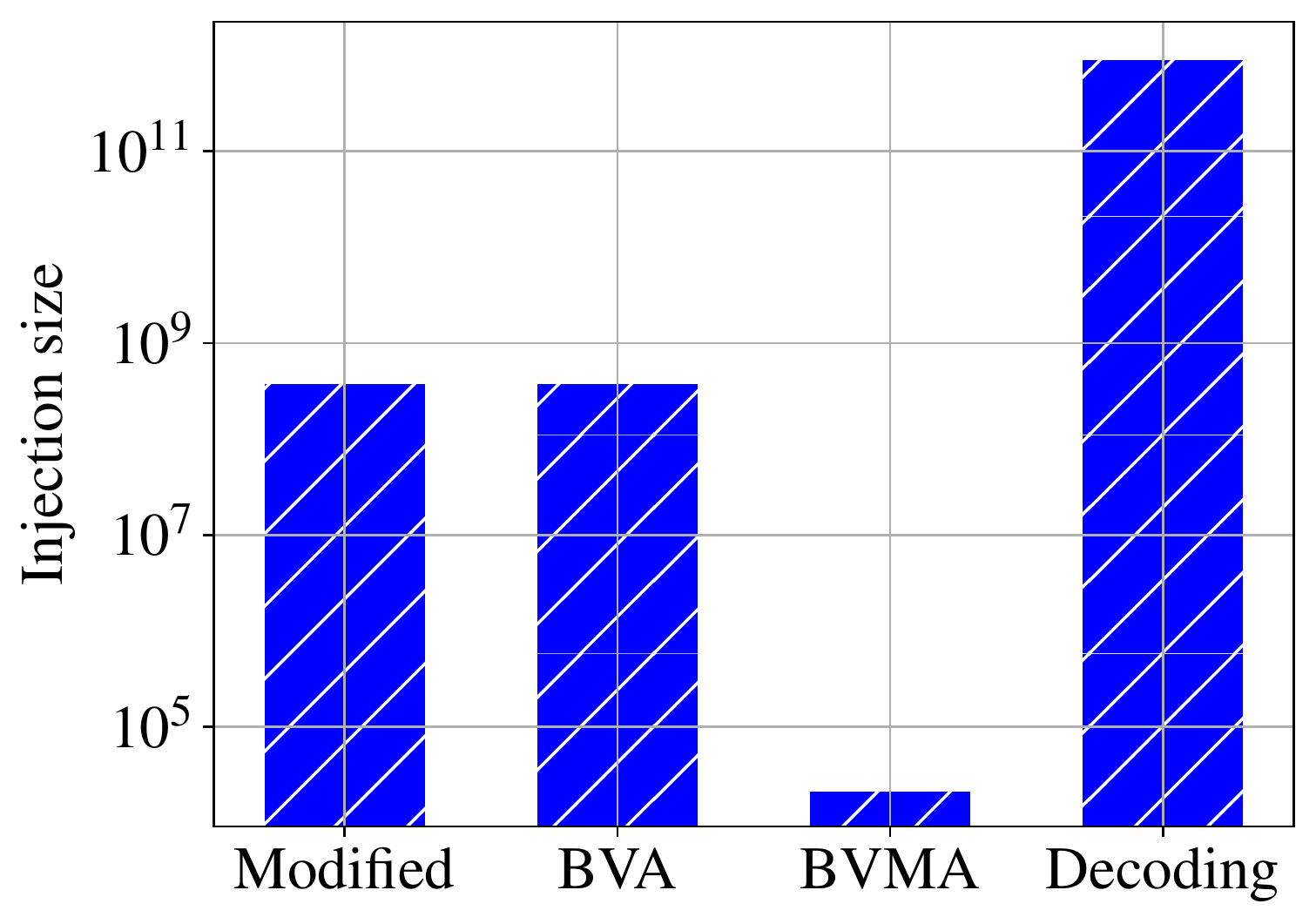}
            \label{ComIsize}
		\end{minipage}
	}
	\caption{Recovery rate and injection volume of certain VIAs. Note we set $\gamma=32\#W$ in the modified attack and BVA and assume each update operation randomly selects between add or delete.}
	\label{Comp}
\end{figure*}

Actually, setting a small $\gamma$ (e.g., $\gamma=c\cdot\#W$, where $c$ is a constant and $c\ll\#W$) can achieve a practical recovery rate.
We implemented the \textit{modified} attack and ran experiments on the Enron dataset with the following measurements. 
\begin{itemize}\vspace{-2mm}
\item We tested the recovery under different $\gamma$ (see Figure \ref{AccWithGamma}) and update percentage $UP$ (see Figure \ref{AccWithUP}, where $UP$ dontes the proportion of updated files in the dataset). We investigated the recovery rate when the user update operations include: (1) all add operations (see Figure \ref{UAWGAdd},\ref{UAWUAdd}); (2) a combination of uniform add and delete operations (see Figure \ref{UAWGUniform},\ref{UAWUUniform}); (3) all delete operations (see Figure \ref{UAWGDelete},\ref{UAWUDelete}). \vspace{-2mm}
\item We evaluated the impact of user updates on other VIAs (see Figure \ref{BVAAccWithUP}, \ref{Comp}). 
Note that different update operations (e.g. adding or deleting) may lead to varying recovery rates.
\end{itemize}

In Figure \ref{AccWithGamma}, the recovery enhances with the increase of $\gamma$. 
When $\gamma=10^{4}$, we achieve $20\%$ recovery; while $\gamma$ increases to $10^{5}$, the recovery rate reaches to about $60\%$.
We say a larger $\gamma$ could further boost the recovery, but it would also increase the injection size. 
In Figure \ref{UAWGDelete}, when $UP=100\%$, our modified attack  provides $100\%$ recovery on all reasonable $\gamma$ (i.e. $\gamma\geq{\#W/2}$). 
This is because the files in the original database and the updated files are all considered to be noise. 
$UP=100\%$ means that the noise is completely eliminated (i.e. all files in the original database are deleted), thus allowing us to obtain a perfect recovery rate. 

From Figure \ref{AccWithUP}, we see that the recovery rate shows varying trends with different update operations. %When $\gamma=8\#W$, we can achieve a recovery rate of no less than $20\%$ even when the user update $100\%$ of files. And $\gamma=32\#W$ is enough to make our attack achieve more than $50\%$ recovery rate. Predictably, a larger $\gamma$ will further increase the recovery rate, but also increase the injection size. 
Figure \ref{UAWUAdd} shows that the increase of add operations can harm the recovery.  
For example, when $\gamma=32\#W
$, the recovery on $UP=0\%$ is about $20\%$ higher than that on $UP=100\%$. 
Figure \ref{UAWUUniform} demonstrates that the random update barely affects the attack performance. 
In the last sub-figure, we prove that delete operation brings advantage to the recovery (e.g., when $\gamma=32\#W$, the recovery on $UP=50\%$ is about $10\%$ higher than that on $UP=0\%$). %Especially when $UP=100\%$, our attack will achieve $100\%$ recovery rate on all reasonable $\gamma$ (i.e. $\gamma\geq{\#W/2}$). This is because in our attack, the files in the original database and the updated files are all regarded as noise. $UP=100\%$ means that the noise is completely eliminated (all files in the original database are deleted), so we can achieve the perfect recovery rate.

We present the recovery rate of BVA under different $UP$ in Figure \ref{BVAAccWithUP}. 
As $UP$ increases, the recovery drops rapidly for any $\gamma$ and update operations. 
For example, a $10\%$ update could fail BVA.  
Note that when $UP=0$, BVA shows the advantage as its recovery is higher than that of the modified attack (e.g., when $\gamma=32\#W\wedge{UP=0}$, the recovery rate of BVA is around $90\%$, while that of the modified attack is $70\%$).
%Note that when $UP=0$, the recovery is higher than that of our modified attack (e.g., when $\gamma=32\#W\wedge{UP=0}$, the recovery rate of BVA is around $90\%$, while that of the modified attack is $70\%$).

In Figure \ref{Comp}, we compare the recovery rate and injection volume of the modified attack with other VIAs. 
Note that we did not put single-round attack into comparison, because its injection length is extremely large ($O(m\cdot\#W$), which could be impractical. 
From Figure \ref{ComAcc}, we see that the user updates do not significantly affect the recovery rate of the modified attack (around $70\%$ recovery rate), but harm others (in particular when $UP\geq10\%$). 
Figure \ref{ComIlen} shows that the modified attack has an injection length of approx. $10^{1}$, which is comparable to BVA but also better than decoding ($>10^{3}$).
The injection size of the modified attack and BVA ($<10^{9}$) is larger than BVMA ($<10^{5}$), but outperforms decoding ($>10^{11}$);  see Figure \ref{ComIsize}.

\section{Discussions}\label{Section DOE}
%We've experimentally tested the performance of our attacks across various databases. 
%There are some additional factors and tests that were not present. 
%This section will clarify these aspects we have not considered previously through discussions and theoretical analysis.

% \subsection{Other Techniques}

\noindent \textbf{Queries with multiple keywords.} 
Our attacks can also apply to those dynamic SSE schemes that enable multiple keywords queries \cite{DBLP:conf/crypto/CashJJKRS13,DBLP:conf/ccs/LaiPSLMSSLZ18}. 
We here present a heuristic solution to support our attacks for conjunctive queries with two keywords.
Given $\mathbf{W}=\{w_{1},...,w_{m}\}$, we combine any two keywords into one search query. 
That means a new search set $\mathbf{\overline{W}}=\{\overline{w_t}|\overline{w_t}={(w_{i},w_{j})},i,j\in [m],i\neq{j})\}$ is generated while a tuple $(w_{i},w_{j})$ is paired as a conjunctive query. 
For each element $\overline{w}_{t}\in\overline{W}$, we inject a file with size of $t\cdot\gamma$ ($\gamma$ is the parameter of BVA) containing the element. %, where $\gamma$ is the injection parameter of BVA.
As a result, we inject ${m(m-1)/2}$ files in total. 
% Similarly, by extending the BVMA, we can generate the injected file $\overline{F_j}$ with size $j\cdot{m}$ for the constant $m$ containing the keyword $\overline{w_j}$, where $j\in{\#\overline{\mathbf{W}}}$
%{\color{blue} [could we also mention that we could have similar extension for BVMA???not easy]}  
It is an open problem to reduce the injection volume for the conjunctive queries. % and we leave this as the future work.
\smallskip

{\noindent \textbf{SEAL's Dynamic Padding.} 
We used a straightforward approach to extend SEAL to support the dynamic padding. 
{The extension can effectively resist VIAs, but its overhead (w.r.t. Inj\&Fill) is extremely impractical, for example, the cost is even higher than simply streaming the entire database to the client.}  
It is an open problem to design a more practical solution to balance the security and efficiency.  
%should better balance its security and efficiency. 
One may consider combining clustering-based schemes (e.g., \cite{DBLP:conf/ccs/CashGPR15}) with the probabilistic padding (e.g., \cite{DBLP:conf/ccs/PatelPYY19}).  
For example, for each batch update, one may divide keywords into several clusters and then fill each keyword in a cluster to $x^{t}+noise$, so that the keyword's \textit{rlp} will contain the same $x^{t}$ and probabilistic (and different) noise, where $x$ is the parameter of SEAL. 
}%, which may provide a better privacy and efficiency tradeoff and counter VIAs.}  
%More specific and detailed consideration will be the content of our future research.} 
%Clustering-based  schemes allocate a subset of keywords to a cluster, so that the keywords volumes cannot be distinguished. By specifying the cluster size, this method can achieve better privacy and efficiency tradeoffs. 
%We think about a 

%may be a worthy directions for SEAL to support dynamic padding so as to counter VIAs.}

%\smallskip
\smallskip
\noindent \textbf{LEAKER.} 
Kamara et al. \cite{LEAKER} proposed an open-source attack evaluation framework LEAKER to test the recovery of passive LAA with different known file rates. 
LEAKER unfortunately cannot cover injection attacks \cite{LEAKER} as it does not capture the necessary information leakage obtained and used by active adversaries. %
%In contrast, all injection attacks require the same prior knowledge (known keyword universe), and the evaluation of injection attacks also focus on the injection volumes of different strategies. 
Extending this evaluation framework to test injection attacks could be an interesting problem.
We note that is orthogonal to the main focus of this work.

\smallskip
%\smallskip
\noindent \textbf{Experiments on Multiple Datasets.}
In Section \ref{AppenPadding}, \ref{AttUpd} and Appendix \ref{AppenBVMASP}, \ref{AppenCLW}, \ref{AppenDYSEAL}, we conducted the experiments in Enron. 
Our attacks performance will not differ significantly in Lucene and Enron as these email datasets share similar scales and keyword distributions.    
The Wikipedia could moderately harm the performance. 
For example, the recovery against static padding will be similar to that of Figure \ref{WikiTrendRerVol} (around $60\%$); whilst dynamic padding could further restrict the attacks performance.  
Under the clients' active updates, the attacks could achieve similar trends and results as in Figure \ref{UpdateFigure}. %the type and scale of the dataset 
%With proper injections, our attacks may work well in Lucene and Wikipedia.  
Interested readers may use our codes to conduct extra experiments.
    
\end{appendices}

\end{document}

%% file: ccs-body-revised.tex
\section{Introduction}

\begin{table*}[!t]

    \begin{threeparttable} 
	    \caption{Comparison with leakage abuse (passive) and injection attacks. $\#\mathbf{W}$ is the number of the known keywords. We use $k$ ($k\geq{2}$) to represent the count of keyword partitions  {for the multiple-round attack}, and $m$ ($m\geq{1}$) refers to the injection constant {of the single-round attack}. We denote $\textit{offset}, \gamma$ (generally, $\textit{offset}\gg{\#\mathbf{W}}$ and $\gamma\geq\#\mathbf{W}/2$) as the basic size of the injected files for the decoding attack and BVA, respectively. The last column shows the minimum rounds required to restore all the observed queries. Other notions and definitions are given in Table \ref{Remark1}, Appendix \ref{AppenDesonPa}.}
		\label{comparison}
		\centering
		%\resizebox{\textwidth}{12mm}
		%\resizebox{\linewidth}{!}
		\setlength{\tabcolsep}{3.4mm}{
        \begin{tabular}{lcccccc}
			\hline
			% \multirow{2}{*}{Name}，2为所占的行数，此语句可以使得内容垂直居中
			% \multicolumn{2}{c|}{Flag}，2为所占的列数，格式由第二个{}控制
			% \cline{2-3}指本行的2,3列画横线
			\multirow{2}{*}{Attack} & \multirow{2}{*}{Leakage} & \multicolumn{2}{c}{Type} & \multicolumn{2}{c}{Injection volume} & \multirow{2}{*}{Round\tnote{2}}\\ 
			\cline{3-6}
			& & Passive & Injection & Length & Size\\ 
			\hline
			IKK \cite{DBLP:conf/ndss/IslamKK12} & ap & \checkmark & $\times$ & -- & -- & --\\
			%\hline
			Count \cite{DBLP:conf/ccs/CashGPR15} & ap,rlp  & \checkmark & $\times$ & -- & -- & --\\
			%\hline
			SelVolAn, Subgraph \cite{DBLP:conf/ndss/BlackstoneKM20} & ap,vp & \checkmark & $\times$ & -- & -- & --\\
			%\hline
			LEAP \cite{DBLP:conf/ccs/NingHPYL0D21} & ap & \checkmark & $\times$ & -- & -- & --\\
			%\hline
			SAP \cite{DBLP:conf/uss/OyaK21} & sp,rlp & \checkmark & $\times$ & -- & -- & --\\
			\hline
			\hline
			ZKP \cite{DBLP:conf/uss/ZhangKP16} & aip & $\times$ & \checkmark & $\mathcal{O}(\log{\boldsymbol{\#W}})$ & $\mathcal{O}(\boldsymbol{\#W}\log{\boldsymbol{\#W}})$ & 1\\
			Multiple-round\tnote{1} \cite{DBLP:conf/eurosp/PoddarWLP20} & rlp & $\times$ & \checkmark & $\mathcal{O}(k{\boldsymbol{\#W}\log_{k}{\boldsymbol{\#W}}})$ & $\mathcal{O}(k{\boldsymbol{\#W}}^{2})$ & $\boldsymbol{\#W}\log_{k}{\boldsymbol{\#W}}$\\
			%\hline
			Single-round \cite{DBLP:conf/eurosp/PoddarWLP20} & rlp & $\times$ & \checkmark & $\mathcal{O}(m{\boldsymbol{\#W}})$ & $\mathcal{O}(m{\boldsymbol{\#W}}^{2})$ & 1\\
			%\hline
			Decoding \cite{DBLP:conf/ndss/BlackstoneKM20} & rsp & $\times$ & \checkmark & $\mathcal{O}({\boldsymbol{\#W}})$ & $\mathcal{O}(\textit{offset}\cdot{\boldsymbol{\#W}}^{2})$ & 1\\
			%\hline
			Search\tnote{1} \cite{DBLP:conf/ndss/BlackstoneKM20} & rsp & $\times$ & \checkmark & $\mathcal{O}({\boldsymbol{\#W}\log{\boldsymbol{\#W}}})$ & $\mathcal{O}({\boldsymbol{\#W}}^{2})$ & $\boldsymbol{\#W}\log{\boldsymbol{\#W}}$\\
			%\hline
			\hline
			\hline
			BVA & rsp  & $\times$ & \checkmark& $\mathcal{O}(\log{\boldsymbol{\#W}})$  & $\mathcal{O}({\gamma}{{\boldsymbol{\#W}}})$ & 1\\
			%\hline
			BVMA & vp, sp\tnote{3} & $\times$ & \checkmark& $\mathcal{O}(\log{\boldsymbol{\#W}})$ & $\mathcal{O}(\boldsymbol{\#W}\log{\boldsymbol{\#W}})$ & 1\\ %\tnote{3}
			\hline
		\end{tabular}
		\begin{tablenotes}    %这行要添加， 从这开始
            \footnotesize               %这行要添加
            \item[1] Unlike those schemes easily restoring multiple queries, search \cite{DBLP:conf/ndss/BlackstoneKM20} and multiple-round \cite{DBLP:conf/eurosp/PoddarWLP20} attacks can only recover a keyword at a time {by running many attack rounds}. This means that the client should make many queries, and the queries must include the target query at each attack round.  The multiple-round is a strategy that depends on query replay, requiring the adversary to evoke the same query repeatedly by controlling the client. 
            These two attacks commit more rounds and injected files (than others in the table) to recover multiple queries.   
            { \item[2] We here say that an attack round consists of (1) an active and complete injection and then (2) an observation within a specific period. We will present a formal definition in Appendix \ref{AppenRound}.}
            { \item[3] {BVMA mainly investigates the vp to recover the queries. Thus, the sp is an optional and non-essential leakage for the attack (see Appendix \ref{AppenBVMASP}).}}
            %{ \item[3] We note that search pattern (sp) for frequency information may be non-essential in BVMA from the experiment in Appendix \ref{AppenUQ}.}%[``continuous injection" --- is this term used in other papers or created by you???] 
        \end{tablenotes}  }
    \end{threeparttable} 
\end{table*}
	
Song et al. \cite{848445} proposed the notion of searchable symmetric encryption (SSE) that enables secure search over an encrypted database. 
Following the notion, researchers have presented various SSE schemes to balance practicability and security \cite{DBLP:conf/eurocrypt/CashT14,DBLP:conf/ccs/BostMO17,DBLP:conf/ccs/ChamaniPPJ18,DBLP:conf/ccs/SunYLSSVN18,DBLP:conf/eurocrypt/KamaraM17,DBLP:conf/esorics/PatelPY18}.
In a standard SSE scheme, a search query is a series of interactions between a client and a server. 
The server can perform matching operations for the query, such as paring a search token given by the client with ciphertexts, and return the query results.      
Unfortunately, the statistics on the responses leak some patterns of the encrypted database.
These patterns seem natural and harmless to data privacy, as they do not directly reveal the files' contents.  
Adversaries can still exploit them to recover the client's query.  
Leakage abuse attacks (LAA) \cite{DBLP:conf/esorics/LambregtsCNL22, LEAKER,DBLP:conf/ndss/IslamKK12,DBLP:conf/ccs/CashGPR15,DBLP:conf/ccs/PouliotW16,DBLP:conf/uss/OyaK21,DBLP:conf/ccs/NingHPYL0D21,LeakageInversion}, a classic type of passive attacks on SSE, enable adversaries to observe the leakage patterns for a continuous query period and then combine prior knowledge to recover the target queries. 
The recovery rate significantly relies on the ``sufficiently large" amount of prior knowledge \cite{DBLP:conf/ndss/IslamKK12,DBLP:conf/uss/OyaK21}, which restricts LAA in practical use.       

Unlike passive attacks, injection attacks \cite{DBLP:conf/ccs/CashGPR15, DBLP:conf/uss/ZhangKP16, DBLP:conf/eurosp/PoddarWLP20, DBLP:conf/ndss/BlackstoneKM20} enable adversaries to actively inject files into the encrypted database.    
An adversary can generate a certain number of injected files containing \textit{known keywords} for the client who packs these encrypted files and sends them to the server. 
It can then recover the keywords with a high probability by observing the leakage patterns of the target queries.  
Surprisingly, %there are massive playgrounds for this type of attacks in practice malicious
many real-world applications are unable to prevent file injection \cite{DBLP:conf/eurosp/PoddarWLP20}. 

Cash et al. \cite{DBLP:conf/ccs/CashGPR15} introduced a seminal active attack against property-preserving encryption \cite{DBLP:conf/sigmod/AgrawalKSX04,DBLP:conf/eurocrypt/BoldyrevaCLO09}, allowing the adversary to ``implant" files into the client's database and then reconstruct partial {plaintexts}.  %\cite{DBLP:conf/sigmod/AgrawalKSX04,DBLP:conf/eurocrypt/BoldyrevaCLO09}. 
Following a similar philosophy, Zhang et al. \cite{DBLP:conf/uss/ZhangKP16} proposed a concrete file injection attack (ZKP) to SSE. 
ZKP enables the adversary to generate and inject files $F_{1},...,F_{\log n}$ by using $n$ known keywords, such that each keyword is in a unique subset of the files.
After the client makes queries, the adversary must identify the exact set of the returned injected files and further reveal the queries. 
The adversary \textit{must} also know the access injection pattern, {i.e., the set of injected files matching the queries.} 
%Current defense mechanisms (e.g., Oblivious RAM (ORAM) \cite{DBLP:conf/crypto/GargMP16,DBLP:conf/uss/DemertzisPPS20}) can easily counter the attack.   

The SSE schemes {protecting the access (injection) pattern, such as those built on ORAM \cite{DBLP:conf/crypto/GargMP16,DBLP:conf/uss/DemertzisPPS20}, can successfully resist ZKP \cite{DBLP:conf/uss/ZhangKP16} and other access pattern based attacks \cite{DBLP:conf/ndss/IslamKK12, DBLP:conf/ccs/CashGPR15, DBLP:conf/ccs/NingHPYL0D21}. 
However, they still leak the \textit{volume pattern (vp)} from the search results. 
We say that the \textit{vp} includes (1) the number of response files, referred to as the \textit{response length pattern (rlp)}, and (2) the word count of returned files, referred to as the \textit{response size pattern (rsp)}. 
In this work, we will exploit these patterns in volumetric injection attacks (VIAs)\footnote{Note recent VIAs naturally leverage either the \textit{rlp} or \textit{rsp}.}. 

Poddar et al. \cite{DBLP:conf/eurosp/PoddarWLP20} proposed injection-based \textit{multiple-round} and \textit{single-round} attacks by only exploiting the \textit{rlp}. 
In each round of the \textit{multiple-round} attack, the adversary divides the candidate keywords into $k$ partitions and generates the same number of empty files. 
For a keyword in the $i$th partition, the adversary adds the keyword to $i$ files {as injection}. 
If there are $i$ more response files to the target query than before (i.e., a previous round), the adversary can narrow the search space of the candidate keyword to the $i$th partition. 
The adversary must run $\log_{k}{n}$ rounds of injections and observations to recover a query, given $n$ known keywords. 
The attack works properly \textit{as long as} the adversary can make the client constantly repeat the same query.  
This restriction  works in some scenarios, e.g., websites using HTTP/1.1 RFC \cite{fielding2014hypertext}.  
% could be impractical in reality.  
%  Web by HTTP/1.1 RFC \cite{fielding2014hypertext}
%The above round complexity and condition may make the attack impractical in reality. 
%
%
The \textit{single-round} attack selects a specific $m$ and enables the adversary to inject $m\cdot{n}$ files for $n$ known keywords in which a keyword $w_{i}$ is in $m\cdot{i}$ files.
When observing a query with the response of $l$ files, the adversary will recover the query as $w_{\lfloor \frac{l}{m} \rfloor}$.  
The $m$ should be much larger than the number of the client’s files containing $w_i$ so that $\lfloor \frac{l}{m} \rfloor$ is equal to $i$ and the adversary can correctly recover the query. % the query can be correctly recovered. 
Compared to the multiple-round technique, this attack reduces the number of interactive rounds and recovers multiple queries. 
{But its injection amount is still large.} %in large-scale databases. %(e.g., $m$ times the known keyword universe) 

Blackstone et al. \cite{DBLP:conf/ndss/BlackstoneKM20} leveraged the \textit{rsp} to propose a \textit{decoding} attack and a \textit{search} attack for multi-query and one query recovery, respectively.  
For each keyword $w_{i}$, the $decoding$ attack generates and injects a file with a size of $i\cdot{\textit{offset}}$. 
Given a query, if the difference of its response sizes before and after injection exceeds $i\cdot{\textit{offset}}$ (i.e., observing $i\cdot{\textit{offset}}$ more after injection), the adversary can infer that the underlying keyword is $w_{i}$. 
A drawback of this attack is that the adversary consumes a considerable amount of resources in calculating \textit{offset} and injections (which are linear w.r.t. the number of keywords) to perform well in query recovery. 
The $search$ attack, recovering one keyword at a time, uses a binary search method. % {\color{blue}({within $n$ rounds for $n$ keywords}).}  
The adversary can inject a file containing half of the candidate keywords at an attack round. 
{It can determine if the injected file is associated with the target query by a follow-up observation.} 
Using this inject-and-observe approach in the following rounds, the adversary can filter the candidate keywords by half until there is only one keyword left.
The attack can only recover a single keyword via massive file injections and attack rounds.

%One may observe from the above discussion that 
Existing VIAs are limited by round complexity and injection amount.
An interesting question thus arises:  

\textit{Could we propose practical injection attacks that provide a high recovery rate with fewer injections and further circumvent popular defenses?}

\noindent \textbf{Our contributions.}
We present an affirmative answer to the above question by proposing two new injection attacks for dynamic SSE schemes and particular defense mechanisms (e.g., ORAM and static padding). 
Our attacks leverage the dynamic coding injection approach with fewer injections than prior works.
We show the comparison among the attacks in Table \ref{comparison}. 
The main contributions are summarized as follows. 
\\
$\bullet$ \textit{Practical binary volumetric attacks.} We propose two practical injection attacks (see Section \ref{Section Attack}) that provide comparable performances to prior attacks, e.g.,  \cite{DBLP:conf/ndss/BlackstoneKM20, DBLP:conf/eurosp/PoddarWLP20}. 
First, we present the binary variable-parameter attack (BVA) by exploiting the \textit{rsp}. 
We use a dynamic injection parameter $\gamma$ to balance a trade-off between injection size and recovery rate. 
Second, we develop the binary volumetric matching attack (BVMA), which is the first injection attack combining the (small) leakage of the \textit{rlp} and \textit{rsp}.
For any queries, the BVMA can filter incorrect keywords, with a small amount of injection, by observing the difference in the response volume before and after injection.  
We may leverage other leakage information(e.g., query frequency) to refine the recovery.
%Thus, the BVMA injection volume (especially injection size) can be significantly reduced with relatively high recovery rate and adaptability to different test datasets. 
We also present a generic method that can transform current VIAs (\cite{DBLP:conf/ndss/BlackstoneKM20, DBLP:conf/eurosp/PoddarWLP20} and ours) to counter the threshold countermeasure (TC)\footnote{Note TC enables SSE schemes to constrain each (encrypted) file's word-count to a small threshold in order to defend against injection attacks.} \cite{DBLP:conf/uss/ZhangKP16}.%, which may be of independent interest.}
\\
$\bullet$ \textit{Comprehensive evaluations.} We compare our attacks with BKM \cite{DBLP:conf/ndss/BlackstoneKM20}, PWLP \cite{DBLP:conf/eurosp/PoddarWLP20}, and ZKP \cite{DBLP:conf/uss/ZhangKP16} in three real-world datasets (see Section \ref{Section Exper}). 
We generate queries by obtaining keyword trends from Google trend \cite{GoogleTrends} and the Pageviews tool \cite{Pageviews}. 
Experimental results show that our attacks can provide a comparable level of recovery rate (e.g., $>80\%$ on average in Enron and Lucene) as the single-round (with $m=\#\mathbf{W}$) and decoding attacks {while requiring fewer injections} (e.g., {saving $> 99\%$ of injection costs given the keyword universe}).  
Our attacks can also practically apply to a large-scale dataset such as Wikipedia, {with around $60\%$ recovery.} 

We evaluate our attacks against defenses (e.g., TC, padding) and client's active update.
Under TC, our attacks require relatively ``lightweight" injections (e.g., $<10^{3}$ files injected by the BVMA in Enron and Lucene). 
In contrast, the single-round and decoding attacks require a serious number of {injections} to work properly, with injection sizes respectively over $10^{4}\times$ and $10^{6}\times$ the cost of the BVMA (particularly in Wikipedia). 
The static padding cannot effectively resist our attacks ($>60\%$ recovery rate on average against SEAL \cite{DBLP:conf/uss/DemertzisPPS20}), while the dynamic padding (ShieldDB \cite{ShieldDB}) requires an expensive overhead ($1.5\times$ {query overhead} more than no padding) to restrict our attacks' performance.   
We show that an optimization of our attacks can yield a high recovery rate against ShieldDB.  %by injecting more files.  
%Even so, it is still possible to break some dynamic padding schemes through more injection files. 
For example, our attacks maintain $>80\%$ recovery by injecting around $600$ files in Enron, under ShieldDB. 
We also demonstrate that our attacks perform well under the client's active updates. 
We also demonstrate that our modified attacks from BVA perform well under client's active updates.
Even if the client commits $100\%$ updates with respect to the dataset, our modified attack can achieve $>50\%$ recovery rate by increasing the injection size (e.g., $O(32\cdot\#\mathbf{W})$) while maintaining $O(\log\#W)$ injection length.

\section{Model Definitions}\label{Section attack model}
A dynamic SSE scheme (see Appendix \ref{SectionSE}) should not directly leak any other information to adversaries except those that can be inferred from setup leakage $\mathcal{L}_{St}$, query leakage $\mathcal{L}_{Qr}$, and update leakage $\mathcal{L}_{Up}$,  throughout interactions between the client and the server. 
{It captures the adaptive security if adversaries can choose the target queries and the corresponding operations adaptively \cite{Curtmola06searchablesymmetric}}. %and inspires our definitions of the leakage and attack models.

\subsection{Leakage Model} \label{Leakage Model}
The operations of a dynamic SSE scheme naturally yield multiple leakage patterns.   
We denote an adaptive semantic secure SSE scheme as $\mathbf{SSE}=(Setup: \lambda\rightarrow\mathbf{D},\ Query: \mathbf{D \times Q \rightarrow RF},\ Update: \mathbf{D \times U \rightarrow \delta})$, in which $\mathbf{D}$ is the encrypted database, $\mathbf{RF}$ is the set of files matching the queries $\mathbf{Q}$, $\mathbf{U}$ is the set of updates containing $op$ and $(w, id)$, and $\delta$ represents the state of the client after the update from $\mathbf{U}$.
We define two types of patterns:
\smallskip
\\
$\bullet$ the $access\ pattern$ is the function family where $\mathbf{ap}:\mathbf{D}\times\mathbf{Q}^t\rightarrow \mathbf{R}^{t}$ such that for a sequence of queries $\mathbf{Q}=(q_{1},\ ...,\ q_{t})$, it outputs the response file identifiers $\mathbf{R}=(ids(q_{1}),\ ...,\ ids(q_{t}))$.
\smallskip
\\
$\bullet$ the $access\ injection\ pattern$ is similar to the $access\ pattern$, except that we need to recognize the injected file identifiers such that $\mathbf{aip}:\mathbf{D}\times\mathbf{Q}^t\rightarrow \mathbf{IR}^{t}$, where $\mathbf{IR}=(iids(q_{1}),\ ...,\ iids(q_{t}))$ represents the injected file identifiers for the queries.

We note that most of LAA schemes \cite{DBLP:conf/ndss/IslamKK12, DBLP:conf/ccs/CashGPR15, DBLP:conf/ccs/NingHPYL0D21} rely on the access pattern; while ZKP \cite{DBLP:conf/uss/ZhangKP16} exploits the access injection pattern for injection attack.  
We also formally define the search and volume patterns used in our attacks.  
\smallskip
\\
%\begin{itemize}
$\bullet$ the $search\ pattern$ is the function family that $\mathbf{sp}:{\mathbf{D}}\times\mathbf{Q}^t\rightarrow \mathbf{M}^{t \times t}$ where for a sequence of queries $\mathbf{Q}=(q_{1},\ ...,\ q_{t})$, it outputs a binary $t\times{t}$ matrix $\mathbf{M}$ such that $\mathbf{M}[i,\ j]=1$ if $q_{i}=q_{j}$ and otherwise, $\mathbf{M}[i,\ j]=0$.
\smallskip
\\	
$\bullet$  the $response$ $length$ $pattern$ is the function family that $\mathbf{rlp}:\mathbf{D}\times\mathbf{Q}^t\rightarrow \mathbf{RL}^{t}$ where, for a sequence of queries $\mathbf{Q}$, it outputs the number of the response files $\mathbf{RL}=(\mathbf{\#D}(q_{1}),\ ..., \mathbf{\#D}(q_{t}))$.
\smallskip
\\	
$\bullet$ the $response$ $size$ $pattern$ is the function family that $\mathbf{rsp}:\mathbf{D}\times\mathbf{Q}^t\rightarrow \mathbf{RS}^{t}$
where, for a sequence of queries $\mathbf{Q}$, it outputs the size of the response files $RS=(\sum_{f\in{D(q_{1})}}|f|_{w},...,\sum_{f\in{D(q_{t})}}|f|_{w})$.

\subsection{Attack Model}

    To capture a general injection attack model, we enable the adversary to actively generate and inject files. 
    We regard the adversary as an \textit{honest-but-curious server} who follows the protocols but can still inject files to the client database\footnote{Or one may regard the adversary as an \textit{active observer} who can generate injected files for the client and observe the query response.}. 
    We divide the whole attack process into three stages (see Figure \ref{AttackModel}).
	\smallskip
	\\
    (1) In the baseline phase, the adversary observes the client’s query leakage $\mathbf{LP}^{*}$ as pre-knowledge (provided the client sends queries to the server).  
    This step is of importance for injection attacks (except for ZKP \cite{DBLP:conf/uss/ZhangKP16} and single-round attack \cite{DBLP:conf/eurosp/PoddarWLP20}). 
    This is because the adversary should obtain the correlation between the keywords and response files from the baseline’s observations. 
    The idea is to compare the volume difference between the response results before (baseline) and after injection for query recovery. 
    %We note some injection strategies may also depend on the baseline observations (e.g., decoding attack \cite{DBLP:conf/ndss/BlackstoneKM20}). 
    \\
    (2) In the injection, the adversary should carefully generate files $\mathbf{F}$ by using its known keywords $\mathbf{W}$ and the information obtained from the baseline.   
    Next, the client encrypts the files and uploads them to the server.
    \\
    (3) During the recovery phase, the adversary obtains the target queries’ leakages $\mathbf{LP}$ and recovers them by combining all the known information, namely $\mathbf{W}$, $\mathbf{LP}^{*}$, $\mathbf{F}$, and $\mathbf{LP}$.
	\smallskip
	
	We let $\mathbf{Q}$ denote the target query set observed by the adversary and $\mathbf{Q}_{r}$ denote the query recovery results. 
	We refer to the recovery rate as 
	${Rer}:\mathbf{\#{CorrectPred(Q_{r})}}/\#\mathbf{Q}$. We denote the number of injected files as ${ILen}:\#\mathbf{F}$ and the word count of the injected files as ${ISize}:|\mathbf{F}|_{w}$, where $\mathbf{F}$ is the set of injected files.
	\textit{The goal of injection attacks is to obtain a high \textit{Rer} and make \textit{ILen} and \textit{ISize} as “small” as possible by only exploiting the known keyword universe and pre-injection leakage.} 

\section{Practical Volumetric Injection Attacks}\label{Section Attack} 
Given the keyword universe $\mathbf{W}$, VIAs can recover the client’s queries {on the encrypted database $\mathbf{D}$ by observing the \textit{vp} from the response results.} 
A well-design and practical injection attack should limit the number and size of injected files. 
%We design two new attacks with less injection volume than prior schemes \cite{DBLP:conf/eurosp/PoddarWLP20, DBLP:conf/ndss/BlackstoneKM20}.  
Our first attack, BVA, uses a dynamic parameter $\gamma$ to flexibly set the size of files to adjust the recovery rate.   
With a slight loss of recovery rate, it can significantly reduce the size of injection.  
Unlike the decoding attack, it does not need to calculate $\gamma$ fully and accurately, reducing the computational cost.
To further optimize the injection size and boost the recovery rate in the worst-case scenario (i.e., $\gamma=\#\mathbf{W}/2$), we propose the second attack called BVMA by $twisting$ both the volume and search pattern.  
{By carefully controlling the size of each injected file, the BVMA can ensure that each known keyword has a distinct injection volume.} 
For any query, it can reveal the underlying keyword according to the difference of the response results before and after injection.
We also provide a generic conversion that enables VIAs (e.g., \cite{DBLP:conf/ndss/BlackstoneKM20,DBLP:conf/eurosp/PoddarWLP20} and our attacks) to counter the TC.

    \begin{figure}[!t]
    	\centering
    	\includegraphics[width=0.9\linewidth]{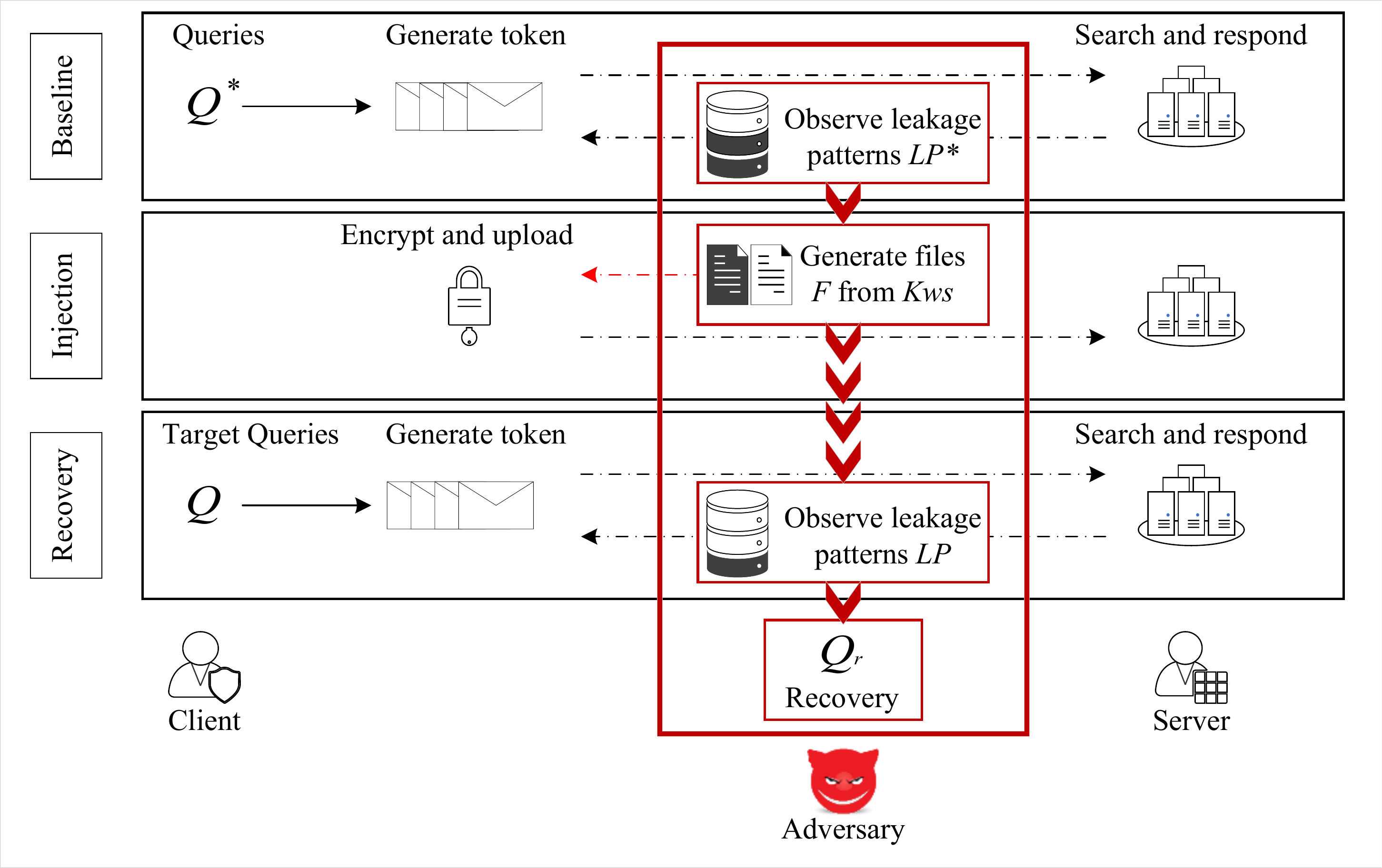}
    	\caption{Attack model.}
    	\label{AttackModel}
    \end{figure}

\subsection{Binary Variable-Parameter Attack}\label{Section BVA}

The BVA (see Algorithm \ref{BVA}) works as follows. 
In the baseline phase, the adversary observes and records the response sizes of unknown queries.
During the injection phase, it generates and injects files according to the injection parameter $\gamma$ in a $binary$ way. 
In the last phase, the adversary observes an additional sequence of the client's queries $\boldsymbol{{Q}}=({q}_{1},...,{q}_{m})$ as the attack targets with response size  $\boldsymbol{RS}=(rs_{1},...,rs_{m})$. 
It aims to recover all the queries in $\boldsymbol{{Q}}$. 
%We describe each phase in detail below. 

\smallskip

\noindent $\mathbf{BVA-Baseline.}$ The adversary observes the response size $\boldsymbol{\widetilde{RS}}$ for a sequence of queries $\boldsymbol{\widetilde{Q}}$ (line \ref{A1L2}).

\smallskip

\noindent $\mathbf{BVA-Injection.}$ {The adversary identifies the keyword universe $\mathbf{W}$ with the set $\{0,...,\#\mathbf{W}-1\}$ and indicates the injected files with set $\mathbf{F}=\{f_1,...,f_{\lceil\log\#\mathbf{W}\rceil}\}$.}  %  which has exactly $\lceil\log\#\mathbf{W}\rceil$ files
First, the adversary adaptively {selects} the injection parameter $\gamma$ satisfying $\gamma\in\mathbb{N}$ and $\gamma\geq\#{\mathbf{W}}/2$ to ensure each file can accommodate half of the keywords (line \ref{A1L7}). 
Second, each file $f_{i}$ has a size of $\gamma\cdot{2^{i-1}}$ and contains keywords whose $i$th bit is equal to $1$ so that when $w_{i}$ is queried, the total response size of injected files is $\gamma\cdot{i}$ (lines \ref{A1L8}-\ref{A1L12}). 

Note an example (see Figure \ref{Our-Decoding}, Appendix \ref{AppenComp}) shows the differences between the decoding attack and ours in terms of injection length and size.

\smallskip

\noindent $\mathbf{BVA-Recovery.}$ The adversary observes  $\boldsymbol{RS}=(rs_{1},...,rs_{n})$ again  for the target queries $\boldsymbol{Q}=(q_{1},...,q_{n})$. For a query $q_{i}\in \boldsymbol{Q}$ whose response size is $rs_{i}\in\boldsymbol{RS}$, it traverses $\boldsymbol{\widetilde{RS}}$ (obtained from the $baseline$ phase) to get a $u\in[\#{\mathbf{W}}]$, satisfying the condition:  
$
   rs_{i}-u\cdot \gamma = \widetilde{rs}_{j}.\nonumber
$
The adversary then recovers the query $q_{i}$ with $w_{u}$ (lines \ref{A1L18}-\ref{A1L21}).

\newcounter{Clactr}
\newenvironment{claim}[1]{\stepcounter{Clactr}\normalfont\bfseries\par\noindent{Claim \arabic{Clactr}.}\space#1}{}%\space#1
\newenvironment{claimproof}[1]{\normalfont\itshape \par\noindent{Proof.}\space#1}{\hfill $\blacksquare$}

\smallskip

Let $P$ be the probability distribution over the keyword universe and $P(w_{u})$ be the probability that the client will query the keyword $w_{u}$. 
We use $\widetilde{rs}_{w_{u}}$ to represent the response size of $w_{u}$ observed in the $baseline$ phase. 
We formalize the probability of incorrect recovery as follows. 
%\smallskip
%
\begin{claim}
    \mdseries\itshape For any query $q_{w_{i}}$ and $\gamma\geq\#\mathbf{{W}}/2$, the probability that $\mathbf{BVA-Recovery}$($q_{w_{i}}$) outputs an incorrect $w_{u}$ is
    \begin{equation}
        Pr(w_{u}\neq w_{i})=\sum_{\substack{u\in [\#\mathbf{{W}}],u\neq i,\\ {rs}_{w_{i}} = \widetilde{rs}_{w_{u}}+ u*\gamma } } {P(w_{u})} \nonumber
    \end{equation}
\end{claim}

\begin{claimproof}
    \normalfont Consider the client made a query $q_{w_{u}}$ with probability $P(w_{u})$, and its response size $\widetilde{rs}_{w_{u}}$ was observed in the baseline.  
    In the recovery phase, the adversary observed that the target query $q_{w_{i}}$'s response size is $rs_{w_{i}}$ after injection. 
    We then have three cases.  
    (1) $rs_{w_{i}}<\widetilde{rs}_{w_{u}}$.  
    %
    %, which means that the adversary will not output a guess $w_{u}$. 
    %This is because 
    The response size (after injection) must be $> \widetilde{rs}_{w_{u}}$ (before injection). 
    In this case, the adversary will not {guess any keywords}. 
    (2) $rs_{w_{i}}\geq\widetilde{rs}_{w_{u}}$ but $\gamma\nmid rs_{w_{i}}-\widetilde{rs}_{w_{u}}$. 
    The adversary {will not} output any guess according to the recovery algorithm. 
    (3) $rs_{w_{i}}\geq\widetilde{rs}_{w_{u}}$ and $\gamma\mid rs_{w_{i}}-\widetilde{rs}_{w_{u}}$. 
    The adversary {will} guess an incorrect $w_{u}$ so that $u=(rs_{w_{i}}-\widetilde{rs}_{w_{u}})/\gamma$ if $u\neq i$; otherwise, it {will} reveal the correct keyword.
\end{claimproof}

We see that only the third case can cause an incorrect recovery, which implies that the recovery rate depends on the selection of $\gamma$. 
A well-selected $\gamma$ can greatly reduce the occurrence of the third case and make the adversary achieve the same recovery rate as the decoding attack. %but indeed with a small injection size. 
Even in the worst case (i.e.,  $\gamma=\#\mathbf{W}/2$), the recovery rate can still maintain $>70\%$ in Enron and Lucene. 
%And the injection is much smaller than prior VIAs. 
We further state that the binary injection approach can restrict the injection length to $\log\mathbf{\#\mathbf{W}}$, which outperforms prior schemes \cite{DBLP:conf/ndss/BlackstoneKM20, DBLP:conf/eurosp/PoddarWLP20}.   
Note more details will be given in the experiments (see Section \ref{Section Exper}). 

%The total number of words, also known as injection size, is another crucial factor for us to consider. 
%
%We show that the injection size can be limited to a level which outperforms previous works \cite{DBLP:conf/ndss/BlackstoneKM20, DBLP:conf/eurosp/PoddarWLP20}.
%

\begin{claim}
    \mdseries\itshape For the keyword universe $\mathbf{W}$ and the injection parameter $\gamma\geq\#\mathbf{{W}}/2$, the total injection size incurred by $\mathbf{BVA-Injection}$($\mathbf{W}$) is $\mathcal{O}(\gamma\#\mathbf{W})$.
\end{claim}
\smallskip

\begin{claimproof}
    \normalfont Through the binary injection, the BVA requires $\log\#\mathbf{W}$ files in total, and the size of each file is $\gamma\cdot2^{i-1}$ for $i\in[\log\#\mathbf{W}]$. Therefore, the total injection size during this phase is $\gamma\cdot(2^{0}+2^{1}+...+2^{\log\#\mathbf{W}-1})=\mathcal{O}(\gamma\#\mathbf{W})$.
\end{claimproof}

The attack can recover multiple queries with a high recovery rate (e.g., averagely $80\%$ in Enron) and a small injection volume by leveraging the \textit{rsp}, even the file contents and co-occurrences of keywords are hidden. 
%  (except {\color{red}the total size of response results - i.e., what??? [meanwhile, is this size related to file contents and co-occurrences of keywords???]})
It does not require any knowledge of query and file distribution.  
The BVA is a more practical VIA compared to prior attacks. 

\subsection{Binary Volumetric Matching Attack}\label{Section WLIA}

%The BVA significantly reduces the time overhead and injection size but slightly sacrifices the recovery rate. 
Unlike the decoding attack relying on $\textit{offset}$, the BVA adjusts a proper $\gamma$ for different datasets. 
This could affect the recovery rate and injection size. 
{For example, setting a small $\gamma$ can decrease the injection size (compared with decoding) but could fail to distinguish some queries.} 
To refine the attack performance, we introduce the BVMA, which is fully independent of any $\textit{offset}$ or $\gamma$. 
The BVMA is the first injection attack that $combines$ the response length and size patterns (as well as the \textit{sp}). 
The combination can filter known candidate keywords precisely with less injection size, so that it enhances the recovery rate.  
See the BVMA in Algorithm \ref{BVMA}. 

\begin{algorithm}[!t]
    \footnotesize
    \caption{BVA.}
    \label{BVA}
    \SetKwFunction{FMain}{Baseline}\label{BVA-Baseline}
    \SetKwProg{Bn}{procedure}{}{end}
    \Bn{\FMain{$\mathbf{\widetilde{Q}}$}}{
        observe the response size $\mathbf{\widetilde{RS}}=(\widetilde{rs}_{1},...,\widetilde{rs}_{m})$ for query in $\mathbf{\widetilde{Q}}=(\widetilde{q}_{1},...,\widetilde{q}_{m})$\; \label{A1L2}
        \KwRet $\mathbf{\widetilde{RS}}$\;
    }
    
    \SetKwFunction{FMain}{Injection}\label{BVA-Injection}
    \SetKwProg{In}{procedure}{}{end}
    \In{\FMain{$\mathbf{W}$}}{
        $\mathbf{F}\gets\emptyset$\;
        select an injection parameter $\gamma=\{\gamma\in\mathbb{N}\cap\gamma\geq\#{\mathbf{W}}/2\}$\; \label{A1L7}
		\For{$i=1 \to {\log{\lceil \#\mathbf{W} \rceil}}$}{\label{A1L8}
			generate the files $f_{i}$ that contains the keywords $w\in\mathbf{W}$ whose $i$th bit is 1\;
			pad $f_{i}$ until its $size = {\gamma\cdot 2^{i-1}}$\;
			$\mathbf{F}=\mathbf{F}\cup f_{i}$\;
		}\label{A1L12}
        \KwRet $\mathbf{F}$\;
    }
    
    \SetKwFunction{FMain}{Recovery}\label{BVA-Recovery}
    \SetKwProg{Rn}{procedure}{}{end}
    \Rn{\FMain{$\mathbf{Q}$}}{
        initialize an empty set $\mathbf{Q}_{r}$\;
		observe the response size that $\mathbf{{{RS}}}=({{rs}}_{1},...,{{rs}}_{n})$ for target query in $\mathbf{{Q}}=({q}_{1},...,{q}_{n})$\; \label{A1L17}
		\For{$i=1 \to \#\mathbf{{{RS}}}$}{\label{A1L18}
			find $u\in \#\mathbf{{W}}$ satisfying that ${{rs}}_{i}-u\cdot \gamma =\widetilde{rs}_{j}$ for some $\widetilde{rs}_{j}\in\mathbf{\widetilde{RS}}$\;
			add ${w}_{u}$ to $\mathbf{Q}_{r}$\;
		}\label{A1L21}
		\KwRet $\mathbf{Q}_{r}$\;
    }
    
\end{algorithm}

\noindent $\mathbf{BVMA-Baseline.}$ This is similar to the BVA-Baseline; but we get queries’ response length $\widetilde{\mathbf{RL}}$, response size $\widetilde{\mathbf{RS}}$, and frequency from the \textit{sp} after a period of observation (line \ref{A2L2}). 
\smallskip

\noindent $\mathbf{BVMA-Injection.}$ We inject files in a binary manner. 
Different from the BVA (where the size of injected files relies on the $\gamma$), we set the size of file $f_{i}$ with $size = {2^{i-1}}+\#\boldsymbol{{W}}/2$ for $i\in[\log\#\mathbf{W}]$ containing the keywords whose $i$th bit is 1 (lines \ref{A2L7}-\ref{A2L11}). 
For the keyword $w_{u}\in\mathbf{W}$, we then record its injection length as $\#file_{u}$ and injection size as equal to $u+\#file_{u}\cdot\#\mathbf{W}/2$. 
Each keyword has a unique pair of injection size and length and thus we can distinguish different keywords with a relatively high probability.

\smallskip

\noindent $\mathbf{BVMA-Recovery.}$ We combine the leakage pattern $\boldsymbol{{LP}}\in\{\boldsymbol{{rlp}}$, $\boldsymbol{{rsp}}$, $\boldsymbol{{sp}}\}$ to filter candidate keywords. 
For a target query $q_{i}\in\mathbf{Q}$, we generate a candidate set $\mathbf{CA}$ by adding all $(j,{w}_{u})$ to it for $\forall  j\in[\#\widetilde{\mathbf{RS}}],u\in[\#{\mathbf{W}}]$ (line \ref{A2L16}). 
In the beginning, we exploit the \textit{rsp} to remove $(j,{w}_{u})$ from $\mathbf{CA}$ such that the pair does not satisfy  ${{rs}}_{i}-\#file_{u}\cdot\#\mathbf{{W}}/2-u=\widetilde{rs}_{j}$ (lines \ref{A2L21}-\ref{A2L23}). 
This condition means that $\widetilde{rs}_{j}$ and $w_{u}$ are not the original response size and underlying keyword of $q_{i}$, respectively. 
In the second filtering stage, we exclude those candidate keywords that dissatisfy ${{rl}}_{i}=\#{file}_{u}+\widetilde{rl}_{j}$ for ${w}_{u}\in \mathbf{CA}$ by exploiting the \textit{rlp} (lines \ref{A2L24}-\ref{A2L26}). 
If the condition does not hold, we confirm that the $\widetilde{rl}_{j}$ and $w_{u}$ are not the response length of $q_{i}$ before injection and the correct recovery keyword, respectively.
We then remove the pair $(j,w_{u})$ from the candidate set. 
Finally, we can use the \textit{sp} to identify the closest frequency of the queries between the $baseline$ and $recovery$ phases from the remaining candidate keywords of $\mathbf{CA}$ (line \ref{A2L28}). 
We note that the \textit{sp} moderately improves the recovery rate. 
Even without using it, the BVMA can still achieve about 80\% recovery, e.g., in Enron (see Appendix \ref{AppenBVMASP}).
%\smallskip
%
\begin{claim}
    \mdseries\itshape For any target query $q_{w_{i}}$, the probability that $\mathbf{BVMA-Recovery}$($q_{w_{i}}$) outputs an incorrect $w_{u}$ is
    \begin{equation}
        Pr(w_{u}\neq w_{i})\leq\sum_{\substack{u\in [\#{\mathbf{W}}],u\neq i,\\ {rs}_{w_{i}}-\widetilde{rs}_{w_{u}}=u+\#{file}_{u}\cdot\#{\mathbf{W}}/2,\\ rl_{w_{i}}-\widetilde{rl}_{w_{u}} = \#{file}_{u} } } {P(w_{u})} \nonumber
    \end{equation}
\end{claim}
\begin{claimproof}
    \normalfont Assume that the adversary observed the response size $\widetilde{rs}_{w_{u}}$ and length $\widetilde{rl}_{w_{u}}$ of a query $q_{w_{u}}$ with probability $P(w_{u})$ during the baseline.
    %For the client query $q_{w_{i}}$, its response size $\widetilde{rs}_{w_{i}}$ is observed during the $baseline$.  
    In the recovery phase, the adversary observed the target query $q_{i}$, whose response length is $rl_{w_{i}}$ and response size is $rs_{w_{i}}$ (after injection). 
    If it determines $w_{u}$ as the underlying keyword, that means the difference of the \textit{rsp} before ($\widetilde{rs}_{w_{u}}$) and after ($rs_{w_{i}}$) injection is equal to the total injection size ($u+\#{file}_{u}\cdot\#\mathbf{W}/2$) and the difference on \textit{rlp} before and after injection is the injection length ($\#{file}_{u}$).
    The adversary chooses $w_{u}$ as an incorrect recovery only when the above two conditions (for response length and size) and $u\neq{i}$ are satisfied. 
    {The probability is no more than $\sum{P(w_{u})}$ as we can use other leakages (e.g., the frequency exploited by using the \textit{sp}) to further eliminate those incorrect keywords.} 
\end{claimproof}

We also provide a claim and its proof for the injection size. 
%\smallskip
\begin{claim}
    \mdseries\itshape For the keyword universe $\mathbf{W}$, the total size output by the $\mathbf{BVMA-Injection}$($\mathbf{W}$) is $\mathcal{O}(\#\mathbf{W}\log\#\mathbf{W})$.
\end{claim}
\smallskip

\begin{claimproof}
    \normalfont Similar to the BVA, the attack requires $\log\#\mathbf{W}$ files but with size of ${2^{i-1}}+\#\boldsymbol{{W}}/2$ for $i\in[\log\#\mathbf{W}]$. Thus, the total injection size is  $(2^{0}+2^{1}+...+2^{\log\#\mathbf{W}-1})+\#\boldsymbol{{W}}\cdot\log\#\mathbf{W}/2=\mathcal{O}(\#\mathbf{W}\log\#\mathbf{W})$.
\end{claimproof}
\smallskip

\begin{algorithm}[!t]
    \footnotesize
    \caption{BVMA.}
    \label{BVMA}
    \SetKwFunction{FMain}{Baseline}\label{BVMA-Baseline}
    \SetKwProg{Bn}{procedure}{}{end}
    \Bn{\FMain{$\mathbf{\widetilde{Q}}$}}{
    	observe and record the baseline response size $\mathbf{\widetilde{RS}}$, the response length $\mathbf{\widetilde{RL}}$, and the frequency $\mathbf{\widetilde{Freq}}$ for query in $\mathbf{\widetilde{Q}}=(\widetilde{q}_{1},...,\widetilde{q}_{m})$\;\label{A2L2}
		\KwRet $(\mathbf{\widetilde{RS}},\mathbf{\widetilde{RL}},\mathbf{\widetilde{Freq}})$\;
    }
    
    \SetKwFunction{FMain}{Injection}\label{BVMA-Injection}
    \SetKwProg{In}{procedure}{}{end}
    \In{\FMain{$\mathbf{W}$}}{
        $\mathbf{{F}}\gets\emptyset$\;
		\For{$i=1 \to {\log{\lceil \#\mathbf{{W}} \rceil}}$}{\label{A2L7}
			generate the file $f_{i}$ containing the keywords $w$ in $\mathbf{{W}}$ whose $i$th bit is 1\;
			pad $F_{i}$ until its $size = {2^{i-1}+\#\mathbf{{W}}/2}$\;
			$\mathbf{{F}}=\mathbf{{F}}\cup f_{i}$\;
		}\label{A2L11}
        \KwRet $\mathbf{F}$\;
    }
    
    \SetKwFunction{FMain}{Recovery}\label{BVMA-Recovery}
    \SetKwProg{Rn}{procedure}{}{end}
    \Rn{\FMain{$\mathbf{Q}$}}{
        initialize an empty set $\mathbf{Q}_{r}$\;
        gather the new observed response size $\mathbf{RS}$, response length $\mathbf{RL}$, and frequency $\mathbf{Freq}$ for victim's target queries $\mathbf{Q}$\; \label{A2L16}
		\For{$i=1 \to \#\mathbf{Q}$}{
			$\mathbf{CA}\gets\emptyset$\;
			\For{$\widetilde{rs}_{j}\in\mathbf{\widetilde{RS}},u\in [\#\mathbf{{W}}]$}{
			    add $(j, {w}_{u})$ to $\mathbf{CA}$\; \label{A2L20}
		        \If{${{{rs}}_{i}-{\#\mathbf{{file}}_{u}}\cdot\#\mathbf{{W}}/2-u\neq\widetilde{rs}_{j}}$} 
		        {\label{A2L21}
		            remove $(j, {w}_{u})$ from $\mathbf{CA}$\;
		        }\label{A2L23}
				   
				\If{${{rl}}_{i}\neq{\#\mathbf{{file}}_{u}}+\widetilde{rl}_{j}$}
				{\label{A2L24}
					remove $(j, {w}_{u})$ from $\mathbf{CA}$\;
					
				}\label{A2L26}

			}
			find the minimum value of $|{{freq}}_{i}-\widetilde{freq}_{j}|$ for $(j, {w}_{u})\in\boldsymbol{CA}$ \label{A2L28} {\tcp*{non essential step}}
		
			add ${w}_{u}$ to $\boldsymbol{Q}_{r}$\;
			
		}
		\KwRet $\mathbf{Q}_{r}$\;
    }
    
\end{algorithm}

We see that the BVMA inherits the advantages of the BVA (except the usage of $\gamma$) and further provides an improvement on injection size, for example, the BVMA outperforms the best case's BVA (i.e., when $\gamma=\#\mathbf{W}/2$, see Figure \ref{EnronTrendRerVol}). % , the BVMA still outperforms the BVA. % (let alone other state-of-the-art volumetric injection attacks).

\subsection{Attacks against Threshold Countermeasure}\label{AttackTC}

Threshold countermeasure (TC) \cite{DBLP:conf/uss/ZhangKP16} can prevent large-size files (e.g., \#word > the threshold $T$) from being injected into the database. 
The $T$ could be set relatively small in practice to counter injection attacks, without seriously affecting the functionality of dynamic SSE. %(e.g., $T=500$ when $\#\mathbf{W}=3000$) 
For example, only $3\%$ of the files in Enron ($\#\mathbf{W}=3,000$) contain more than 500 words. 
If employing TC with $T=500$, we could skip these $3\%$ of files from indexing. 
Under this setting, a dynamic SSE can effectively resist prior attacks (e.g., \cite{DBLP:conf/ndss/BlackstoneKM20,DBLP:conf/eurosp/PoddarWLP20} and ours).
%The TC defenses effectively in previous attacks and prevents both our attacks.
%{\color{red}[we many need to explain how it affects???]} 
This is because each injected file must contain a considerably large number of words (e.g., $\geq\#\mathbf{W}/2$ words), 
in particular, the single-round \cite{DBLP:conf/eurosp/PoddarWLP20} and decoding \cite{DBLP:conf/ndss/BlackstoneKM20} attacks force an injected file to contain $\mathcal{O}(m\#\mathbf{W})$ and $\mathcal{O}(\textit{offset}\cdot\#\mathbf{W})$ words. 
% Note that TC  severely harm our attacks performance. 

We design a generic transformation method to “protect” VIAs from TC (see Algorithm \ref{TCPadding}). 
The transformation takes the threshold $T$ and the set of large files $Basic\_F$ as input and runs the \textit{cutting} and \textit{refilling},  where $Basic\_F$ is the output of a concrete VIA algorithm, e.g., BVA.  
%The transformation includes two phases: \textit{cutting} and \textit{refilling}.
In Algorithm \ref{TCPadding}, we cut large files (with \#word>$T$) into smaller ones to satisfy the threshold of TC. 
To eliminate the impact of file cutting on recovery rate, we keep the injection length $\#file_{u}$ or size $|file_{u}|_{w}$ of each keyword $w_{u}\in\mathbf{W}$ consistent before and after Algorithm \ref{TCPadding}.
For example, in the single-round attack \cite{DBLP:conf/eurosp/PoddarWLP20}, % strongly relies on the difference of response length before and after injection. 
for any keyword $w_{u}$, we keep its injection length $\#file_{u}$ unchanged (after cutting) in order to avoid bring any impact to the recovery phase. 
In the decoding \cite{DBLP:conf/ndss/BlackstoneKM20} and our attacks, we keep keywords' injection size consistent (by refilling).

During the \textit{cutting} phase (lines \ref{A3L7}-\ref{A3L9}), for each $f\in{Basic\_F}$, we extract all keywords $C\_W$ from $f$. 
We then generate $\lceil \#C\_W/T \rceil$ files for $Cut\_F$ by following 1) the size (word count) of each file is $T$; and 2) each file contains up to $T$ distinct (non-overlapped) keywords from $C\_W$ so that each keyword $w_{u}$'s injection length $\#{file_{u}}$ is 1 (note the number remains the same before and after cutting).  
We finally have sets of small-size files to bypass TC while maintaining the consistency of the keywords' injection length.

In the \textit{refilling} phase (lines \ref{A3L11}-\ref{A3L15}), for each file $c\_f\in{Cut\_F}$, we generate two file sets $Ref\_F_{1}$ and $Ref\_F_{2}$. 
$Ref\_F_{1}$ is constructed by refilling $|f|_{w}/T-2$ files, in which each file is with size of $T$ and it contains all the keywords in $c\_f$.
Next, we generate as few files as possible (for $Ref\_F_{2}$) to accommodate all the keywords in $c\_f$, with each file having a size of $|f|_{w}-(\lceil |f|_{w}/T \rceil - 1)\cdot{T}$. 
We ensure that the injection size of keywords in $c\_f$ is still $|f|_{w}$ after the refilling. 
%We provide the following claim and proof for the consistency of the keywords' injection size.

\smallskip
\begin{claim}
    \mdseries\itshape For a large-size file set $Basic\_F$ output by ${Injection}(\mathbf{W}$) and a threshold $T$, the injection size of each keyword $w_{u}$ output by the algorithm ${Injection\_shard}(T, Basic\_F)$ is the same as that of ${Injection}(\mathbf{W}$), where ${Injection}(\mathbf{W}$) is the injection phase of a VIA (e.g., \cite{DBLP:conf/eurosp/PoddarWLP20,DBLP:conf/ndss/BlackstoneKM20}, BVA, BVMA).
\end{claim}
\begin{claimproof}
    \normalfont During the $cutting$ phase, for a file $f$ from the set ${Basic\_F}$, we produce multiple small files $Cut\_F$ containing different keywords with size $T$, so the injection size of each keyword is $T$. 
    In the $refilling$ phase, we generate $\lceil |f|_{w}/T \rceil -2$ files, in which each file is with size $T$ and contains all the keywords in a small file $c\_f\in{Cut\_F}$. 
    For each keyword $w_{u}$, we generate another file with size of $|f|_{w}-(\lceil |f|_{w}/T \rceil - 1)\cdot{T}$ to include the keyword. 
    After the above steps, the injection size of each keyword is $size=T+(\lceil |f|_{w}/T \rceil -2)\cdot{T}+|f|_{w}-(\lceil |f|_{w}/T \rceil - 1)\cdot{T}=|f|_{w}$, which is the same as the amount before the transformation.  
\end{claimproof}
%\smallskip
%
 
We say that the injection length strongly relies on the word count of each file $f$ and the number of files in ${Basic\_F}$. 
Our attacks, requiring fewer injected files than others, can naturally provide a practical performance against TC. 
This is proved by the experiments (see Figure \ref{TCFigure}). 
For example, in Enron with $\#\mathbf{W}=3,000$, setting $T=500$, $\gamma=\#\mathbf{W}/2$ for the BVA, and $m=\#\mathbf{W}/2$ for the single-round attack,  
we see that the lower bounds of the injected files for our attacks are approximately $10^{5}$ and $10^{3}$ which are at least $10^{2}\times$ less than that of the single-round ($10^{7}$) and decoding ($10^{9}$) attacks.

\section{Evaluation}\label{Section Exper}

We compared our attacks with the multiple-round and single-round attacks~\cite{DBLP:conf/eurosp/PoddarWLP20}, the search and decoding attacks~\cite{DBLP:conf/ndss/BlackstoneKM20}, and ZKP \cite{DBLP:conf/uss/ZhangKP16}, under various metrics in the real-world datasets. 
We also evaluated our attacks against well-studied defenses (e.g., TC, padding) {and client's active update.} 
We used Python 3.5 to implement the experiments and run the codes in Ubuntu 16.04 of 64-bit mode with 16 cores of an Intel(R) Xeon(R) Gold 5120 CPU(2.20GHz) and 64 GB RAM.
The codes for the BVA and BVMA are publicly available in  \href{https://github.com/Kskfte/BVA-BVMA}{https://github.com/Kskfte/BVA-BVMA}.

\begin{algorithm}[t] 
	\footnotesize
    \caption{Injection Attacks against TC.}
    \label{TCPadding}
    \SetKwFunction{FMain}{Injection\_shard}\label{Injection-shard}
    \SetKwProg{In}{procedure}{}{end}
    \In{\FMain{$T, Basic\_F$}}{
    	${{Shard\_F}}\gets\emptyset$\;
    	\tcp{Note $|f|_{w}$ is the word count of file $f$}
    	\For{$f\in {{Basic\_F}}$}{
            \eIf{$|f|_{w}>T$}{
                \tcp{Cutting}
                extract keyword set ${C\_W}$ from $f$\;\label{A3L7}
                cut $f$ into $\lceil \#{C\_W}/T \rceil$ smaller files ${{Cut\_F}}$ with each size of $T$, and each contains \textit{different} keywords from ${C\_W}$\; \label{A3L8}
                add ${{Cut\_F}}$ to ${{Shard\_F}}$\;\label{A3L9}
                \tcp{Refilling.}
                \For{$c\_f\in {{Cut\_F}}$}{\label{A3L11}
                    generate $\lceil |f|_{w}/T \rceil - 2$ files ${{Ref\_F_{1}}}$ with each size of $T$ and each containing all keywords from $c\_f$\;\label{A3L12}
                    generate as fewer files $Ref\_F_{2}$ as possible containing all keywords from $c \_f$, and each size is $|f|_{w}-(\lceil |f|_{w}/T \rceil - 1)\cdot{T}$  \; \label{A3L13}
                    add $Ref\_F_{1}$ and $Ref\_F_{2}$ to ${{Shard\_F}}$\;\label{A3L14}
                } \label{A3L15}
                 
            }
            {
                add $f$ to ${{Shard\_F}}$\;
            }
    	}
		\KwRet ${Shard\_F}$\;
    }
    
\end{algorithm}

\subsection{Experimental Setup}
	
\noindent \textbf{Datasets.} We used three real-world datasets with different scales.  
The first one is the Enron email corpus \cite{Enron} between 2000-2002, which contains 30,109 emails. 
The second one is the Lucene mailing list between 2001-2020, with about 113,201 emails from Apache Foundation \cite{Lucene}. 
The last dataset is from Wikipedia \cite{Wikipedia}. 
We extracted the contents of Wikipedia (in 2020) into a subset with 6,154,345 files by using an extraction algorithm in \cite{PlainTextWikipedia}. 
We also applied Python's NLTK corpus \cite{NLTK} to obtain a list of all English words without stopwords and then selected the most frequent words to build the keywords set. 
We set the total extracted keyword universe $\mathbf{W}$ for Enron, Lucene, and Wikipedia as 3,000, 5,000, and 100,000, respectively. 

\begin{table}[!t]
	\centering
	\caption{Descriptions on datasets}
	\label{Datasets}
	
	\begin{tabular}{lccc}
		\hline
		 & Enron & Lucene & Wikipedia\\
		\hline
		\#Keyword & 3,000 & 5,000 & 100,000\\
		\hline
		\#File & 30,109 & 113,201 & 6,154,345\\
		\hline
		QI & GTrend \cite{GoogleTrends} & GTrend & Pageview \cite{Pageviews}\\
		\hline
		Coverage & 260 weeks & 260 weeks & 75 months\\
		\hline
	\end{tabular}
\end{table} 
We used Google Trends \cite{GoogleTrends} of 260 weeks trends between October 2016 and October 2021 to {simulate the real query trends for Enron and Lucene}.
We applied the Pageviews, Toolforge \cite{Pageviews} containing 75 months of page views from July 2015 to September 2021 to generate monthly query trends for Wikipedia. 
We assumed that the client performs 1,000 queries weekly for Enron (Lucene) and 5,000 queries monthly for Wikipedia. 
We regarded the queries generated by the client (within ten weeks/months) as the target in the recovery phase. 
{We put the dataset information in Table \ref{Datasets}, in which} QI represents the source of query trends, and ``coverage" is the time interval of QI.

In the experiments, we made 30 runs for each test and further output the average result.   
We measured the query recovery rate for the compared attacks, in which the rate represents how much percentage of the client's queries the attacks could recover correctly.  % and true negative???if we don't have false positive or true negative, we just say ``recover" - meaning they can recover??? 
% We include error bars against padding except in the BVA (against padding) as the standard deviation of other attacks is relatively small ($<1\%$ on average).  

\noindent \textbf{Keyword leakages.} %Unlike the LAA relying on numerous known data, the injection attacks only leverage known keywords for query recovery. 
We used \textit{keywords (Kws) leakage percentage} to measure the prior knowledge of the adversary. 
We say that the keyword universe could be easily and partially obtained for several reasons: (1) a tiny amount of files could contain a large number of keywords; (2) a public database known by the adversary, which shares similar distribution with the target database, could contain partial target keywords \cite{DBLP:conf/uss/Damie0P21}; (3) some expired or spam emails could leak keywords.  
To test our statements, we present the number of known keywords with file leakage in Figure \ref{KeywordsNo.}. 
We assume that Enron is the client database and Lucene (rather than Wikipedia) is the public database known by the adversary. 
We state that Enron and Lucene are both email datasets and thus, they have similar keyword distributions; and also, they share similar keyword universes. 
In Enron, $0.5\%$ leakage files can lead to the exposure of half of the keyword universe; while obtaining $10\%$ of the leakage files, the adversary can reveal the entire universe. 
Given Lucene as the similar database, the adversary easily reveals $>80\%$ keywords from Enron. 

\noindent \textbf{Observation periods.} Different datasets could require different observation periods. 
For Enron and Lucene, which contain several thousand keywords, we only used a few weeks to observe queries to obtain the statistical information of most keywords. 
We spent more time (dozens of months) on Wikipedia than others to collect sufficient leakage information.
To help the reader to understand the relationship between the occurrence of new queries and observation period, we show some examples in Figure \ref{NewQueries}.
We observed 1,000 queries per week for Enron and Lucene, and 5,000 per month for Wikipedia. 
The number of new queries in Figure \ref{NewQueries} decreases dramatically with the extension of observation period under different datasets and query distributions.
In particular, there are $<10\%$ new queries per week for Enron (resp. Lucene) after the observation lasts $>8$ (resp. 16) weeks. 
We say that 8 (resp. 16) weeks are sufficient for the adversary to observe queries leakage from these datasets. 
Even for Wikipedia, a 32-month observation is practical to obtain statistics information.
Note we also see that the occurrence probability of new queries performs similarly under different query distributions (i.e., real-world and uniform). 
We set the observation period to 8 weeks, 16 weeks and 32 months respectively for Enron, Lucene and Wikipedia.
We further randomly selected the start period for the observation from the coverage given in Table \ref{Datasets}.
\begin{figure}[!t]
	\centering
	
    \includegraphics[width=.56\linewidth, height=0.4\linewidth]{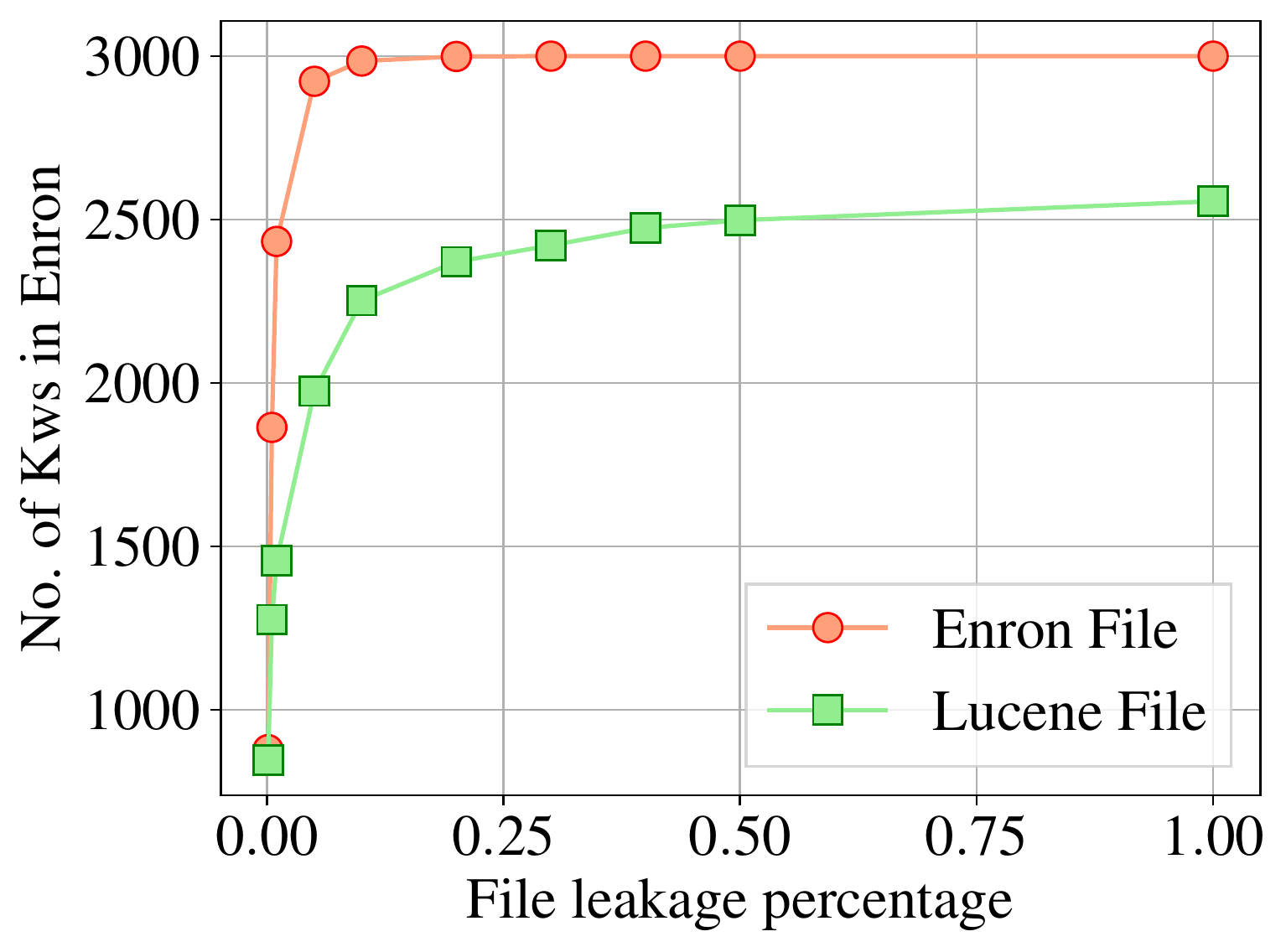}

	\caption{The no. of Enron keywords corresponding to the file leakage. We refer Enron as the client database and Lucene as the public database with similar distributions.} 
	\label{KeywordsNo.}
\end{figure}

\begin{figure*}[!t]
	\centering
	
	\subfigure[Enron]
	{
		\begin{minipage}{.24\linewidth}
			\centering
			\includegraphics[width=\linewidth]{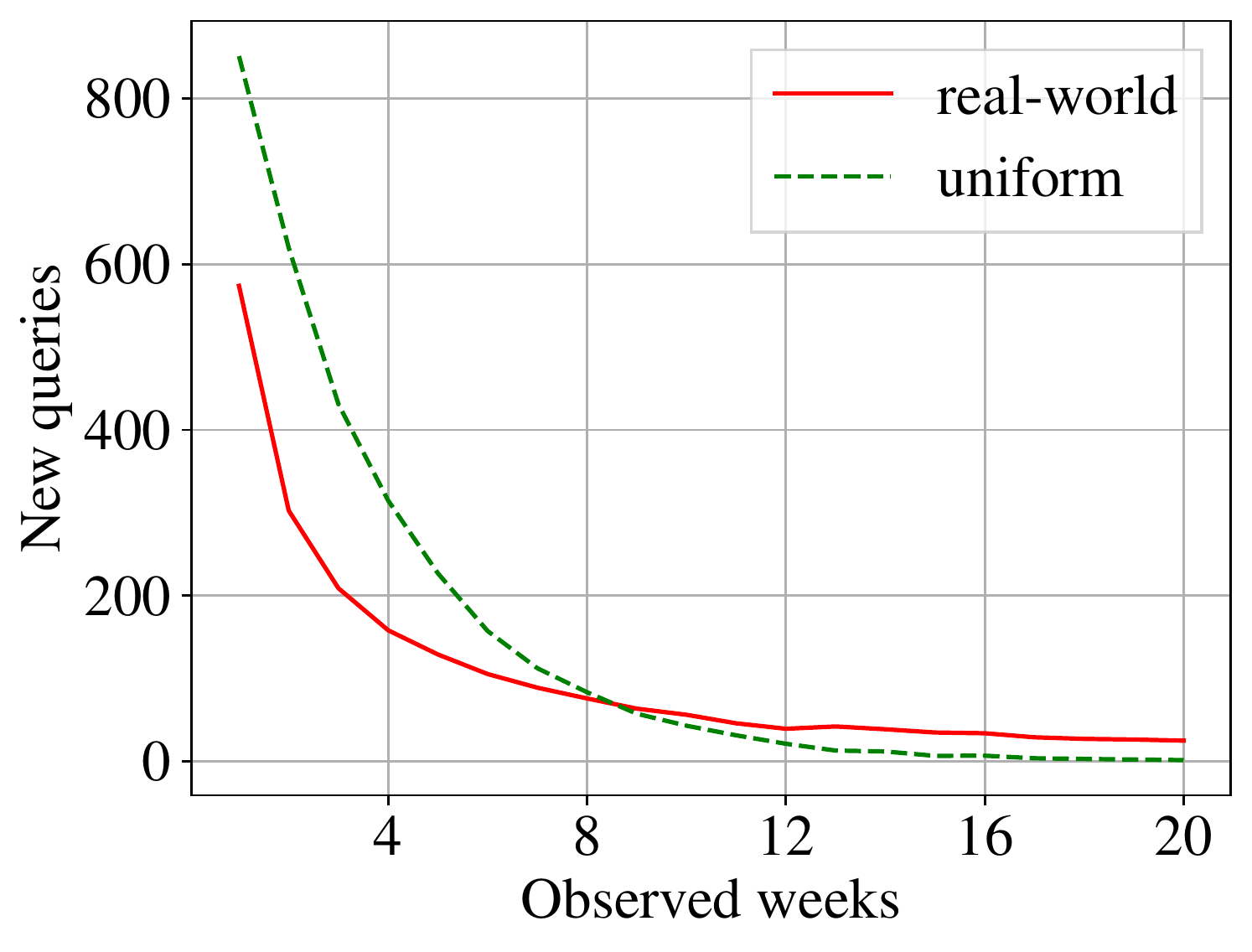}
		\end{minipage}
	}
	\subfigure[Lucene]
	{
		\begin{minipage}{.24\linewidth}
			\centering
			\includegraphics[width=\linewidth]{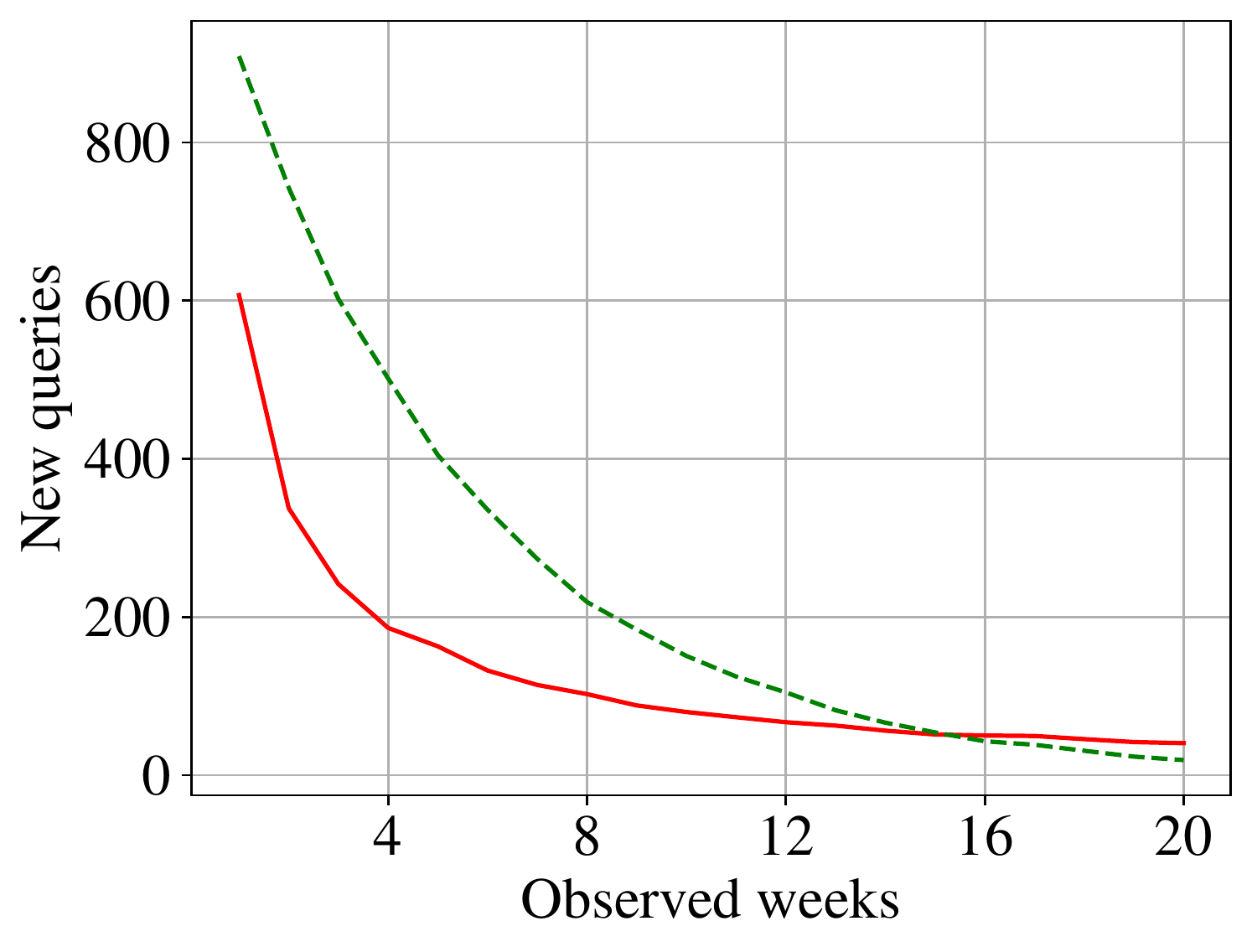}
		\end{minipage}
	}
	\subfigure[WikiPedia]
	{
		\begin{minipage}{.24\linewidth}
			\centering
			\includegraphics[width=\linewidth]{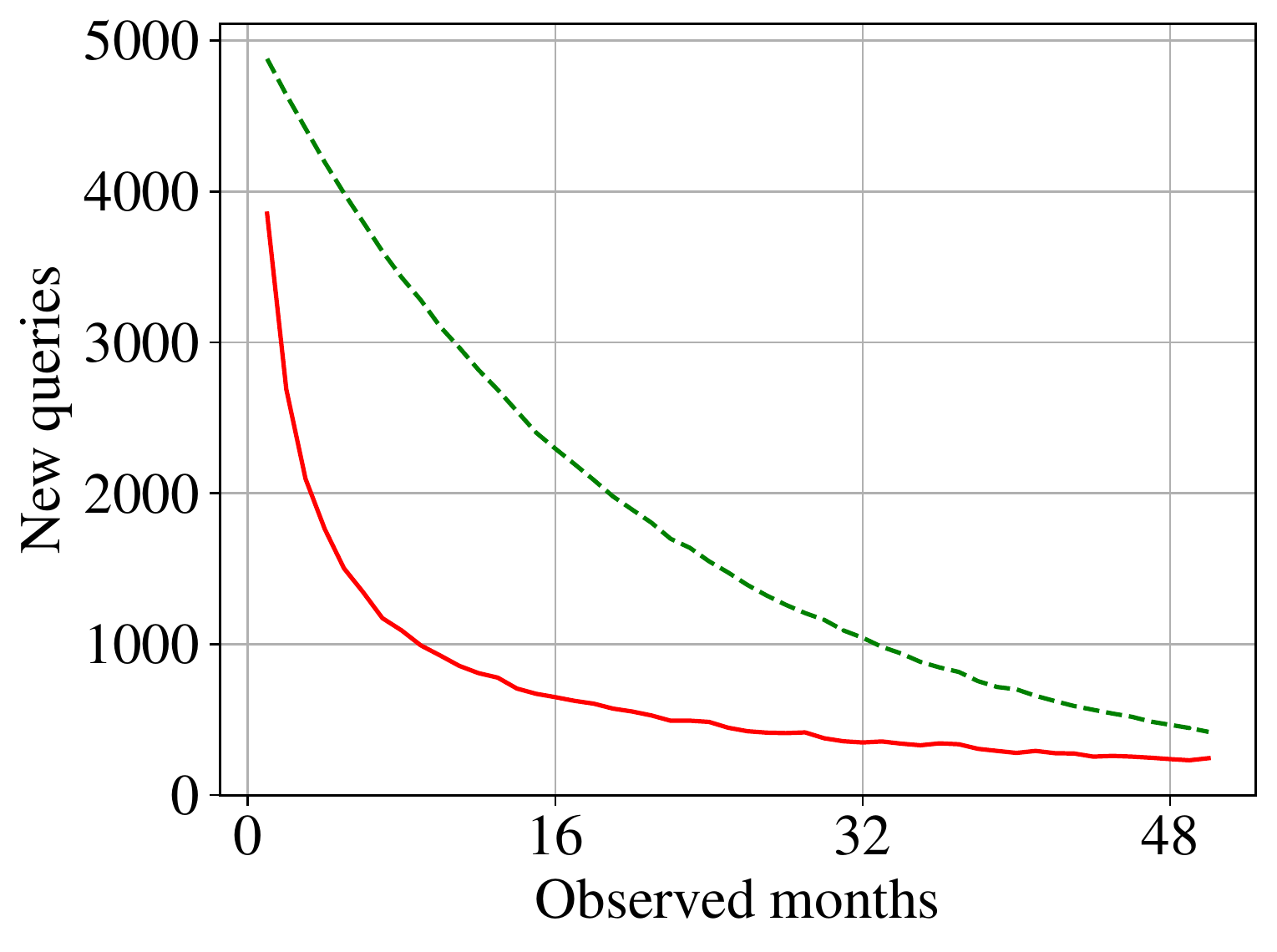}
		\end{minipage}
	}
	\caption{The occurrence probability of new queries as the observation period increases.}
	\label{NewQueries}
\end{figure*}

\begin{figure*}[!t]
	\centering

	\subfigure[Enron]
	{
		\begin{minipage}{.29\linewidth}%
			\centering
			\includegraphics[width=\linewidth]{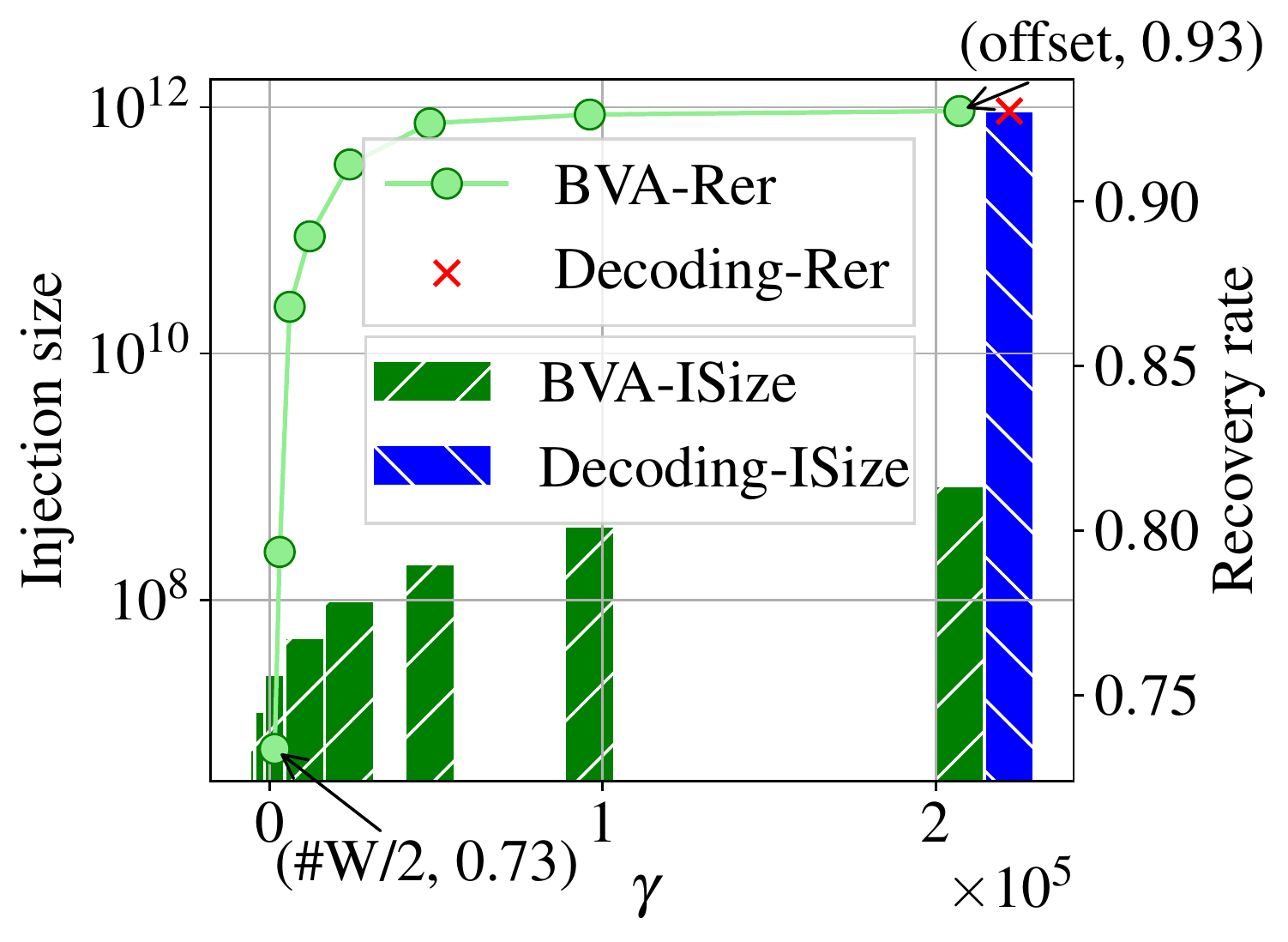}
		\end{minipage}
	}
	\subfigure[Lucene]
	{
		\begin{minipage}{.29\linewidth}
			\centering
			\includegraphics[width=\linewidth]{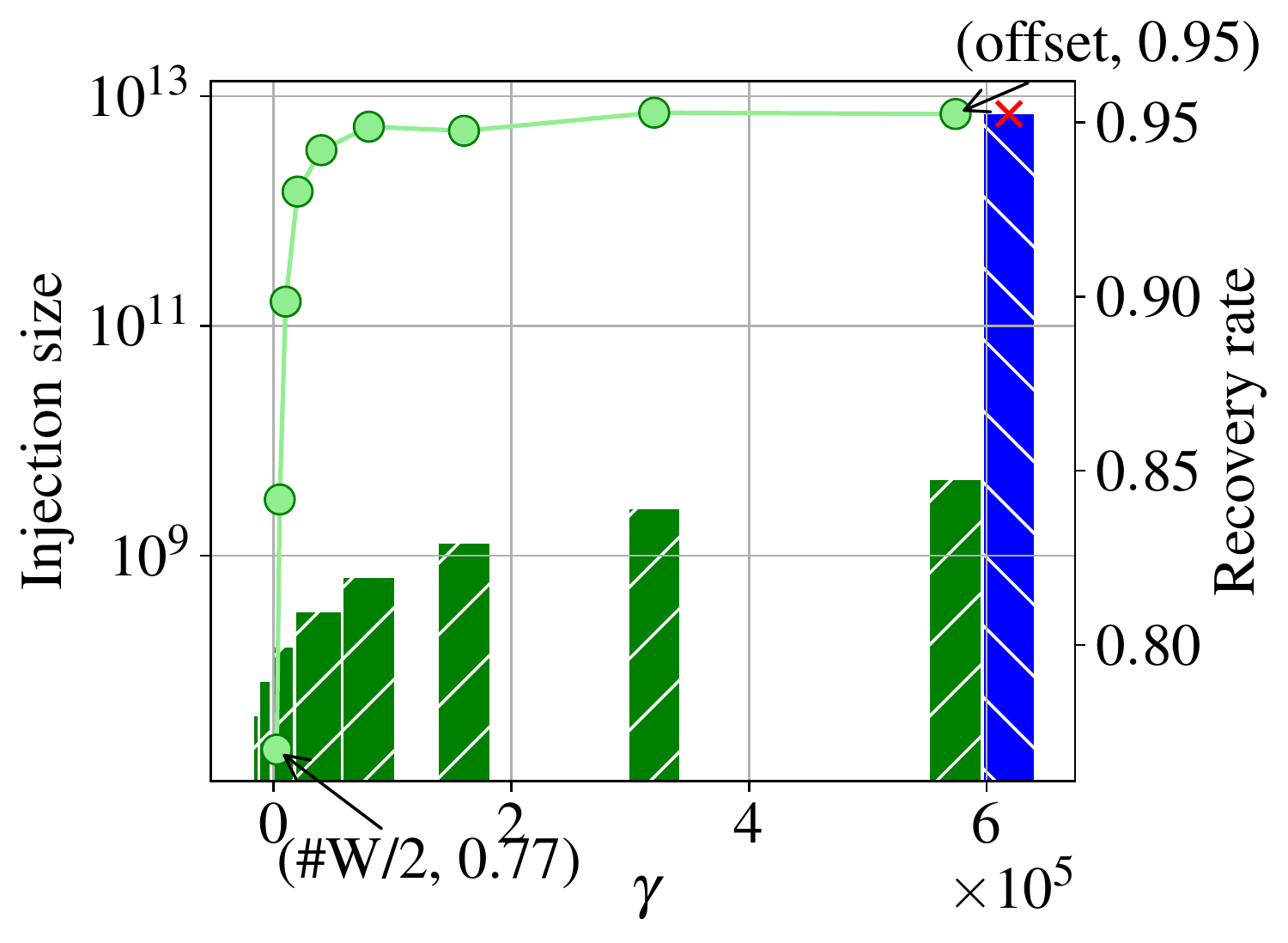}
		\end{minipage}
	}
	\subfigure[WikiPedia]
	{
		\begin{minipage}{.29\linewidth}
			\centering
			\includegraphics[width=\linewidth]{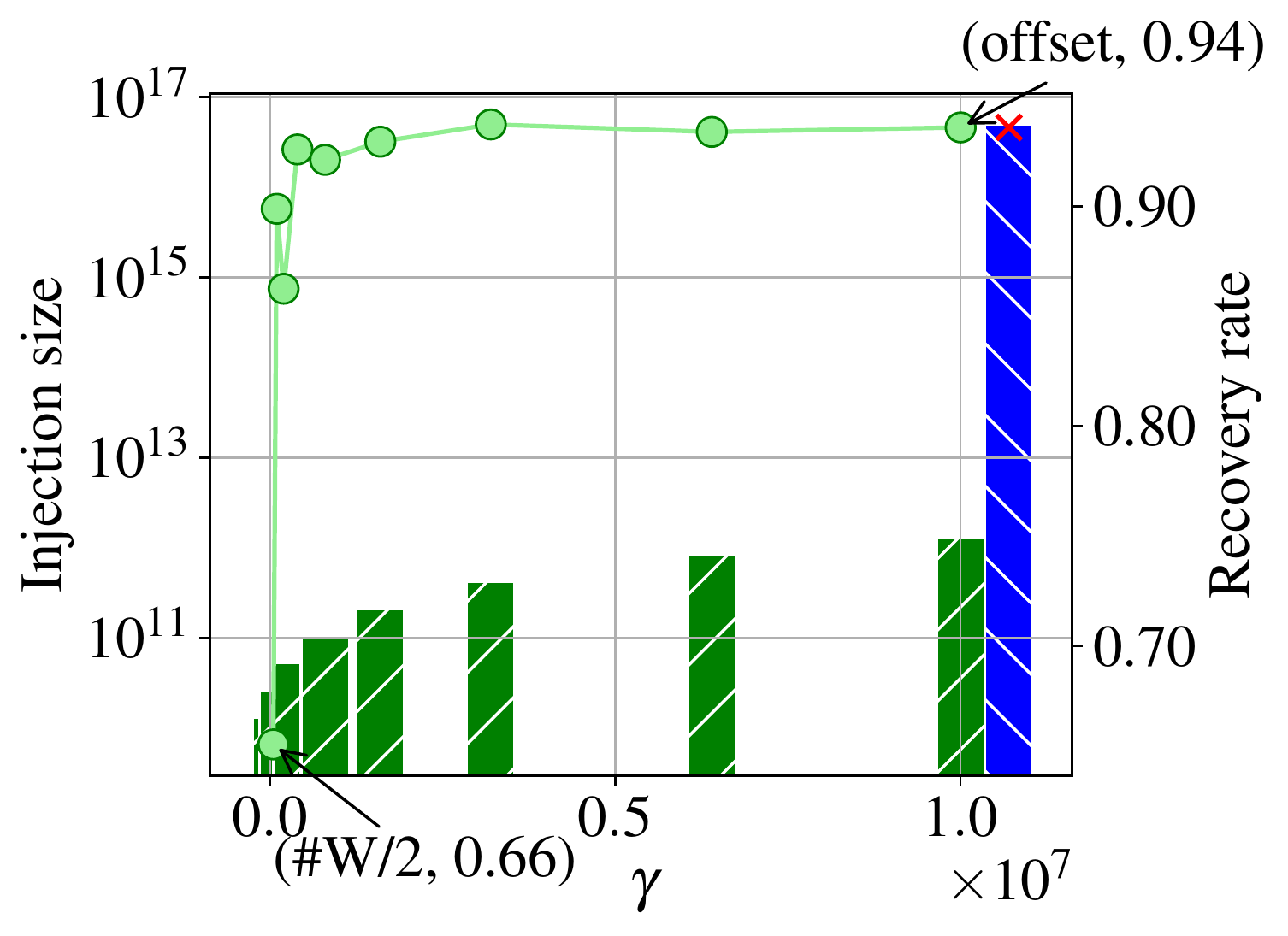}
		\end{minipage}
	}
	\caption{Comparison between the BVA and decoding attack under $\gamma$.}
	\label{BVAExperiment}
\end{figure*}

\subsection{Comparison with Other Attacks} \label{ExBVA}
We compared our attacks with prior injection attacks in three real-world datasets. 
Note we do not consider padding strategies in this section but will test the attacks against them later.
We tested the performance of the BVA and decoding attack under different parameters since they exploit the \textit{rsp} with similar recovery approaches.  
We also performed empirically evaluations on the BVA, the BVMA and other attacks. Note that we did not put ZKP into further comparisons (except in Figure \ref{TCFigure} against TC). 
The crucial difference between our attacks and ZKP is on the leakage pattern. 
Our attacks exploit the \textit{vp} while ZKP focuses on the \textit{aip}. 
There are both pros and cons for the attacking strategies. 
For example, our attacks could bypass ORAM (around $60\%$ against both static padding and ORAM on average, see Figure \ref{SEALFigure}), while ZKP cannot; but ZKP requires less injection size than ours (see Figure \ref{TCFigure}).
%, such as, the recovery effect under different keyword leakage rates, the injection volume for a single query recovery, and the performance against TC.}

\smallskip

\noindent \textbf{Injection Parameter for BVA and Decoding.} 
Recall that the BVA uses $\gamma$ to flexibly control the injection volume. 
We evaluated the recovery rate and injection size while $\gamma$ varies from $\#\mathbf{W}/2$ to the \textit{offset} (see Figure \ref{BVAExperiment}).   
Like the setting in the decoding attack\cite{DBLP:conf/ndss/BlackstoneKM20}, 
we assume that all queried keywords fall in the adversary's known keyword universe. 
As the $\gamma$ increases, the recovery rate of the BVA abruptly rises to the maximum (providing the same recovery as the decoding attack) and meanwhile, the injection size performs steady. 
One can observe that even \textit{in the worst case}, where $\gamma=\#{\mathbf{W}}/2$, the BVA can still achieve practical recovery rates ($>70\%$ in Enron and Lucene, $>60\%$ in the large-scale Wikipedia).
It shows $20\%$ decline on average in Enron and Lucene (note $28\%$ in Wikipedia) at the recovery in this case, as compared to the decoding attack. 
But its injection size is approximately five orders of magnitude less than that of the decoding attack. % (note the logarithmic left axis).
We also see that setting $\gamma=\textit{offset}/4$ can sufficiently ensure the BVA to achieve the similar recovery rate ($<5\%$ gap) as the decoding attack. 
A further increase on $\gamma$ could not produce significant improvement on the recovery rate.
Based on the results, we confirm that a small but reasonable $\gamma$ (e.g., $\gamma=\#\mathbf{W}/2$) can guarantee a practical recovery rate ($>60\%$) with a relatively small injection size (around $10^{8}$ in Enron and Lucene, $10^{11}$ in Wikipedia) as compared to the decoding attack ($>10^{12}$ in Enron and Lucene, $10^{17}$ in Wikipedia).

\smallskip 

\noindent \textbf{Overall Comparison with Decoding and Single-round.} % \label{ExDS}
We compared our attacks with the single-round \cite{DBLP:conf/eurosp/PoddarWLP20} and decoding \cite{DBLP:conf/ndss/BlackstoneKM20} attacks. 
We tested the recovery rate $Rer$, the injection length $ILen$, and the injection size $ISize$ under different keyword leakage ratios in the fixed observation period. 
We evaluated the BVA with error bars by varying $\gamma\in[\#\mathbf{W}/2,\textit{offset}/4]$.
We used a constant $m\geq{1}$ to control the recovery rate and simulated two particular cases where $m=1$ and $m=\#\mathbf{W}$ in the single-round attack.     
We put the running time in Table \ref{RunningTimeEnron} and the recovery rate in Figure \ref{EnronTrendRerVol} (and Table \ref{RunningTimeLucene}-\ref{RunningTimeWiki}, Figure \ref{LuceneTrendRerVol}-\ref{WikiTrendRerVol} in Appendix \ref{AppenCLW}). % for the evaluations on different datasets.

Table \ref{RunningTimeEnron} shows the time cost of restoring $10\times{1,000}$ queries (provided that the adversary knows the keyword universe).  
Most of the attacks take $<3s$, while the BVMA is slower than others (about $19s$). 
This is because it merges multiple leakages for keyword filtering, which naturally increases the time complexity.

\begin{table*}[!t]
	\caption{Running time of recovery for $10\times1,000$ queries in Enron. {Note the cost of the BVA is within the time range by varying $\gamma\in[\#\mathbf{W}/2, \textit{offset}/4]$}.} 
	\label{RunningTimeEnron}
	\centering
	% {\color{red} [what does this mean???average???]}
	\setlength{\tabcolsep}{3.6mm}{\begin{tabular}{lccccc}
		\hline
		 & Decoding & Single-round $(m=1)$ & Single-round $(m=\#{W})$ & BVA & BVMA\\
		\hline
		Running time $(s)$ & 2.48 & 0.01 & 0.02 & (1.81, 2.46) & 19.16\\
		\hline
		
	\end{tabular}}
\end{table*}

\begin{figure*}[!t]
	\centering
	\subfigure[Recovery accuracy]
	{\label{EnronTrendRerVol:Rer}
		\begin{minipage}{.25\linewidth}
			\centering
			\includegraphics[width=\linewidth]{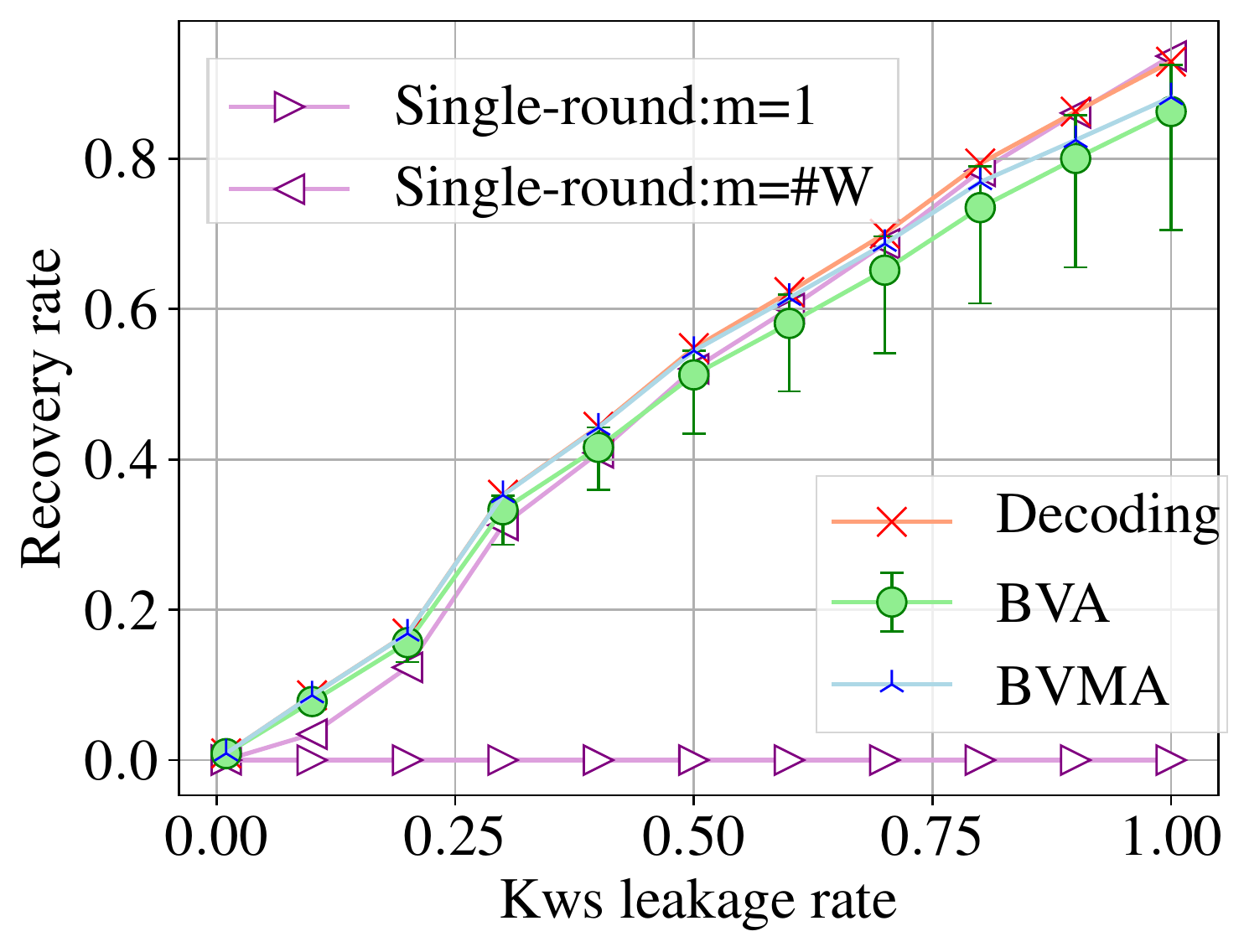}
		\end{minipage}
	}
	\subfigure[Injection length]
	{\label{EnronTrendRerVol:ILen}
		\begin{minipage}{.25\linewidth}
			\centering
			\includegraphics[width=\linewidth]{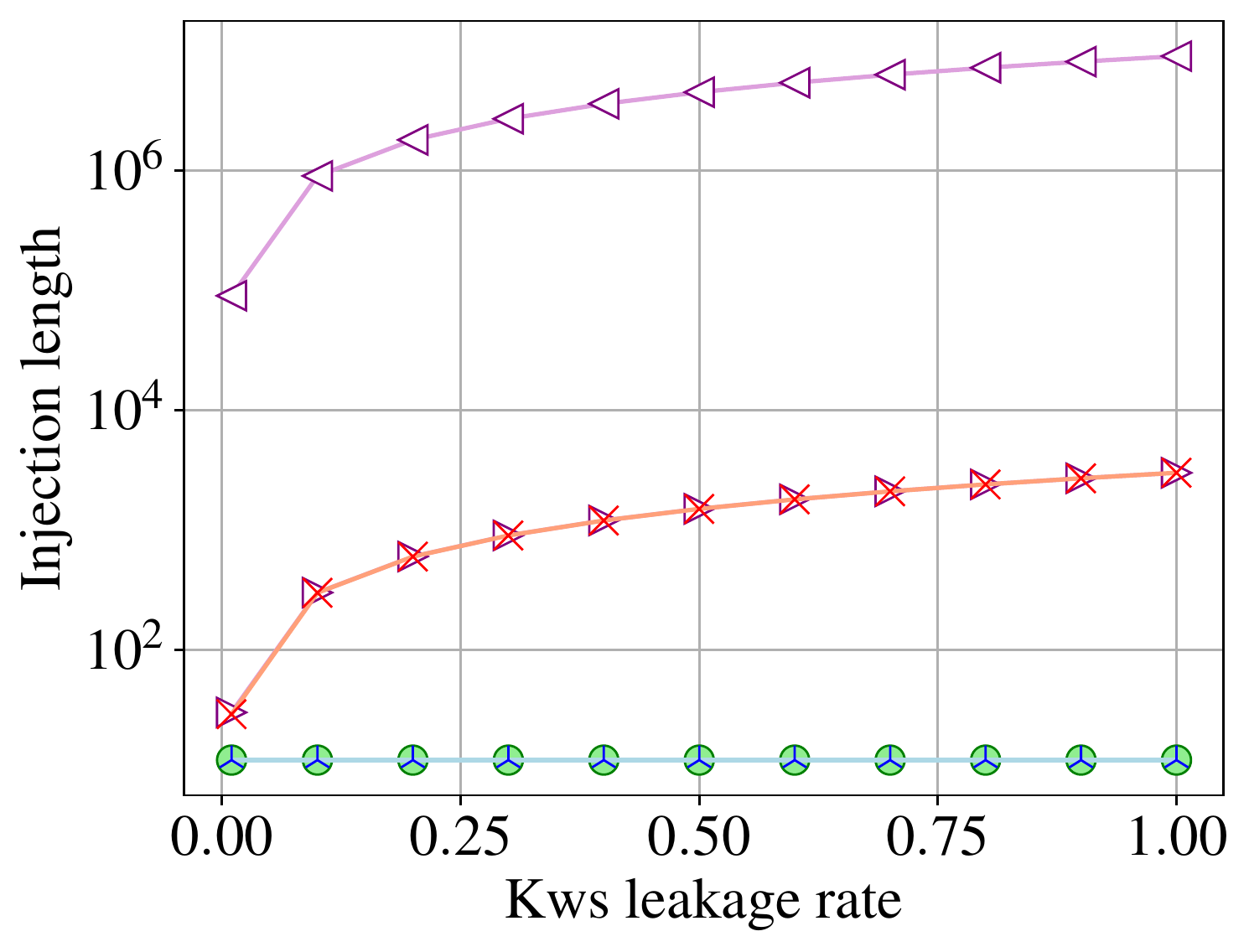}
		\end{minipage}
	}
	\subfigure[Injection size]
	{\label{EnronTrendRerVol:ISize}
		\begin{minipage}{.25\linewidth}
			\centering
			\includegraphics[width=\linewidth]{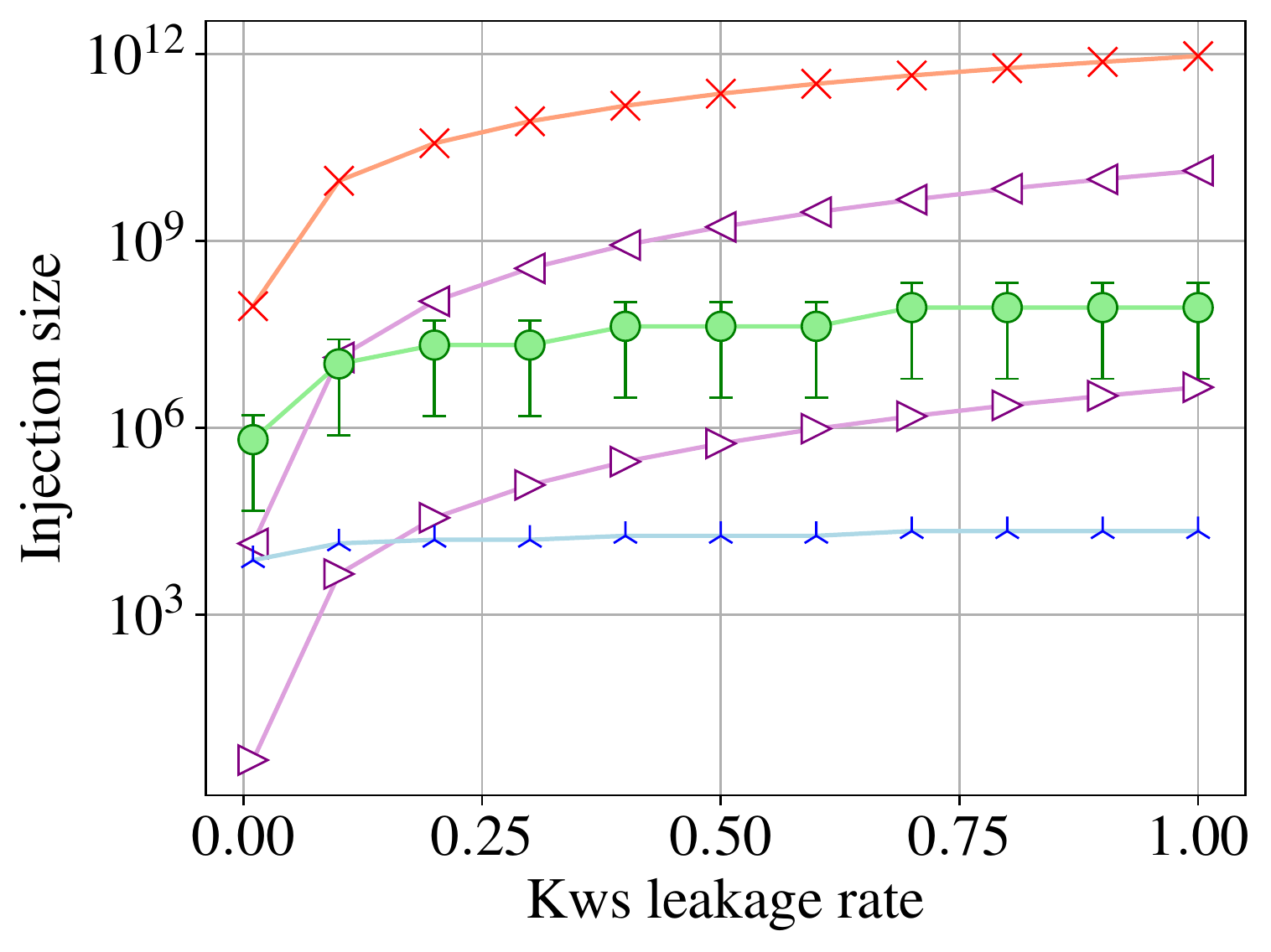}
		\end{minipage}
	}
	\caption{Comparison on the recovery rate, injection length, and injection size with different keywords (Kws) leakages in \textbf{Enron}.}
	\label{EnronTrendRerVol}
\end{figure*}

Figure \ref{EnronTrendRerVol} illustrates that as the number of leaked keywords increases, $Rer$, $ILen$ and $ISize$ present an upward trend. 
More concretely, $Rer$ delivers a quasi-linear growth, while the incline of $ILen$ and $ISize$ gradually flattens.  
From the recovery rate, we see that the single-round attack with $m=1$ can only restore $<1\%$ queries, which performs the worst. 
The BVA and BVMA share a small gap (e.g., $<10\%$ on average in Enron and Lucene) with the decoding and the single-round ($m=\#\mathbf{W}$) attacks. 
We also notice that the BVMA surpasses the minimum recovery rate of the BVA (see the error bar in Figure \ref{EnronTrendRerVol:Rer}) by almost $10\%$ when the keyword leakage reaches $100\%$.  
This is reasonable as the BVMA combines different leakage patterns (e.g., \textit{rlp} and \textit{rsp}) to filter candidate keywords, which can bring advantage on recovery rate. 

The results of $ILen$ (see Figure \ref{EnronTrendRerVol:ILen}) show that both BVA and BVMA (only injecting $<20$ files) require much fewer injections (at least $50 \times$) than the decoding attack. % (with nearly $10^{5}$ files injection). 
The single-round attack with $m=1$ takes the same injection length as the decoding attack but with a poor recovery rate (see Figure \ref{EnronTrendRerVol:Rer}).
We note one may set $m=\#\mathbf{W}$ to produce a practical recovery in the single-round attack. 
But this requires a massive amount of (> $10^{6}$) injected files.

As for the metric $ISize$, the decoding attack gives the worst performance.
This is because the attack strongly relies on $\textit{offset}$.   
%
%We give an example in Table \ref{offset} to explain the performance of the decoding attack. 
% 
Table \ref{offset} illustrates that $\textit{offset}$ grows abruptly with the expansion of keyword universe. 
When $\#{\boldsymbol{W}}=3,000$, each injected file $f_{i}$ {must contain $207,015\times i$ words (where $i \in [\#\mathbf{{W}}] \nonumber$)},  %[confusing???inject these words into the injected file???inject them into other files???]}, 
%In the large-scale databases (e.g., $\#{\boldsymbol{W}}=100,000$), the cost climbs up to $\textit{offset}>10^{7}$, which could be extremely impractical in reality. 
which is extremely impractical in reality.  
Similarly, the performance of the single-round attack is restricted by $m$.  
To achieve a $>80\%$ recovery, the attack (with $m=\#\mathbf{W}$) requires a relatively large injection size, around $10^{10}$ in Enron.  

\begin{table}[!t]
	\caption{$\textit{offset}$ with different sizes of keyword universe. Assume the adversary can observe the response of all keywords.}
	\label{offset}
	\centering
	
	\begin{tabular}{lccccc}
		\hline
		$\#\boldsymbol{{W}}$ & 30 & 300 & 1,000 & 3,000 & 100,000\\
		\hline
		$\textit{offset}$ & 157 & 6,615 & 48,447 & 207,015 & $>10^{7}$\\
		\hline
		
	\end{tabular}
\end{table}

Fortunately, the BVA and BVMA are independent of the $\textit{offset}$ (and $m$) and require fewer injections.  
For example, in Enron (resp. Lucene) with 3,000 (resp. 5,000) keywords (see Figure \ref{EnronTrendRerVol}, \ref{LuceneTrendRerVol}), they achieve $>80\%$ recovery rate on average by only taking nearly $10^{8}$ and $10^{4}$ injection size, respectively. 
The costs are multiple orders of magnitude less than those of the decoding attack ($10^{12}$) and single-round attack with $m=\#\mathbf{W}$ ($10^{10}$). 
In Wikipedia (see Figure \ref{WikiTrendRerVol}) including 100,000 keywords, the injections of the BVA and BVMA are only $10^{10}$ and $10^{6}$ with approx. $60\%$ recovery rate. 
To maintain the same level of recovery, the single-round and decoding attacks must take respectively $10^{4}\times$ and $10^{6}\times$ costs.  

We conclude that our attacks can provide: 1) practical recovery rate, and 2) much fewer injections (length and size) than other attacks. 
Our attacks (the BVA in particular) can be also applicable to the large-scale dataset (see Appendix \ref{AppenCLW}). 
\smallskip

\noindent \textbf{Comparison for Single Query.}
The multiple-round \cite{DBLP:conf/eurosp/PoddarWLP20} and search attacks \cite{DBLP:conf/ndss/BlackstoneKM20} only recover one query at a time. 
In Figure \ref{SGFigure}, we show how the average injection size (per query) varies with the increasing number of target queries. 
We ignored the recovery rate in the experiments, because the recovery performances of the BVA and BVMA with a single query is similar to those given in Figure \ref{EnronTrendRerVol:Rer} (and Figure \ref{LuceneTrendRerVol:Rer}, \ref{WikiTrendRerVol:Rer} in Appendix \ref{AppenCLW}). 
We note the multiple-round and search attacks can provide $100\%$ recovery rate by taking {sufficiently large (e.g., $\#\mathbf{W}\log\#\mathbf{W}$) injection volumes and attack rounds.}
% [sufficiently large???give a degree to sufficient???]

With the increase in the target queries, the multiple-round and search attacks give constant straight lines on injection size, while our attacks provide a sharp decline.  
The cause of the drop is that we can reuse injections on the previous queries to recover the following queries.  
At the beginning, when no. of queries is 10, the cost of the BVMA (e.g., slightly $>10^{4}$) is close to that of the multiple-round and search attacks. 
After $100$ queries, the BVMA outperforms others. 
In Enron and Lucene, the BVA yields the similar results as the multiple-round and search attacks, around $10^{4}$, when no. of queries is up to $2,000$. 
But its cost is roughly $10^{2}\times$ larger than that of the BVMA.  
Our attacks do show a noticeable advantage on injection size with the increase of query number.   

\smallskip

\noindent \textbf{Comparison against TC.}
To circumvent TC, we proposed a transformation Algorithm \ref{TCPadding} (see Section \ref{AttackTC}). % for VIAs (e.g., the single-round, decoding and our attacks). 
We note there is also a variant for the ZKP~\cite{DBLP:conf/uss/ZhangKP16} that can counter TC.  
We evaluated the number of injected files caused by the Algorithm \ref{TCPadding} and the ZKP variant with different thresholds $T$ (see Figure \ref{TCFigure}).  
We limited $T$ to be no more than the total number of keywords (i.e., $\#\mathbf{W}$) in the experiments, because 1) when $T$ reaches to $\#\mathbf{W}$, the number of injected files approaches to stable, in particular, for the ZKP and BVMA; and 2) the word count of a file (in a real-world dataset) {is normally less than the total number of keywords. For example, in Enron, the file with the largest size contains roughly $2,000$ words, which is smaller than the size of $\mathbf{W}$ ($3,000$ keywords).}  
%[the number of keyword of number of universe??? or just the number of keyword in the keyword universe???],[are we sure???any evidence???]

\begin{figure*}[!t]
	\centering
	\subfigure[Enron]
	{
		\begin{minipage}{.24\linewidth}
			\centering
			\includegraphics[width=\linewidth]{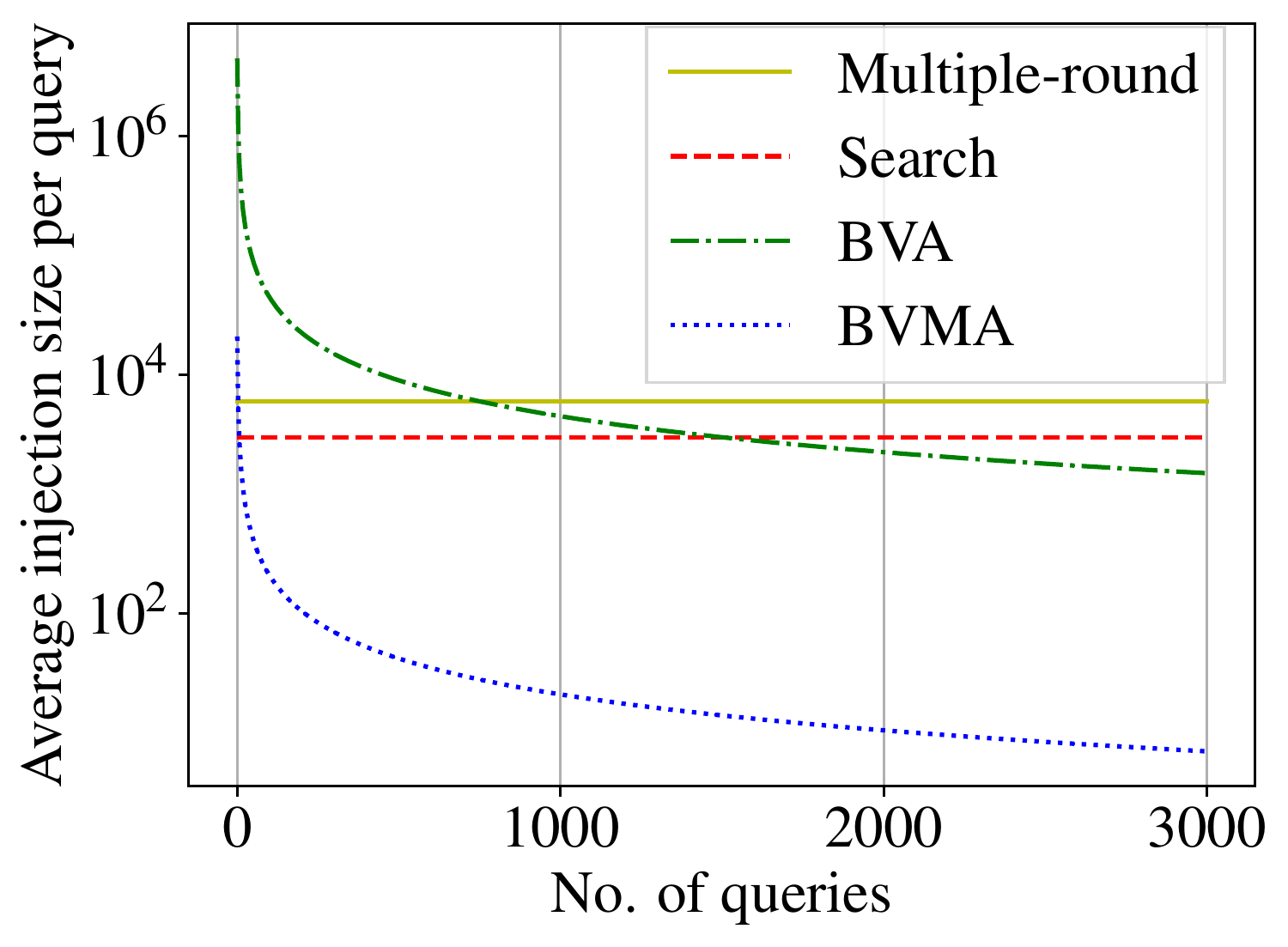}
		\end{minipage}
	}
	\subfigure[Lucene]
	{
		\begin{minipage}{.24\linewidth}
			\centering
			\includegraphics[width=\linewidth]{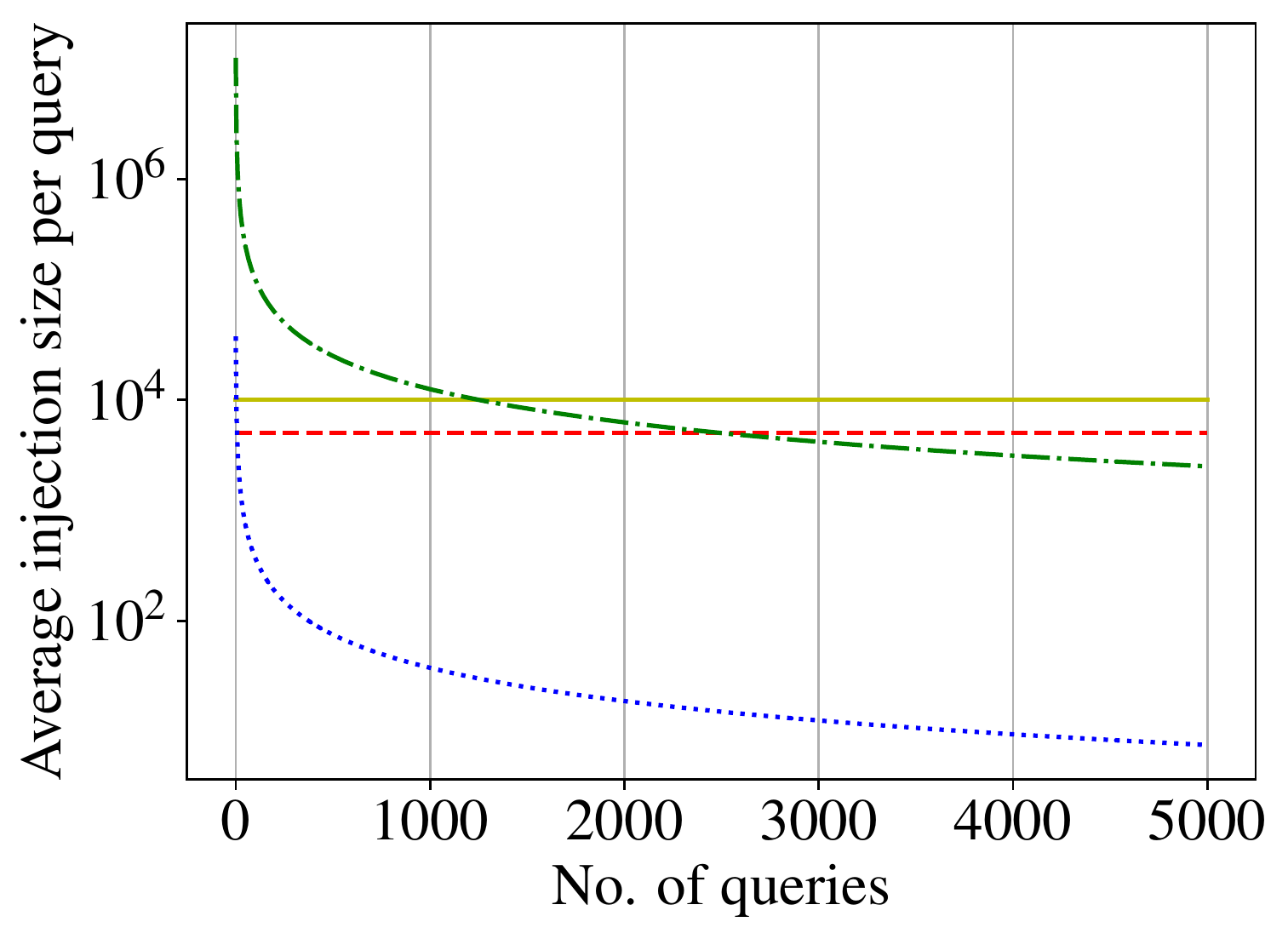}
		\end{minipage}
	}
	\subfigure[Wikipedia]
	{
		\begin{minipage}{.24\linewidth}
			\centering
			\includegraphics[width=\linewidth]{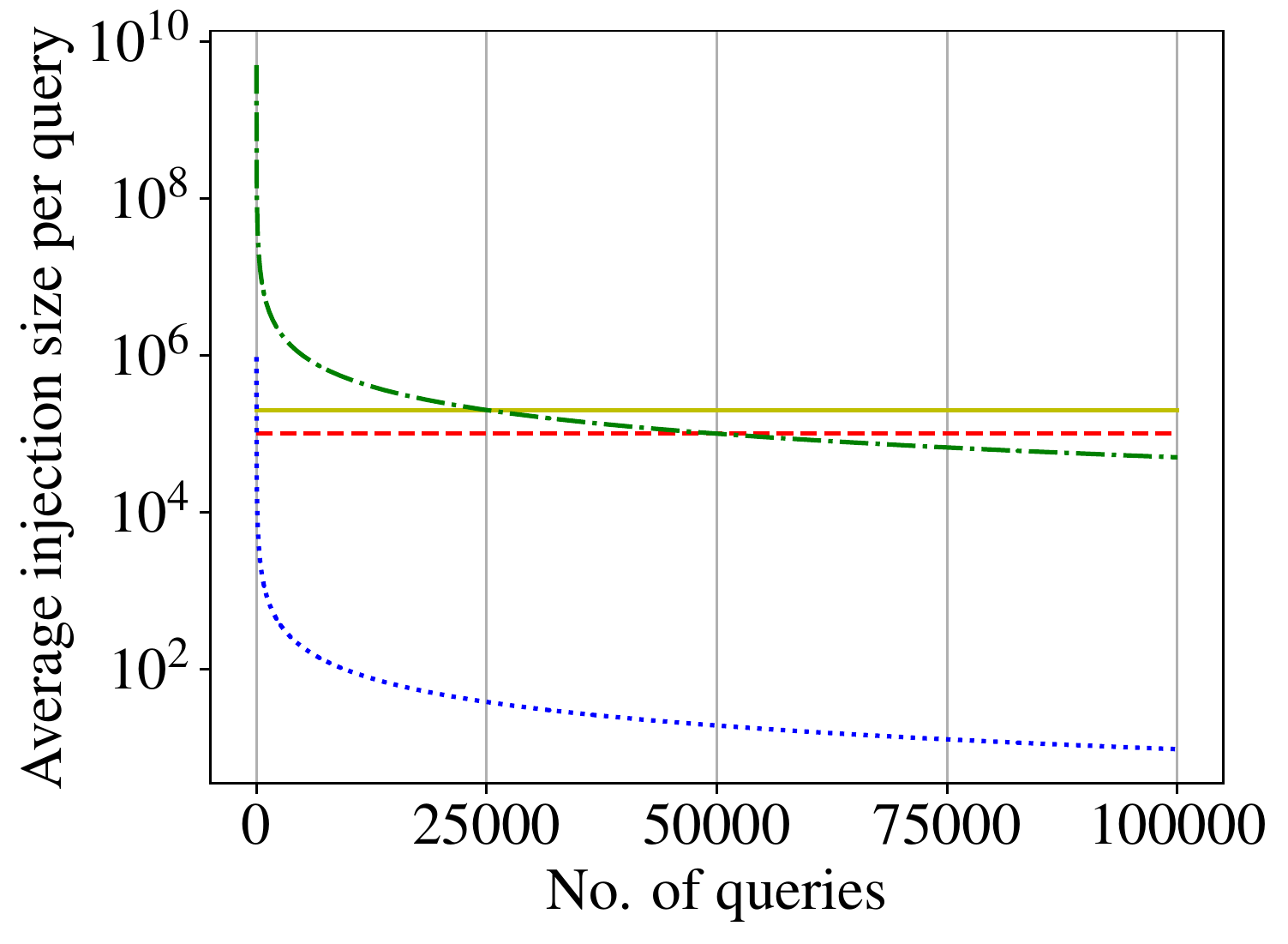}
		\end{minipage}
	}
	%\includegraphics[width=2.5in]{pic/Figure_1.png}
	% where an .eps filename suffix will be assumed under latex, 
	% and a .pdf suffix will be assumed for pdflatex; or what has been declared
	% via \DeclareGraphicsExtensions.
	\caption{Average injection size with number of queries (we set the keyword partition $k=2$ for multiple-round).}
	\label{SGFigure}
\end{figure*}

\begin{figure*}[!t]
	\centering
	\subfigure[Enron]
	{
		\begin{minipage}{.25\linewidth}
			\centering
			\includegraphics[width=\linewidth]{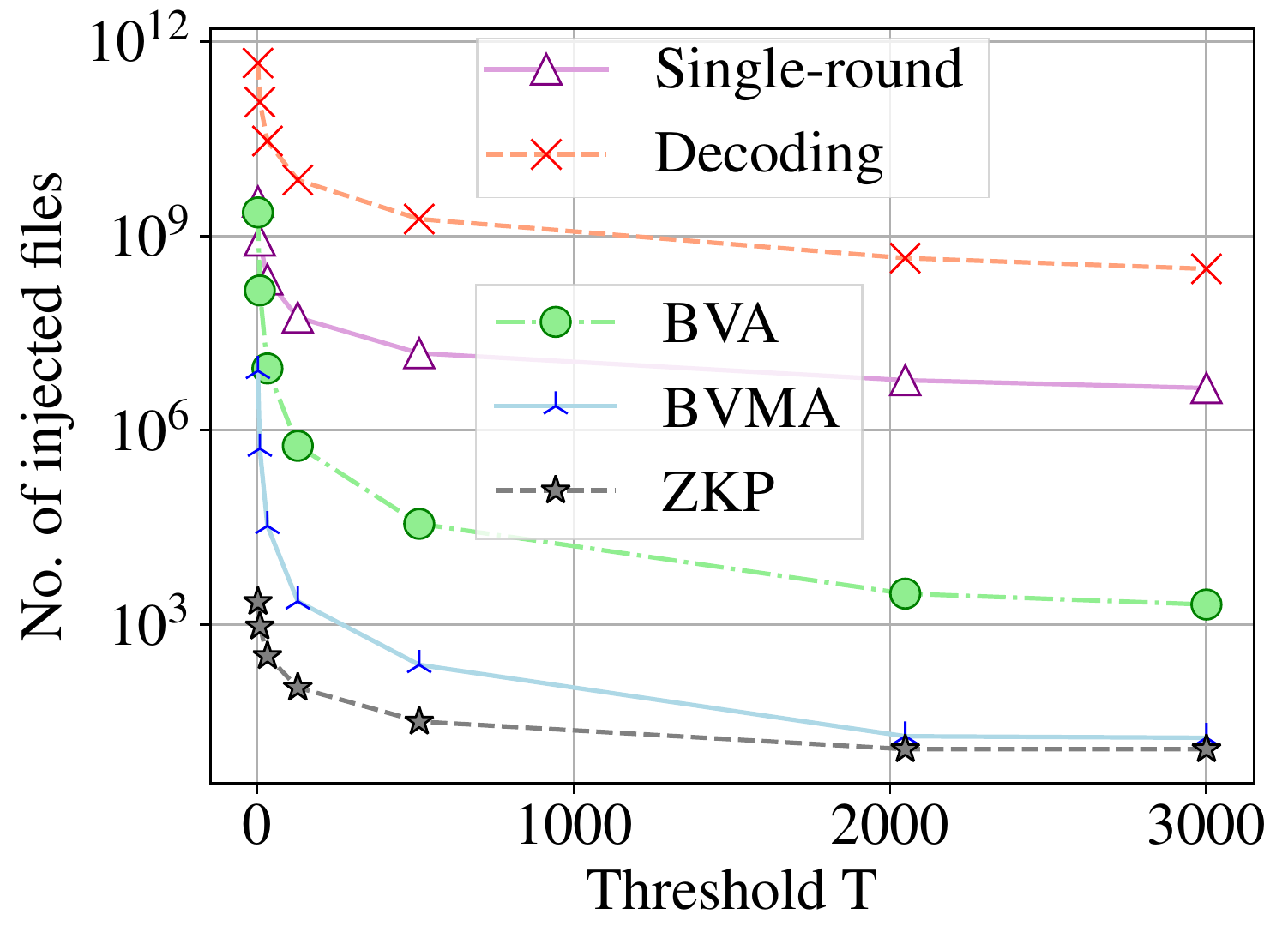}
		\end{minipage}
	}
	\subfigure[Lucene]
	{
		\begin{minipage}{.25\linewidth}
			\centering
			\includegraphics[width=\linewidth]{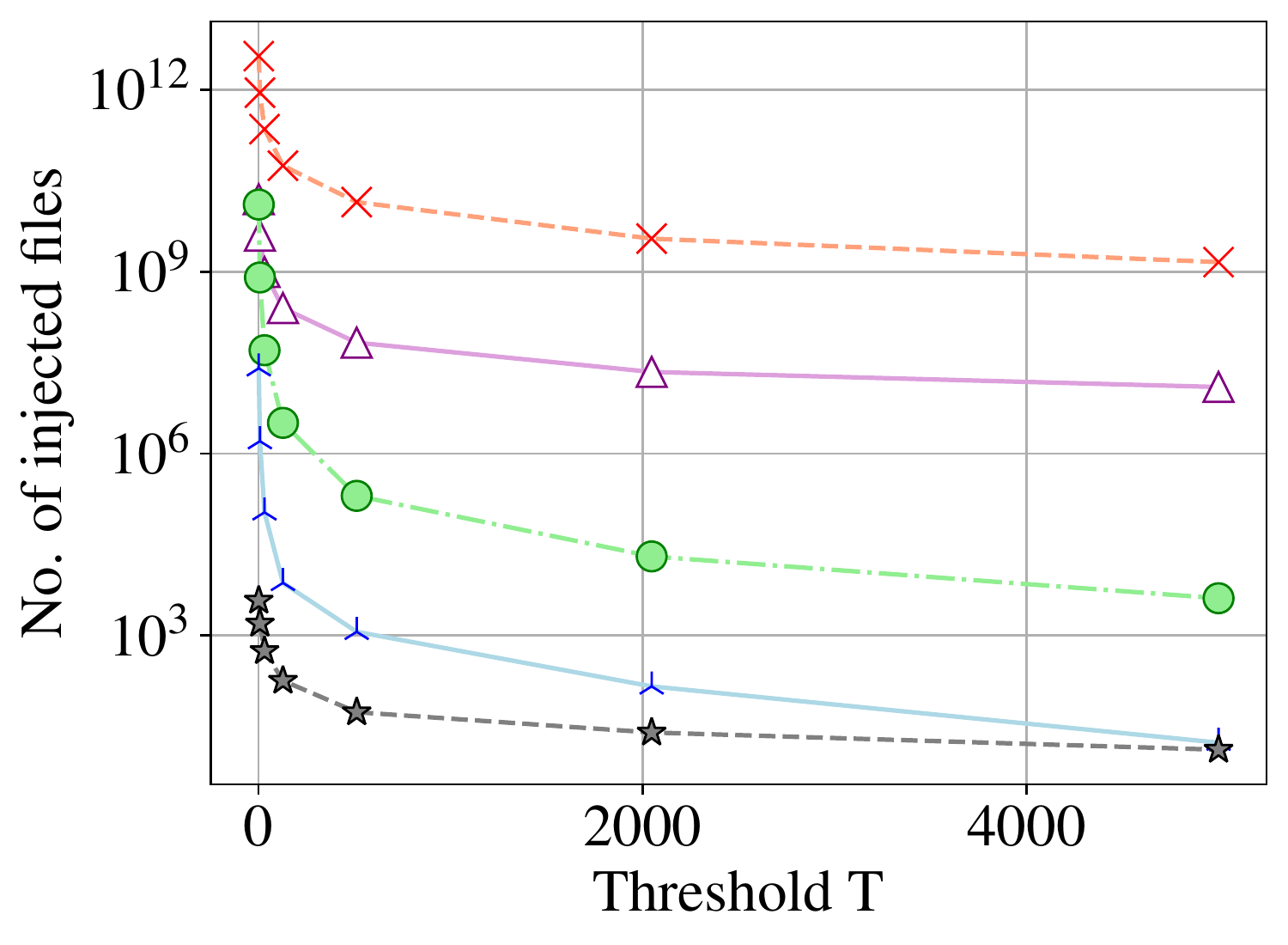}
		\end{minipage}
	}
	\subfigure[Wikipedia]
	{
		\begin{minipage}{.25\linewidth}
			\centering
			\includegraphics[width=\linewidth]{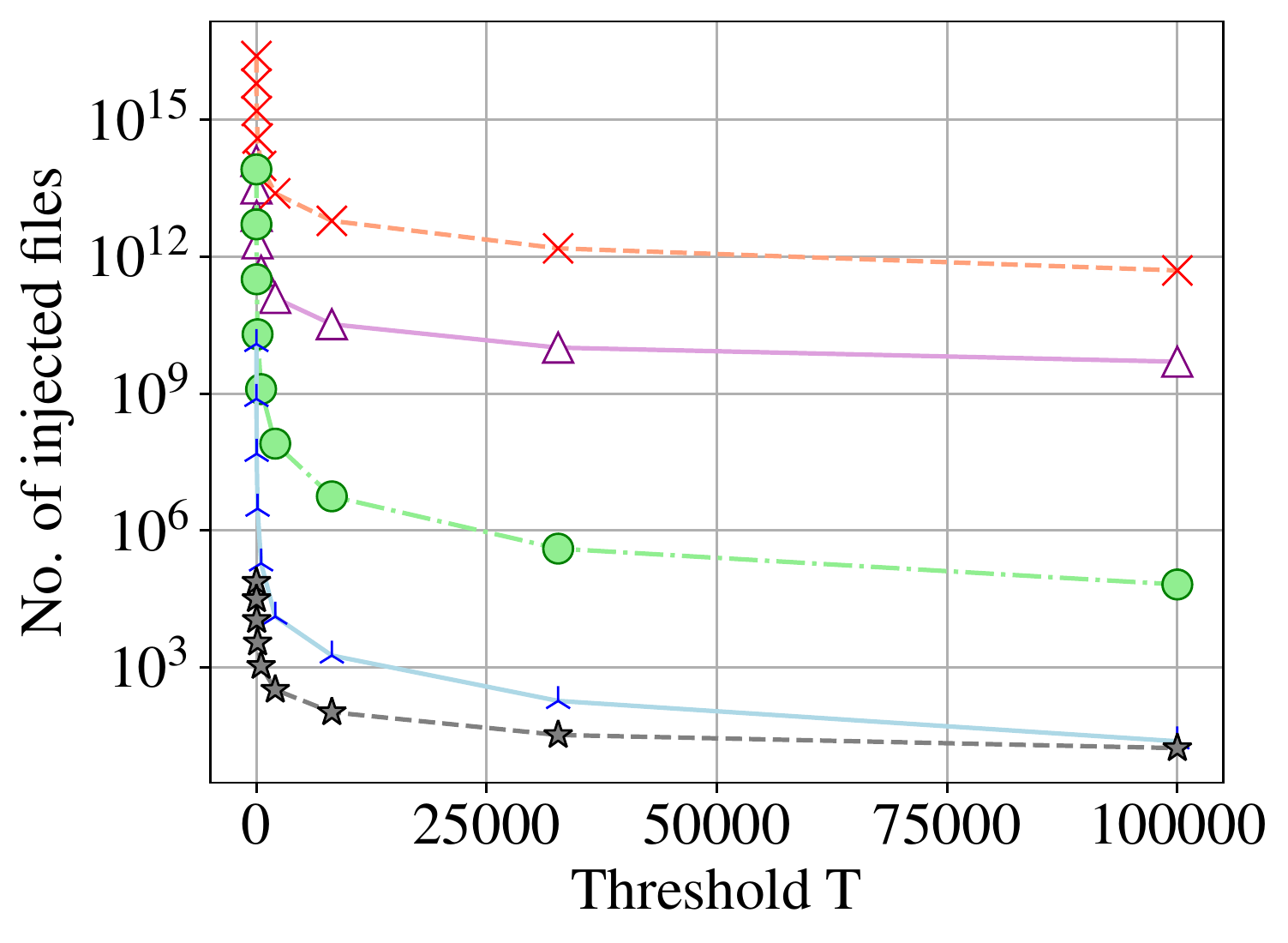}
		\end{minipage}
	}
	\caption{The number of injected files against TC (we set $m$ and $\gamma$ to $\#\mathbf{W}/2$ for the single-round attack and BVA).}
	\label{TCFigure}
\end{figure*}

In Figure~\ref{TCFigure}, increasing $T$ leads to the decline of the injection length. 
The ZKP variant requires the least number of injection. 
This is because it uses a unique approach {(instead of just the volume information)} to recognize the injected files. 
In contrast, other attacks leverage injections to increase the differences in the volume of each keyword. 
There is a small gap on injection between the BVMA and the ZKP variant.   
The injection amount of the former drops fast and ultimately approaches to that of the latter.  
Given a small and reasonable $T$ (e.g.,$T=500$), the injection length of the BVMA is around $10^{3}$ in Enron and Lucene, which is still practical.  
Under a large threshold (e.g., $T=2,000$), the single-round and decoding attacks take >$10^{6}$ injected files in Enron and Lucene which are at least $10^{3}\times$ and $10^{5}\times$ larger than our costs, respectively. 
In Wikipedia (under $T>25,000$), they require at least $10^{4}\times$ and $10^{7}\times$ more injections than ours.
% respectively

Our attacks (especially the BVMA) outperform most of existing VIAs against TC in terms of injection length. 
The BVMA and the ZKP variant deliver similar results under a proper $T$ (e.g., $<10^{2}$ injected files in Enron with $T=2,000$). 
The former is more applicable than the latter if the adversary prefers to exploit the \textit{vp}. 

\subsection{Attacks against Padding}\label{AppenPadding}

Padding \cite{ShieldDB, BostCLuster, DBLP:conf/ccs/CashGPR15, TwoLayer, DBLP:conf/uss/DemertzisPPS20,DBLP:conf/infocom/ChenLRZ18,DBLP:conf/ccs/PatelPYY19} can protect the volume pattern. %It may be static \cite{BostCLuster, DBLP:conf/uss/DemertzisPPS20,DBLP:conf/infocom/ChenLRZ18,DBLP:conf/ccs/PatelPYY19} or dynamic \cite{ShieldDB, TwoLayer}. 
One may apply static padding \cite{BostCLuster, DBLP:conf/uss/DemertzisPPS20,DBLP:conf/infocom/ChenLRZ18,DBLP:conf/ccs/PatelPYY19} upon establishing a database, while using dynamic padding \cite{ShieldDB, TwoLayer, DBLP:journals/iacr/AmjadPPYY21, DBLP:journals/iacr/ZhaoWL21} both in the setup and update stages.
Efficient padding strategies use \textit{keyword clustering} to balance security and padding overhead. 
Keyword clustering technique assigns keywords into different clusters so that the keywords in a cluster share the same or computationally indistinguishable volume after padding. 
    
We tested our attacks against both the static and dynamic padding strategies in Enron. 
We used SEAL \cite{DBLP:conf/uss/DemertzisPPS20} as the static padding solution. 
This is because it requires {less query bandwidth} than  \cite{DBLP:conf/infocom/ChenLRZ18,DBLP:conf/ccs/PatelPYY19} and it can mitigate well-known attacks ({e.g., \cite{DBLP:conf/ndss/IslamKK12,DBLP:conf/ccs/CashGPR15}}). 
As for the dynamic padding, we chose ShieldDB \cite{ShieldDB} {(which is a more practical encrypted database than \cite{DBLP:journals/iacr/AmjadPPYY21, DBLP:journals/iacr/ZhaoWL21})} supporting keyword search with a dynamic countermeasure against query recovery attacks. 
In our experiments, we assumed that each dummy file employed dummy keywords not present in the database, utilizing the padding modules of ShieldDB and SEAL for random file padding to ensure that the file's size is within $[min\_file\_size, max\_file\_size]$ as obfuscation, where $min\_file\_size$ (resp., $max\_file\_size$) is the smallest (resp., largest) file size in the original client database.
Note when using the ORAM module in SEAL, we assumed that the system uses blocks with the same size for filling (instead of dummy files with random sizes).
We tested the padding overhead of SEAL and ShieldDB besides the recovery rate. 
We defined the padding overhead as $N_{padding}/N_{no-padding}-1$, which reflects a ratio between the increased overhead (i.e., number of files) of padding and ``no padding". 
We calculated the storage overhead from setup $(Setup\&Fill)$ and injection\&fill $(Inj\&Fill)$, and the bandwidth overhead from queries-after-setup $(S$-$Query)$ and queries-after-$Inj\&Fill$ $(I$-$Query)$. 
%{\color{blue}Note we evaluated our attacks against different padding strategies in Enron.}   
     % it is still a future work to extend SEAL to support dynamic padding actions. 
%It is natural to apply our attacks to these two padding systems. 
    %Thus, it is very suitable to evaluate the effect of our attacks on padding. 
    %Note that the experiments were conducted on the Enron dataset. {\color{blue}From Figure \ref{LuceneTrendRerVol:ILen}, \ref{WikiTrendRerVol:ILen}, we infer that the type and scale of the database will not greatly affect the experiment. With proper injections, our attacks on Lucene and Wikipedia may have the similar results as Enron dataset.}
    
    %Padding may not be affected by the scale of the database. With proper injections (from Figure \ref{LuceneTrendRerVol:ILen}, \ref{WikiTrendRerVol:ILen}), we infer that using these strategies in Lucene and Wikipedia, our attacks still achieve a practical recovery rate.%It should be noted that SEAL is mainly focus on the static padding scenarios. Even though SEAL introduced the possibility in dynamic scenes (using $SD_{a}$ \cite{DBLP:conf/ndss/DemertzisCPP20}), it still did not present a concrete dynamic padding method.
    \smallskip

    \noindent \textbf{Static Padding by SEAL.} 
    %A static padding indexes dummy files to obfuscate the volume pattern in the setup, rather than the update algorithm (see Section \ref{SectionSE}), before uploading the encrypted database to the server. 
    %Note that padding in the update phase may seriously increase the search time as we need to inverse the whole indexes in the query phase.
    %Moreover, this padding mode also causes extra storage overheads on dummy files and the corresponding indexes to the server.
    %Thus, we only used the padding in the setup in the experiments. Our attacks mainly depend on vp, so we only consider its padding module.
%
    %Demertzis et al. \cite{DBLP:conf/uss/DemertzisPPS20} proposed a well-defined system SEAL with an adjustable leakage consisting of 
    SEAL has two core modules.  
    One is the quantized ORAM module \cite{DBLP:conf/ccs/StefanovDSFRYD13} hiding the \textit{ap} and \textit{sp}, and the other is the parameterized padding module concealing the \textit{vp}.
    We assume the response is $D_{q}$ for a query $q$, where $D$ is the encrypted database stored in the form of ORAM blocks.   
    Note each block in ORAM shares the same size (e.g., with $B$ keywords). 
    The number of blocks \textit{rlp} for a query $q$ is $\#{D_{q}}$, and the total response size \textit{rsp} is $\#{D_{q}}\cdot{B}$. 
    In this case, the \textit{rsp} and \textit{rlp} are somewhat ``mixed", which indicates that the \textit{rlp} can be derived from \textit{rsp} and vice versa,  
    so that the BVMA ``degenerates" to the BVA only relying on the \textit{rsp}.  
    We thus only evaluated the BVA against SEAL (with both padding and ORAM, see Figure \ref{SEALFigurePaddingORAM}).  

    In the ORAM module, for a file $f$, if its size is smaller than the block size $B$ (i.e., $|f|_{ w}<B$), the system will pad the file to size $B$ and then store the resulting file in a random block; if $|f|_{w}>B$, the system will split and store the file in $\lceil |f|_{w}/B \rceil$ blocks. 
    Due to ORAM, each query will obliviously access to the related blocks. 
    In the padding module, the system can add dummy blocks with the underlying keyword of $q$ until the query's response length $\#\widetilde{D_q}$ is the next power of $x$:  
    $
        \#\widetilde{D_q} = \min \{x^{k}: x^{k}>\#D_{q}, k\in [\log {\#D}]\}.\nonumber
    $
    %SEAL fills the response length of each query to the closest power of $x$. 
    
    %We notice that SEAL does not explain how to generate the dummy files.  
    %We assume that each (dummy) file utilizes dummy keywords for random padding to ensure that {\color{blue}the file's} size is within $[min\_file\_size, max\_file\_size]$ as obfuscation, where $min\_file\_size$ (resp., $max\_file\_size$) is the smallest (resp., largest) file size in the (original) dataset. 

    In the experiments, we set $B$ as the average size of files in Enron. 
    We selected a minimum $\gamma$ satisfying $\gamma\geq\#\mathbf{W}/2\wedge{B|\gamma}$ to ensure that each injected file $f$ can be exactly divided into $|f|_{w}/B$ blocks.
    %, so as to eliminate the influence of ORAM module on injection files
    We tested the BVA and BVMA against SEAL with only padding module (see Figure \ref{SEALFigurePadding}) and also investigated the BVA applying both padding and ORAM modules (see Figure \ref{SEALFigurePaddingORAM}). 
    {Note we tested SEAL's padding overhead instead of ORAM cost, as it was hard to evaluate the ORAM's complexity which relies on various factors (e.g., block variables, access method, bucket size).}
    % Figure \ref{SEALFigurePadding} also includes the padding overhead. 
    % We defined the (padding) overhead as $N_{padding}/N_{no-padding}-1$, which reflects a ratio between the increased overhead (i.e., number of file) of padding and "no padding". 
    % We captured the overheads from setup $(Setup)$, query-after-setup $(S-Query)$, injection $(Injection)$ and query-after-injection $(I-Query)$.
    %Note we did not test the complexity of ORAM in Figure \ref{SEALFigurePaddingORAM}
    We could regard the overhead as the lower bound of SEAL's cost. 
    This makes sense as ORAM definitely yields extra and noticeable cost. 
    %However, it is certain that the addition of ORAM will further increase SEAL's cost compared with padding only.
    
    In Figure \ref{SEALFigurePadding}, the increase of $x$ (in the padding module) has little influence over the recovery rate.
    The recovery remains at about $70\%$. 
    But the padding cost, especially in the setup and query, increases by $200\%$ and $400\%$ (when $x$ is up to 16), which seriously affects the practicability. 
    Figure \ref{SEALFigurePaddingORAM} shows that padding and ORAM modules disturb the stability of the BVA's performance.  
    For example, when $x=16$, the BVA achieves $70\%$ recovery in the best case and only $10\%$ in the worst case. 
    The negative impact incurred by SEAL on the average recovery rate is still limited.  
    Under various $x$, the BVA can maintain $60\%$ recovery on average.   
    We note that one may keep increasing $x$, but this will make SEAL become less and less practical.  % more the recovery 
    %Predictably, with the further increase of $x$, our attacks still pose a great threat to query privacy, and the overhead of SEAL may become more unacceptable.
%    
    The results indicate that the static padding is not the best countermeasure to our attacks.  
    %We will use the dynamic padding in the following subsection.

    \begin{figure}[!t]
	\centering
	\subfigure[Padding.]
	{
		\begin{minipage}{.58\linewidth}
			\centering
			\includegraphics[width=\linewidth]{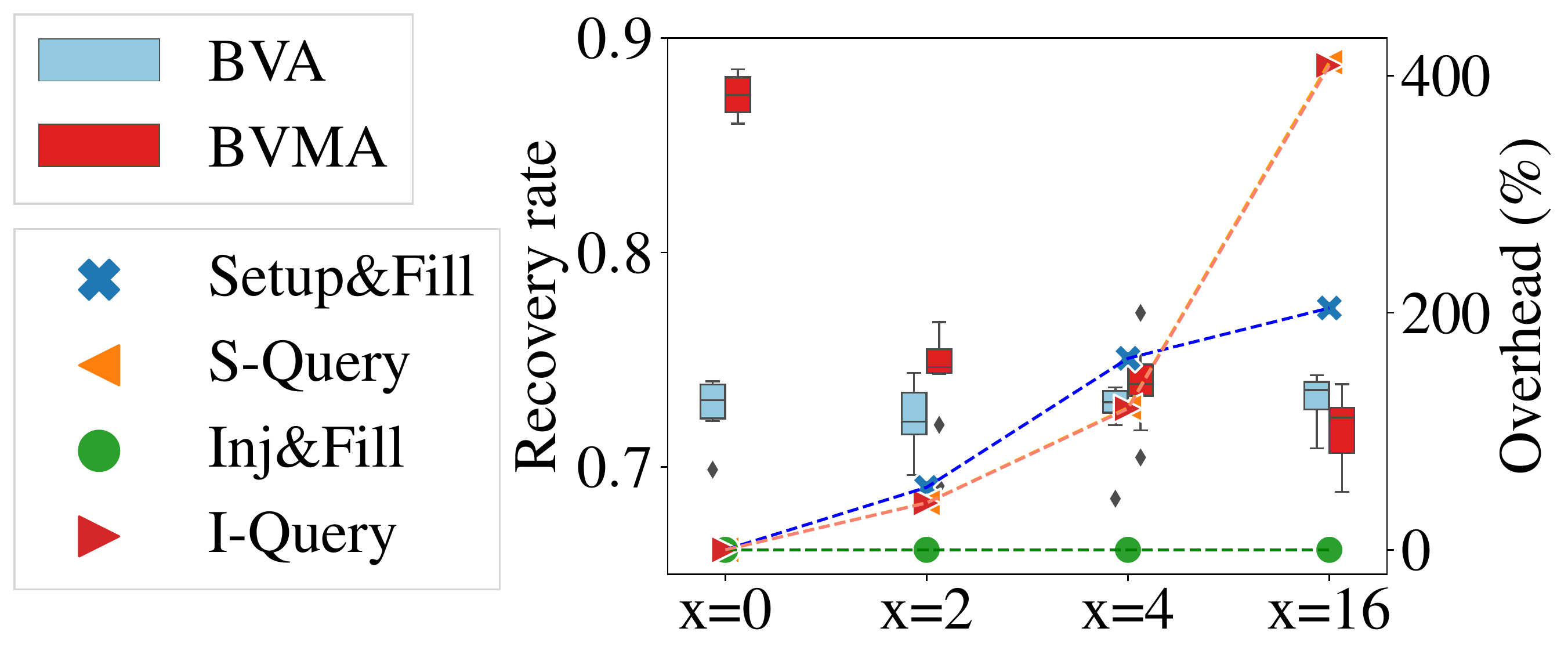}
			\label{SEALFigurePadding}
		\end{minipage}
	}
	\subfigure[Padding \& ORAM.]
	{
		\begin{minipage}{.34\linewidth}
			\centering
			\includegraphics[width=\linewidth]{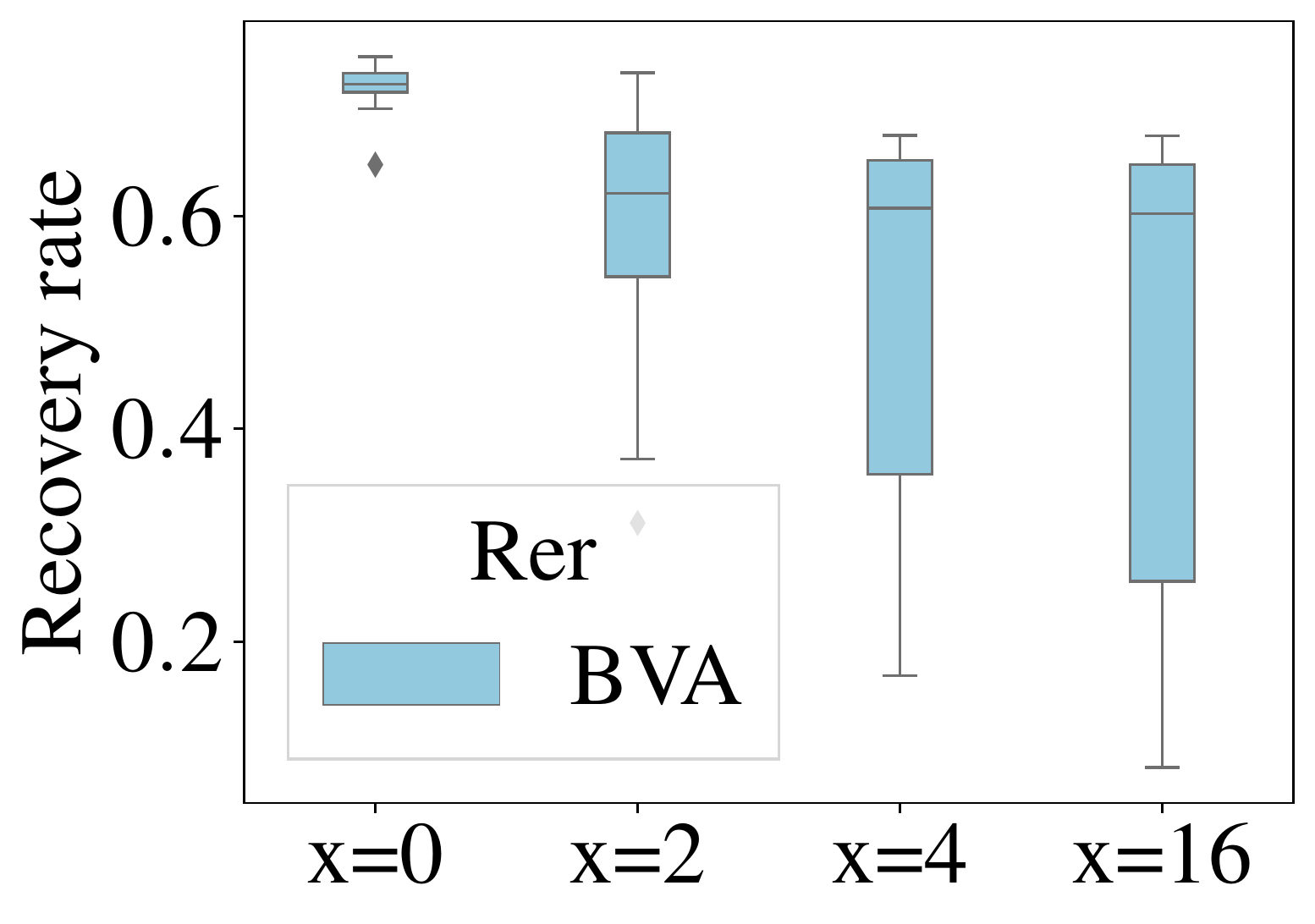}
			\label{SEALFigurePaddingORAM}
		\end{minipage}
	}
	\caption{BVA against static (pre-injection) padding\&ORAM (SEAL). Note ``x=0" is the baseline (``no defense"). %{\color{blue}The test here includes the recovery rate of BVA\&BVMA and overhead of the padding module of SEAL.}}
	}
	\label{SEALFigure}
    \end{figure}

    \smallskip

    %we choose Enron dataset as an example to illuminate how well our attacks fight against padding. 
    %We can infer from Figure \ref{TCFigure} that Lucene and Wikipedia may own the similar numbers of injection files when meeting our attacks.
    % SEAL uses the parameter $x$ to fill the rlp of each keyword to the smallest $x^{t}$ at the setup phase
    \noindent \textbf{Extend SEAL to Dynamic Padding.}
    {We notice that SEAL describes a possible extension for dynamic updates (e.g., by handling the update operation using $SD_{a}$ \cite{DBLP:conf/ndss/DemertzisCPP20}); but it is still unknown how we could properly extend SEAL to support dynamic padding.
    A straightforward idea\footnote{Following the SEAL's padding (as well as settings), the idea supports the strategy during (dynamic) update.} could be to fill the total \textit{rlp} of the corresponding keyword into $x^{t}$ after every batch update.
    We designed an extra experiment (see Table \ref{DynamicSEALT}, Appendix \ref{AppenDYSEAL}) to evaluate our attacks against such a dynamic variant.
    We assume that after a batch injection (by our attacks), the padding module recalculates the \textit{rlp} of each keyword and then fills it to the proper $x^{t}$. 
    The padding can appropriately resist VIAs ($<1\%$ recovery rate) but at the same time produce extreme overheads. 
    More concretely, given $x=2$ (i.e., the minimum padding), as compared to ``no padding", the query-after-injection overhead raises by nearly $200\%$. 
    Meanwhile, the number of extra files for padding expands by $>10^{3}\times$. 
    The padding requires approx. $33,000$ extra files against injections and this number is {even larger than} the total files of the entire database ($30,109$ files).}
    %
    %Moreover, the experiment implies that the straightforward method may be vulnerable to denial-of-service (Dos) attacks, where an adversary can inject a few files (but cause costly padding) to deny service to honest clients. 
    %
    %On the contrary, ShieldDB has considered dynamic padding, which has better-balanced bandwidth overhead and security in DSSE scenarios. Therefore, we will further evaluate the effect of ShieldDB in the following subsection.
    \smallskip
    
    \noindent \textbf{Dynamic Padding by ShieldDB.} ShieldDB \cite{ShieldDB} uses the parameter $\alpha$ to measure the size (i.e., number of keywords) of each cluster. In the update phase, there are two $cases$ for upload according to if the keywords (of clusters) have been stored in the server.

    \begin{figure}[!t]
	\centering
	\subfigure[Query-cluster location accuracy.]
	{
		\begin{minipage}{.35\linewidth}
			\centering
			\includegraphics[width=\linewidth]{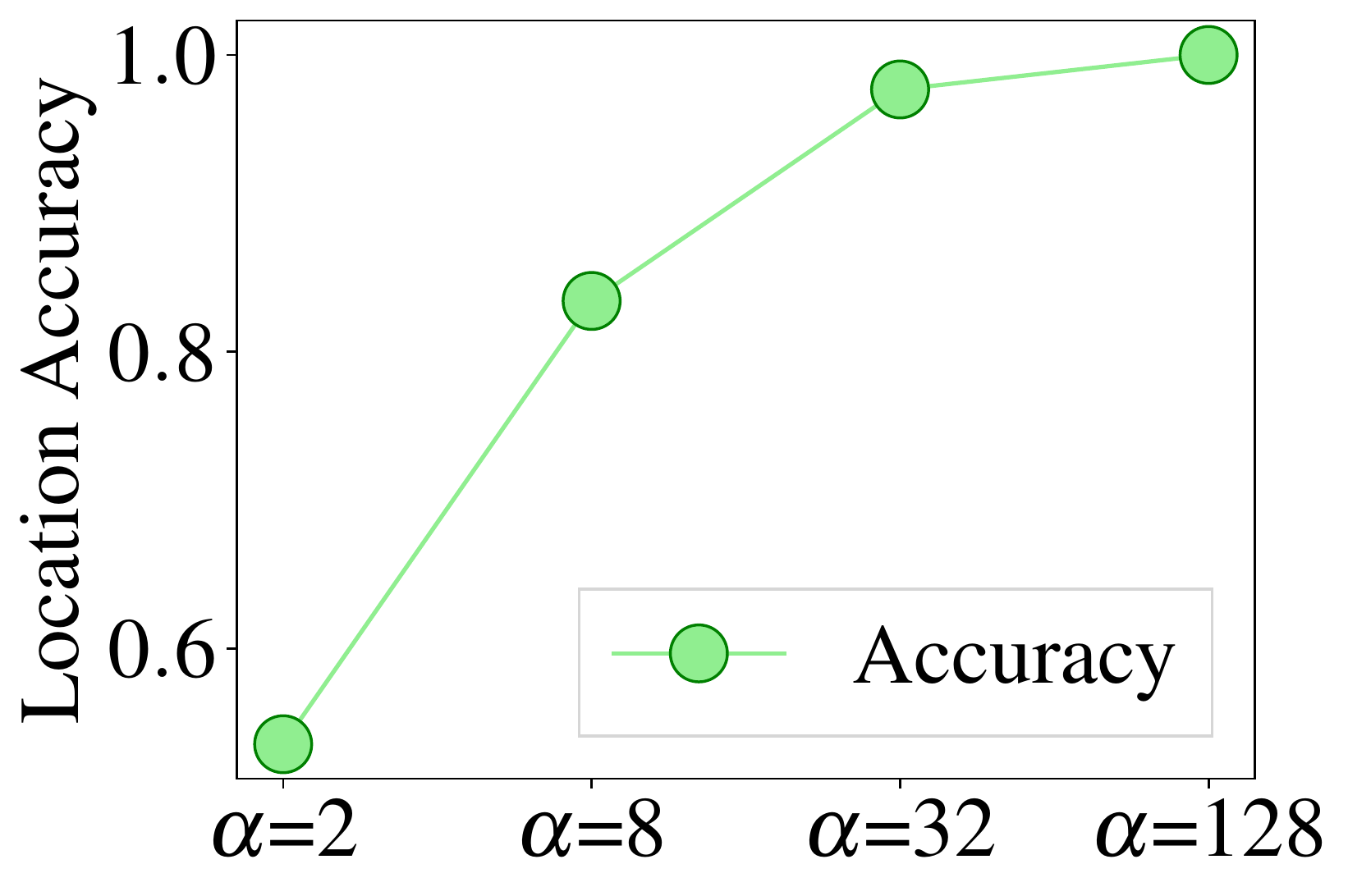}
			\label{ShieldDBLA}
		\end{minipage}
	}
	\subfigure[Recovery rate and overhead.]
	{
		\begin{minipage}{.55\linewidth}
			\centering
			\includegraphics[width=\linewidth]{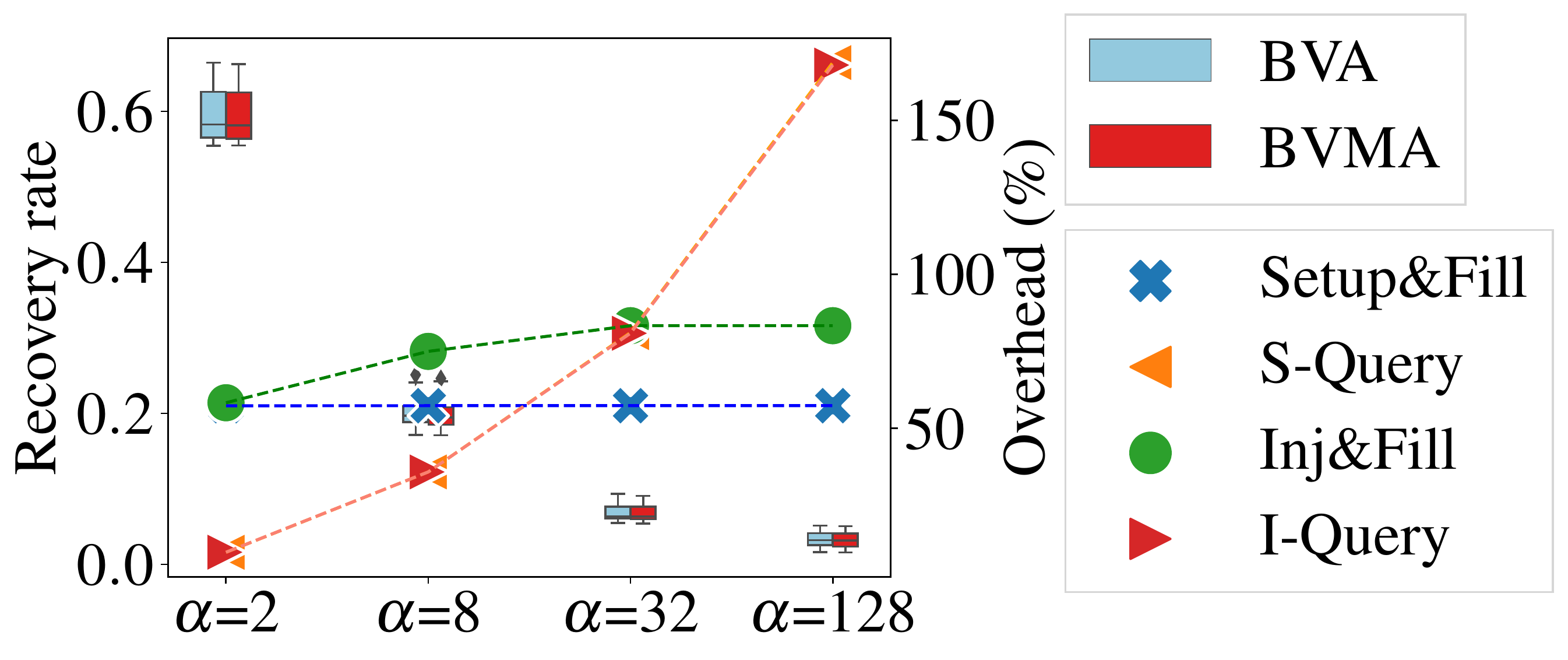}
			\label{ShieldDBRerBO}
		\end{minipage}
	}
	\caption{Attacks against dynamic (post-injection) padding (ShieldDB).}
	\label{ShieldDBFigure}
    \end{figure}

    \begin{itemize}
        \item \textit{Case 1:} The server's current database has not yet contained any keywords of a cluster. 
    In this case, only after all the existing keywords of the cluster have been (updated and) stored in the local cache, they are filled and uploaded. 

        \item \textit{Case 2:} All the keywords in the cluster have already been stored in the server's database. 
    There are two triggers for the uploads: 1) if the interval between the last and the current upload time slots exceeds a specific threshold (i.e. \textit{Tthreshold}); 2) if the number of keyword-file (i.e. (w, id)) pairs of the cluster stored locally exceeds a concrete size (i.e. \textit{Cthreshold}). 
    \end{itemize}

    Once keywords meet the corresponding conditions, the system fills the keywords to the same volume (which is usually the largest volume, called high mode in ShieldDB) and uploads to the server. 
    Concretely, assume the client added files is $U_{w}$ for a keyword $w$. After padding, the update length $\#\widetilde{U_{w}}$ of $w$ is: 
    $
        \#\widetilde{U_w} = \max \{\#{U_{w_{m}}}, \forall w_{m}\in C\}.\nonumber
    $ 
%
    % We consider the persistent adversary with high mode (PH), where the adversary can continuously observe the query result. 
    For simplicity, {we considered two batch paddings in our experiments. 
    The first padding is to cluster the keywords in the setup phase, fill them in each cluster, and upload to the server; while the second is in the injection phase, where we injected specific files containing all the keywords and further used the padding strategy to obfuscate the volume.} 
    
    The adversary takes the following three steps for query recovery:  
    (1) guess each keyword's cluster; (2) relate the target queries to the correct clusters; and (3) use attack algorithms (BVA and BVMA) to recover the queries from the clusters.

    \begin{figure}[!t]
	\centering
	\subfigure[Recovery rate for different $\alpha$. We set $t=\alpha$ in this case.]
	{
		\begin{minipage}{.415\linewidth}
			\centering
			\includegraphics[width=\linewidth]{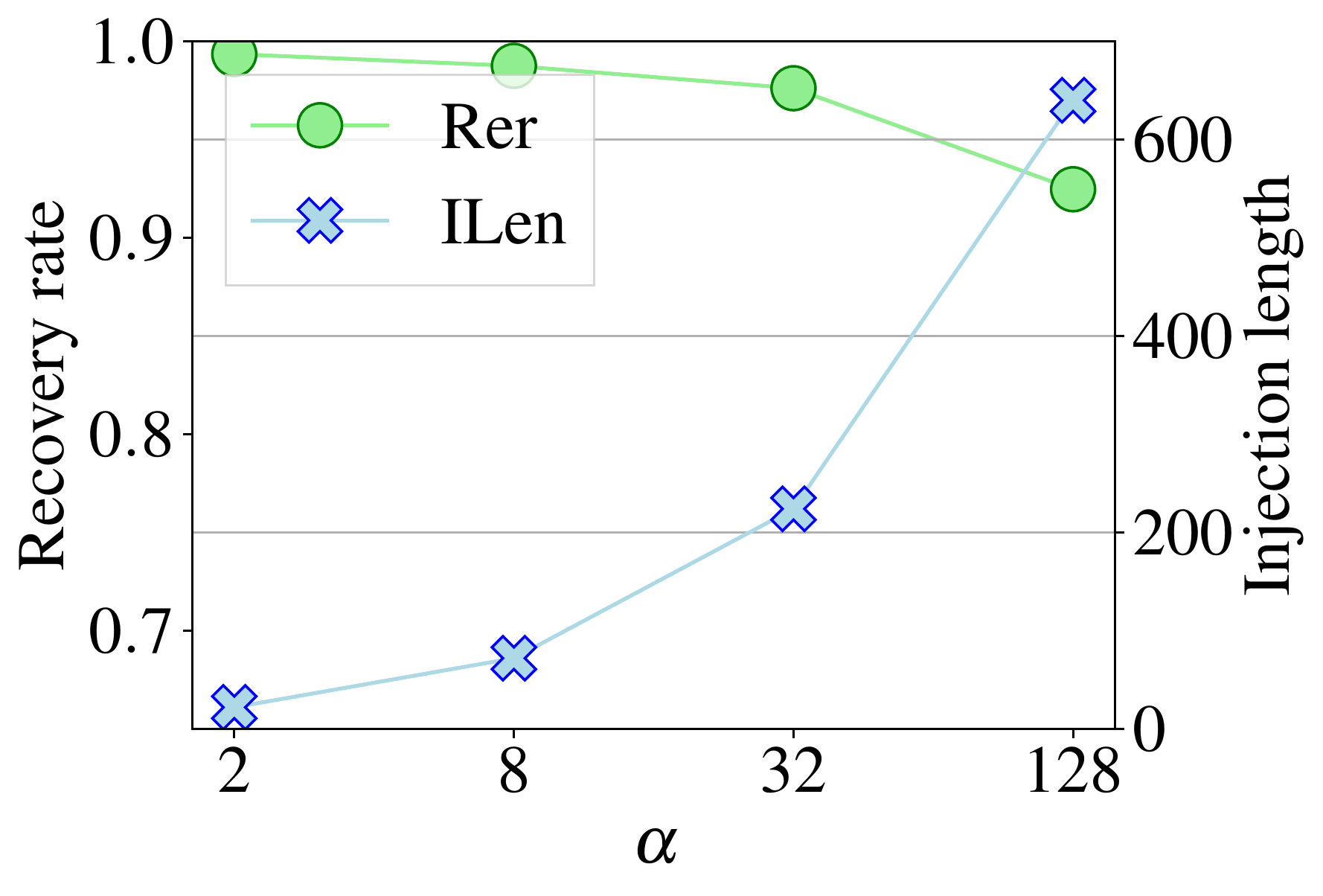}
			\label{ABVIAA}
		\end{minipage}
	}
	\subfigure[Recovery rate for different $t$. We set $\alpha=128$ in this case.]
	{
		\begin{minipage}{.415\linewidth}
			\centering
			\includegraphics[width=\linewidth]{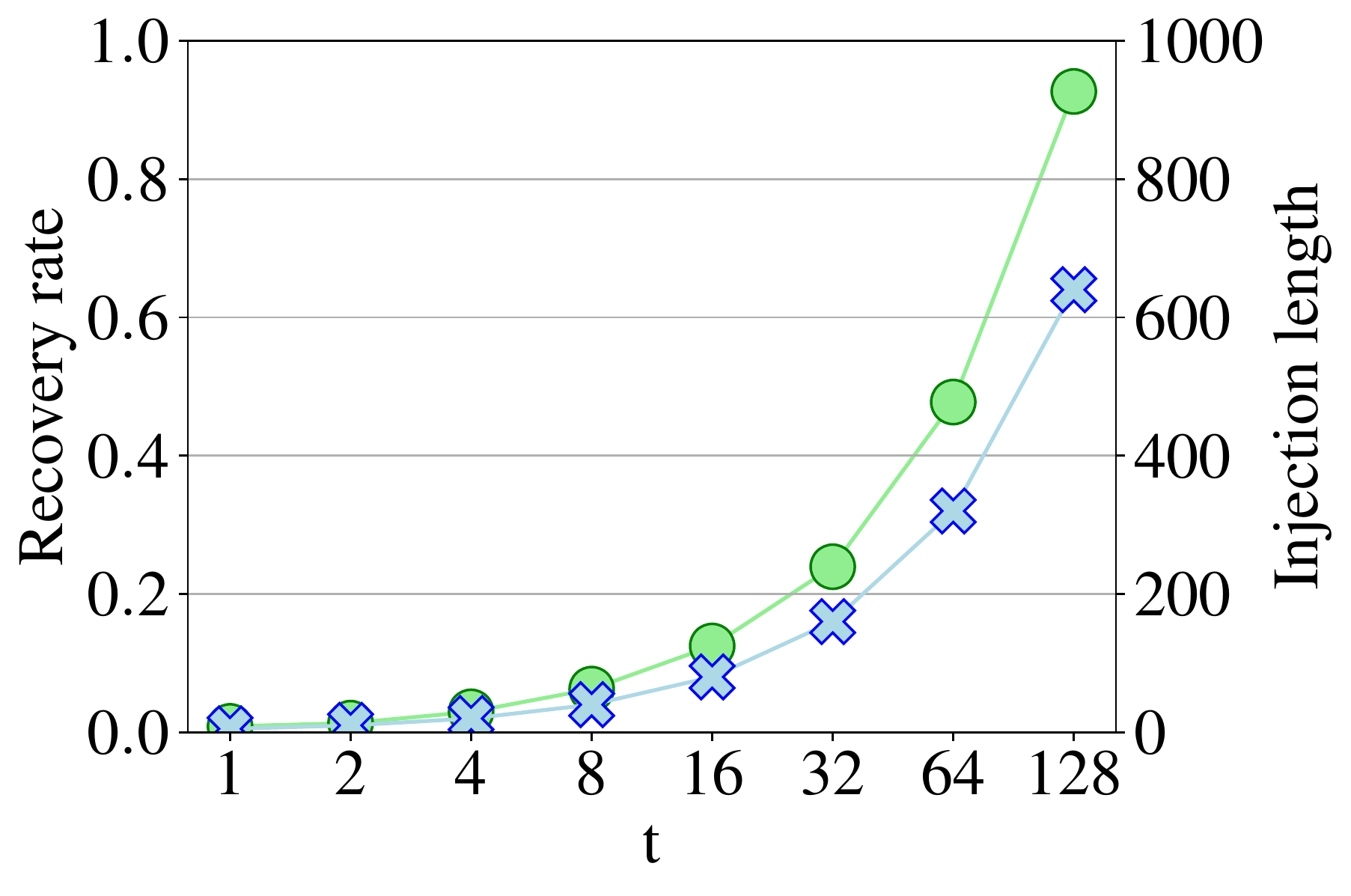}
			\label{ABVIAT}
		\end{minipage}
	}
	\caption{Optimization against dynamic (post-injection) padding (ShieldDB).}
	\label{ABVIA}
    \end{figure}
    
    The adversary can use the observe-inject-observe approach to guess the cluster for $w$.  
    First, one can inject a file containing all keywords and wait for a time period $>$ \textit{Tthreshold} to ensure that the injection file has been uploaded to the server. 
    It ensures that all keywords have been included in the server, and the follow-up uploads (and injections) follow \textit{Case 2}. 
    Then given the keyword $w$, to guess its cluster, it first observes the volume of all queries, sends a injection file containing only $w$ to the user, waits for a time period $>$ \textit{Tthreshold}, and observes the volume again. 
    In this case, only the cluster containing $w$ will be uploaded to the server (in which none of the keywords in other clusters is updated). 
    This cluster with the ``changed" volume after injection is the one that $w$ belongs to. 
    We can see that the above also requires $O(\#W)$ injection length and attack rounds.
    %For example, to confirm the cluster of $w$, it first observes the volume of all queries, injects a file containing all other keywords except $w$, and later observes the volume again.  
    %{By checking the clusters before and after injection, it can easily identify a ``non-modified" cluster which associates with $w$.}  
    %It can learn the clustering information of all keywords through $O(\#W)$ {file injections and attack rounds}. 
    %
    The adversary can also properly guess the clusters by using extra prior knowledge.  
    Recall that ShieldDB uses a training dataset to cluster keywords in the real database.  
    The training dataset could not be completely confidential. 
    The adversary can leverage other public datasets with similar distributions (i.e., similar data attacks) to determine the cluster.   
    In the second step, the adversary should identify the cluster according to the response length of each query. 
    We present the accuracy of the ``identification" under various $\alpha$ in Figure \ref{ShieldDBLA}.  
    With the growth of $\alpha$, the number of clusters (i.e., $\#\mathbf{W}/\alpha$) gradually decreases, which increases the probability of correctly locating queries to the clusters. 
    %Accordingly, the accuracy naturally rises up to 100\%.  
    For example, we can achieve around $60\%$ accuracy when $\alpha=2$, and nearly $100\%$ when $\alpha\geq{32}$. 
    At last, we used the BVA and BVMA to recover the queries from the identified clusters. 

    We tested the query recovery rate and the related overhead under different parameters (see Figure \ref{ShieldDBRerBO}). 
    As the $\alpha$ increases, ShieldDB is more effective against VIAs but at the same time, it yields more overheads. 
    When $\alpha=2$, its total overheads incurred by the setup, query, and update phases are still $<80\%$.  
    But it cannot resist our attacks (with nearly $60\%$ recovery rate). 
    Whilst our attacks only achieve around $3\%$ recovery (under $\alpha=128$), the costs of the injection approach to about $100\%$, and the query bandwidth increases by $>150\%$.    
    More concretely, without padding, the server returns $1,150 KB$ of data \textit{per query} in Enron; after padding by ShieldDB (under $\alpha=128$), it responds (averagely) $2,920 KB$ of data (\textit{per query}), which is $>1.5\times$ extra bandwidth.% , which may not be practical enough in real scenarios.

    %The experiments reveal the importance of dynamic padding.  
    A traditional dynamic (encrypted) database usually leaks more information (such as the update pattern \cite{DBSYNC}) than a static one.   
    Dynamic padding can mitigate the leakage and resist our attacks to some extent by sacrificing the bandwidth overheads (e.g., $>150\%$ query bandwidth).  
    It is an open problem to design a cost-effective countermeasure to VIAs. 
    \smallskip
    
    \noindent \textbf{Optimization against ShieldDB.}Clustering-based padding pads the keywords in the same cluster to make them indistinguishable. 
    Most current strategies are deterministic, which means that they fill the keywords to a \textit{fixed} \textit{rlp}. 
    We can leverage this to develop an optimization on our attacks (see Appendix \ref{AppenOA}). %The adversary may exploit this. % This may be exploited by the adversary.  
    Upon injection, we can use ``multi-group" coding injections and keep the keywords in the cluster to have the same \textit{rlp} so as to {circumvent the padding.}     
    %We present the optimization in .  

    In Figure \ref{ABVIA}, we show the recovery rate and injection length of the optimization against ShieldDB in Enron under different parameters $(\alpha, t)$, where $\alpha$ is the parameter of ShieldD and $t (t\leq\alpha)$ is the number of groups of keywords.
    The $\alpha$ now has little influence on the recovery rate, in particular, when $\alpha=128$, the recovery can still remain $>90\%$.  
    To mitigate the impact of padding, we set the number of keyword groups ($t$) to be the same as that of clusters ($\alpha$) in ShieldDB. 
    %The no. of injection files jumps rapidly with the increase of $\alpha$ (650 files, when $\alpha=128$). 
    We also see that the injection length increases steadily, from around 20 to 200, and finally up to around 600. 
    Figure \ref{ABVIAA} and \ref{ABVIAT} indicate that by increasing the number of injected files (i.e., enhancing the attack ability), we can still counter the dynamic padding.  
    %We specifically confirm the above statement in Fig. \ref{ABVIAT}, for example, see the recovery rate when $\alpha=128$.  
    The results show that ``more" injections can definitely lead to a ``higher" recovery.
    % {\color{blue}Note that our optimization algorithm can mitigate or even circumvent the impact of ShieldDB dynamic padding. Therefore, even in Lucene or Wikipedia, we can infer that the optimized attack still achieve good performance through more injections.}
    
    %Specifically, when $t=32$, less than 200 files can cause a recovery rate of more than 20\%, which cannot be ignored.
    
    %In conclusion, the above experiments show that a deterministic padding is not a proper countermeasure to our attacks, although it can increase the adversary's attack cost;  
    %while the probabilistic padding strategies \cite{DBLP:conf/ccs/PatelPYY19,DBLP:journals/popets/AgarwalHKM19,DBSYNC} may be a better idea. In this case, the adversary cannot know the exact padding files and thus can resist VIAs. Next, we will explain other mitigation strategies and discuss potential effective solutions.
    \smallskip
    % On the other hand, by setting a certain error item \cite{TwoLayer}, a passive adversary may still achieve an effective recovery rate (around $20\%$ \cite{DBLP:conf/uss/OyaK21}).

\begin{figure}[!t]
	\centering

	\includegraphics[width=.5\linewidth]{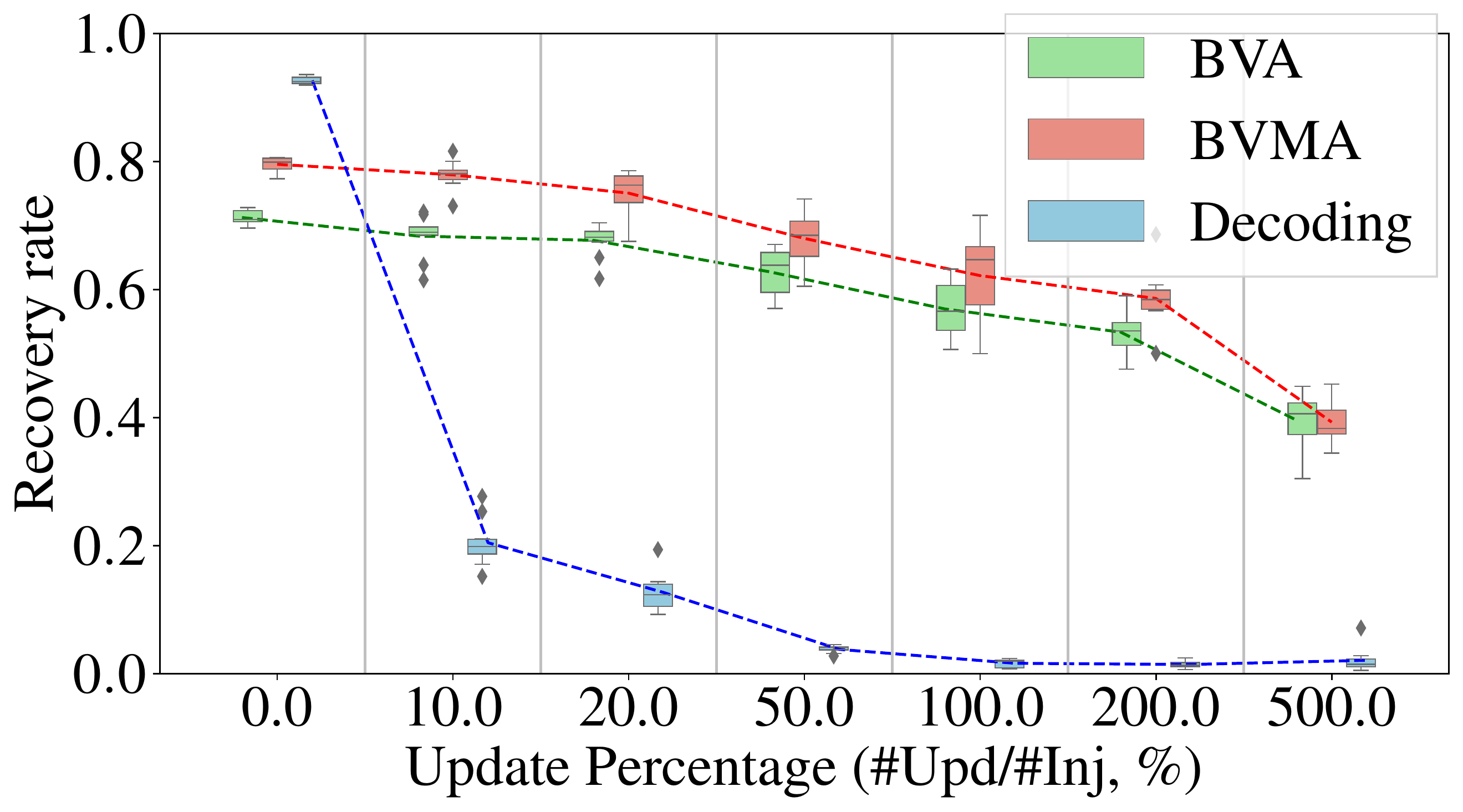}
	\caption{Performance under client's active updates.}
	\label{UpdateFigure}
\end{figure}

\subsection{Injections with Client Active Update}\label{AttUpd}

%A more reasonable scenario for VIAs is that clients may actively update databases, which can affect the recovery rate. In this subsection, 
We evaluated our attacks performance under the client's active updates in Enron.   
We used half of the files in Enron (about 15k) as the real database and the rest as the auxiliary database for the ``add" file operations (which means that each newly added file is from this auxiliary database).  
We assume that the client can randomly select an operation from $[add, delete]$ for each update. 
For an $add$ operation, the client randomly selects a file from the auxiliary database and further creates the keyword-file indexes; 
as for a $delete$, it deletes a file and the keyword-file pairs randomly from the real database. 
We present the results in Fig. \ref{UpdateFigure}, in which the $x$-axis is the ratio $(\%)$ between the number of updates and the number of (adversary) injections, and the $y$-axis is the recovery rate. 

%More injections may cause more client updates or occur during frequent client updates [cannot see from the figure]. 
%Therefore, evaluating the recovery rate of different attacks under the different update percentages can better show the effect of these attacks [don't understand]. 
%Specifically, 

The updates are equivalent to ``adding noise" in the adversary's view, which can ``blur" the observations.  
We see that the decoding attack slides into the biggest drop after the client commits $10\%$ update, for example, its recovery rate falls sharply from $90\%$ to $20\%$; and after $50\%$ update, the recovery is below $5\%$.  
In contrast, as the ratio increases, our recovery rates undergo a downward drift. 
Our attacks still remain $>60\%$ recovery after $50\%$ update. 
Even in the worst case ($500\%$ update), our performances still stand at $40\%$. %, which is still a considerable threat to DSSE.

We note the experiment did not include the single-round attack. 
This is because the attack requires an extreme amount of injected files, $O(\#{W^{2}})$. 
More concretely, it could need 9 million injection emails, which is seriously impractical for Enron. 
We conclude that the client's updates can easily depress the recovery rate of prior attacks, but they have limited impact on the BVA and BVMA. 

For frequent file updates, {e.g., $1,500$ file updates (i.e. $10\%$ files in the Enron dataset)}, when these file updates (covering almost all keywords) happen between baseline phase and recovery phase of VIAs, our attacks without optimisation ($O(\log\#W)$ injection length) and decoding attack ($O(\#W)$ injection length) could easily fail.
The single-round attack could survive only if we set its constant $m$ to be considerably large (e.g., m $\gg$ the number of updated files). 
Recall that its injection length is $O(m\#W)$. 
Setting a large $m$ could make the single-round attack more impractical w.r.t. injection cost. In Appendix \ref{Section AFU}, we present a \textit{modified} algorithm with $O(\log\#W)$ files to deal with users' frequent updates.

\section{Conclusion}\label{Section Conclusion}
We proposed new attacks against dynamic SSE using a binary volumetric injection strategy. 
Unlike existing attacks which require a considerable amount of prior knowledge (e.g., LAA \cite{DBLP:conf/ndss/IslamKK12,DBLP:conf/ccs/CashGPR15}) and injections (e.g., \cite{DBLP:conf/ndss/BlackstoneKM20, DBLP:conf/eurosp/PoddarWLP20}), our attacks can offer a high recovery rate with fewer injections and pose a non-trivial threat to current defenses (e.g., padding). 
%We explored our attacks by injecting files in a binary manner only exploiting volume pattern. 
%Especially with our last attack, which combines the information making it more stronger against most of schemes.
%
%
We provided empirical evidence to confirm the performance of our attacks.  
%The experiments show that our attacks outperform others, by achieving a relatively high recovery rate ($80\%$) in the mail datasets while reducing huge amount of injection size compared with others (e.g., \cite{DBLP:conf/ndss/BlackstoneKM20,DBLP:conf/eurosp/PoddarWLP20}). 
%Testing with the popular defenses, we concluded that our attacks still maintain acceptable effect.
%We hope this work could inspire more efforts for practical and secure DSSE systems.  